\documentclass[journal ]{new_aiaa}
\usepackage[utf8]{inputenc}
\usepackage{textcomp}

\usepackage{graphicx}
\usepackage{amsmath}
\usepackage[version=4]{mhchem}
\usepackage{siunitx}
\usepackage{longtable,tabularx}
\usepackage{bm}
\usepackage[final]{changes}

\title{Understanding the instability-wave selectivity of hypersonic compression ramp laminar flow}

\author{Peixu Guo\footnote{Research Assistant Professor, Department of Aeronautical and Aviation Engineering,\added{ 11 Yuk Choi Road, Hung Hom, Kowloon, Hong Kong,} peixu.guo@outlook.com.}, Jiaao Hao\footnote{Assistant Professor, Department of Aeronautical and Aviation Engineering,\added{ 11 Yuk Choi Road, Hung Hom, Kowloon, Hong Kong,} jiaao.hao@polyu.edu.hk.} and Chih-Yung Wen\footnote{Chair Professor, Department of Aeronautical and Aviation Engineering,\added{ 11 Yuk Choi Road, Hung Hom, Kowloon, Hong Kong,} and Associate Fellow AIAA.}}
\affil{The Hong Kong Polytechnic University, Kowloon, Hong Kong SAR}

\begin{document}

\maketitle

\begin{abstract}
The hypersonic flow stability over a two-dimensional compression corner is studied using \replaced{Resolvent}{resolvent} analysis, linear stability theory (LST) and parabolised stability equation (PSE). The authors find that the interaction between upstream convective-type disturbances and the laminar separation bubble can be divided into two regimes\deleted{, whose behaviour can be well explained by comparative research}. First, two-dimensional (2-D) high-frequency Mack (second, third, ...) modes neutrally oscillate with the presence of alternating stable and unstable regions inside the separation bubble\replaced{, which arises from}{. These discontinuous unstable regions are generated by} repeated synchronisations between discrete modes with evolving branches. \replaced{Meanwhile, }{Through a modal synchronization analysis, we report that }the second modes upstream and downstream of the separation bubble can be differ significantly from each other\deleted{, since they originate from different branches of discrete modes due to flow separation. Favourable agreement between LST, PSE and resolvent analysis is reached}. Second, the 2-D low-frequency `shear-layer mode' is \replaced{stable}{found to be stable in the separation bubble by LST}, whereas multiple unstable three-dimensional (3-D) eigenmodes are identified by LST. \deleted{In general, three significant modes are dominant successively near the separation point, in the separation bubble and near the reattachment point.} These modes are found to be sensitive to the streamline curvature effect. The locally dominant modes agree with the \replaced{Resolvent}{resolvent} response in terms of \added{the preferential spanwise wavenumber, }the disturbance shape and the growth rate of energy. \deleted{The preferential spanwise wavenumber in the resolvent analysis, i.e., the wavenumber of the most amplified oblique wave, falls well in the unstable wavenumber range of LST.} Thus, a combination of global and local analyses demonstrates that the separation bubble tends to selectively amplify low-frequency 3-D disturbances and `freeze' high-frequency Mack-mode disturbances in an explainable manner. These findings facilitate the understanding of the early evolution of low- and high-frequency instabilities in hypersonic separated flows. 
\end{abstract}
\newpage
\section*{Nomenclature}

{\renewcommand\arraystretch{1.0}
\noindent\begin{longtable*}{@{}l @{\quad=\quad} l@{}}
$(x,y,z)$ & Cartesian coordinates in streamwise, wall-normal and spanwise directions\\
$(\xi,\eta,z)$ & orthogonal body-fitted coordinates\\
$(u,v,w)$ & velocity components in Cartesian coordinate system\\
$\rho$ & density\\
$T$ & temperature\\
$p$ & pressure\\
$M$ & Mach number\\
$Re$ & Reynolds number\\
$\bm{\phi}$ & vector of primitive variables\\
$\bm{\psi}$ & vector of modal shape function for primitive variables\\
$\alpha$ & complex streamwise wavenumber\\
$\beta$ & spanwise wavenumber\\
$\omega$ & angular frequency\\
$f$ & physical frequency\\
$K$ & streamline curvature\\
$\tilde{c}$ & complex phase velocity, $\tilde{c}=\omega/\alpha$\\
$\sigma$ & local growth rate\\

\multicolumn{2}{@{}l}{Subscripts}\\
$\infty$ & freestream quantity\\
$w$ & wall quantity\\
$r$	& real part of complex number, or reattachment position\\
$i$	& imaginary part of complex number\\
0 & initial position\\
\multicolumn{2}{@{}l}{Superscripts}\\
$\ast$ & dimensional quantity\\
$'$ & fluctuating quantity
\end{longtable*}}

\section{Introduction}
\lettrine{T}{he} shock-wave/boundary-layer interaction (SWBLI) is of both fundamental and engineering importance, which can induce boundary-layer separation and significant unsteadiness. Over simple two-dimensional (2-D) configurations such as a compression corner, the laminar separation bubble can support an absolute or convective instability depending on the flow and geometric conditions, where the separation bubble is called an oscillator or a noise amplifier, respectively. Experimental and numerical efforts have demonstrated that these instabilities can trigger the laminar-turbulent transition and severe aerodynamic heating  \cite{roghelia2017experimental,dwivedi2022oblique,Lugrin2021,benitez2024separation}. Therefore, it is of interest to investigate how instability waves interact with the separation bubble.

With regard to the convective mechanism, the recently popular \replaced{Resolvent}{resolvent} analysis provides a framework to evaluate responses (outputs) of absolutely stable dynamical systems to time-periodic external forcings (inputs). In general, by solving an optimisation problem transformed from the linearised Navier--Stokes equation (LNSE), the \replaced{Resolvent}{resolvent} analysis seeks the small-amplitude harmonic response which undergoes the strongest energy growth. The identified harmonic response is also called the optimal disturbance. For configurations such as the 2-D compression ramp \cite{hao2023response,cao2023} and the double wedge \cite{dwivedi2022oblique}, a stationary streak may be the optimal disturbance and closely related to the reattachment streak observed under certain experimental conditions. However, recent experiments on hypersonic SWBLI flows \cite{butler2021interaction,benitez2023measurements} reported the significance of travelling waves by measurements such as surface pressure sensors and high-speed Schlieren. Coexistence of high-frequency instability waves with hundreds of kilohertz as well as low-frequency disturbances with tens of kilohertz was detected. The two dominant types of travelling instability waves were believed to be associated with Mack modes and the `shear-layer' mode, respectively. Combinations of computational and experimental studies on hypersonic flows over a cone-cylinder-flare configuration also reported that there were at least two fundamental convective-type instability mechanisms in the recirculation bubble \cite{Caillaud2024Separation,benitez2024separation}. The detected dominant signals of two different natures were found to have resonant-type structures similar to the higher-order Mack modes and the shear-induced first mode, respectively. It is inferred that the travelling wave may overtake the stationary streak in experimental conditions for the sake of a larger upstream amplitude considering the incoming disturbance spectrum, surface roughness and receptivity. It merits an in-depth exploration on how these two different types of instabilities are affected by a sudden presence of flow separation and selected by the separation bubble in the SWBLI problem.

In terms of the nature of Mack modes, it is more extensively investigated and better understood in a hypersonic boundary layer without flow separation. The classical Tollmien--Schlichting mode appear in low-speed boundary layers, and extend to be the first mode in supersonic boundary layers. As Mach number is further increased, a generalised inflection point appears for a supersonic flat-plate boundary layer, which establishes the sufficient condition for the occurrence of inviscid instabilities \cite{lees1946investigation}. As a consequence, Mack found by local analysis that multiple unstable modes appear in hypersonic boundary layers, among which the 2-D second mode mostly has the largest growth rate \cite{mack1984boundary}. The second mode was subsequently understood as trapped acoustic waves between the wall and the relative sonic line \cite{fedorov2011transition}. Kuehl \cite{kuehl2018thermoacoustic} and the authors \cite{tian2021growth,chen2023unified} provided further insights into how the second mode is amplified from the energy analysis and the resonant phase analysis, respectively. In hypersonic boundary layers, Mack modes with different frequencies can be successively amplified in different streamwise locations due to the feature of the unstable region. However, the observation that Mack modes with a broad range of frequency nearly encounter a neutral state in a separation bubble remains poorly understood. Although the second mode was reported to be nearly neutral in the separation bubble by the stability analysis of Balakumar et al. \cite{balakumar2005stability}, the stability diagram and modal synchronisation were not detailed, and the reason for the neutral state remained unclear. With regard to the `shear-layer mode', a comprehensive eigenmodal analysis and a comparison with other theoretical tools are currently lacked and needed. Most existing studies ascribed the growth of the `shear-layer mode' to the Kelvin--Helmholtz (K--H) instability mechanism, since energy growth is mainly concentrated along the separated shear layer. However, \replaced{one}{it} must be careful before making a judgment on the relevance between the `shear-layer mode' and the conventional K--H mode. This is partly because the compressibility in a hypersonic separated flow can significantly suppress the classical 2-D K--H mode which is active in incompressible free shear layers \cite{Karimi2015thesis}. In high-speed flows, the shear-induced unstable modes may tend to appear in the form of three-dimensional (3-D) travelling waves \cite{Karimi2015thesis}. Caillaud et al. \cite{Caillaud2024Separation} utilised an energy budget analysis to show that the `shear-layer mode' is pronounced along the shear layer and behaves similarly to the first mode. An in-depth comparative research is required to clarify the growth mechanism of the `shear-layer mode'.

Currently, it is not yet well understood how the hypersonic compression ramp flow selects the upstream low-frequency or high-frequency disturbances from a stability analysis perspective. There are two questions which merit considerations: 1) what is the eigenmodal interpretation of the neutral Mack modes in the separation bubble? 2) what is the growth mechanism of the observed `shear-layer mode'? To address the concerns above, this paper uses the \replaced{Resolvent}{resolvent} analysis to seek the most amplified response of the compression ramp flow to external forcings. Furthermore, linear stability theory (LST) and parabolised stability equation (PSE) are employed to clarify the eigenmodal mechanism. A combination of global and local analyses would facilitate a more solid investigation on the proposed fundamental problem.

\section{Problem descriptions}
A compression corner configuration is considered in this work, which has a flat plate of length $L^{\ast}=100$ mm with a sharp leading edge and an adjacent ramp of length 80 mm. The asterisk denotes dimensional quantities. The ramp angle is set to 12$^\circ$, where no global instability exists in the laminar base flow under the investigated conditions \cite{hao2023response}. The freestream conditions of the hypersonic Aachen Shock Tunnel TH2 \cite{roghelia2017experimental} are considered and given as follows: Mach number $M_{\infty} =7.7$, static temperature $T^{\ast}_{\infty} = 125$ K and unit Reynolds number $Re^{\ast}_{\infty} =4.2\times10^6$  m$^{-1}$.  It should be noted that this work does not intend to reproduce the results and interpret the flow mechanisms under the same experimental configuration. The instability and transition of the experimental flow with a ramp angle of 15$^\circ$, proved to support the onset of absolute instabilities, have been explained by some of our previous studies \cite{hao2021occurrence,cao2023}. In the present work, instead, another ramp angle is selected to avoid the presence of absolute instabilities. The purpose is to examine the interaction behavior between the separation bubble and upstream convective-type instability waves from fundamental viewpoints. The scope of this paper is to address the questions raised in the introduction, i.e., the selectivity mechanism of instability waves.

\begin{figure*}
	\centering{ \includegraphics[height=4cm]{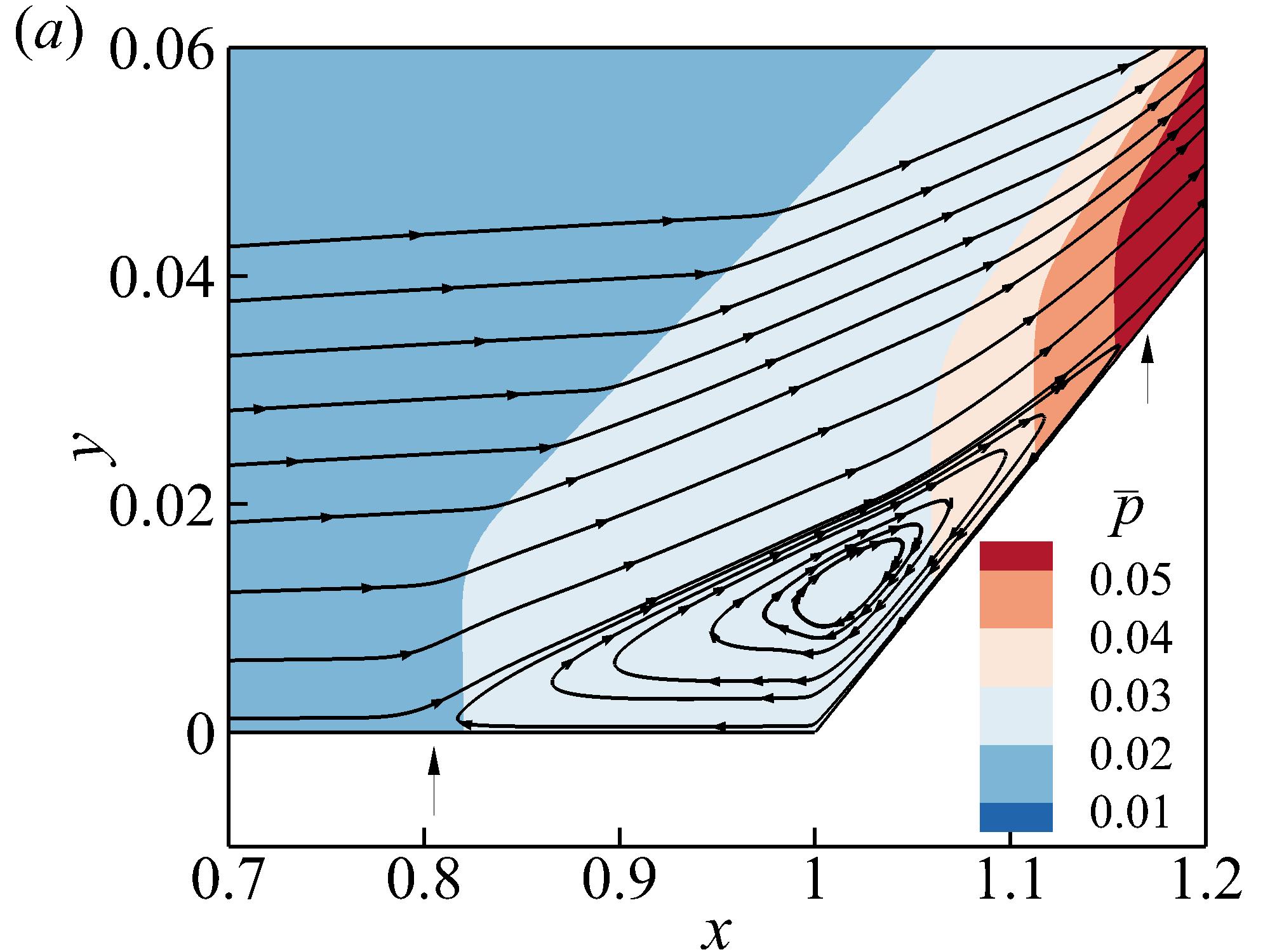}}
	\centering{ \includegraphics[height=4cm]{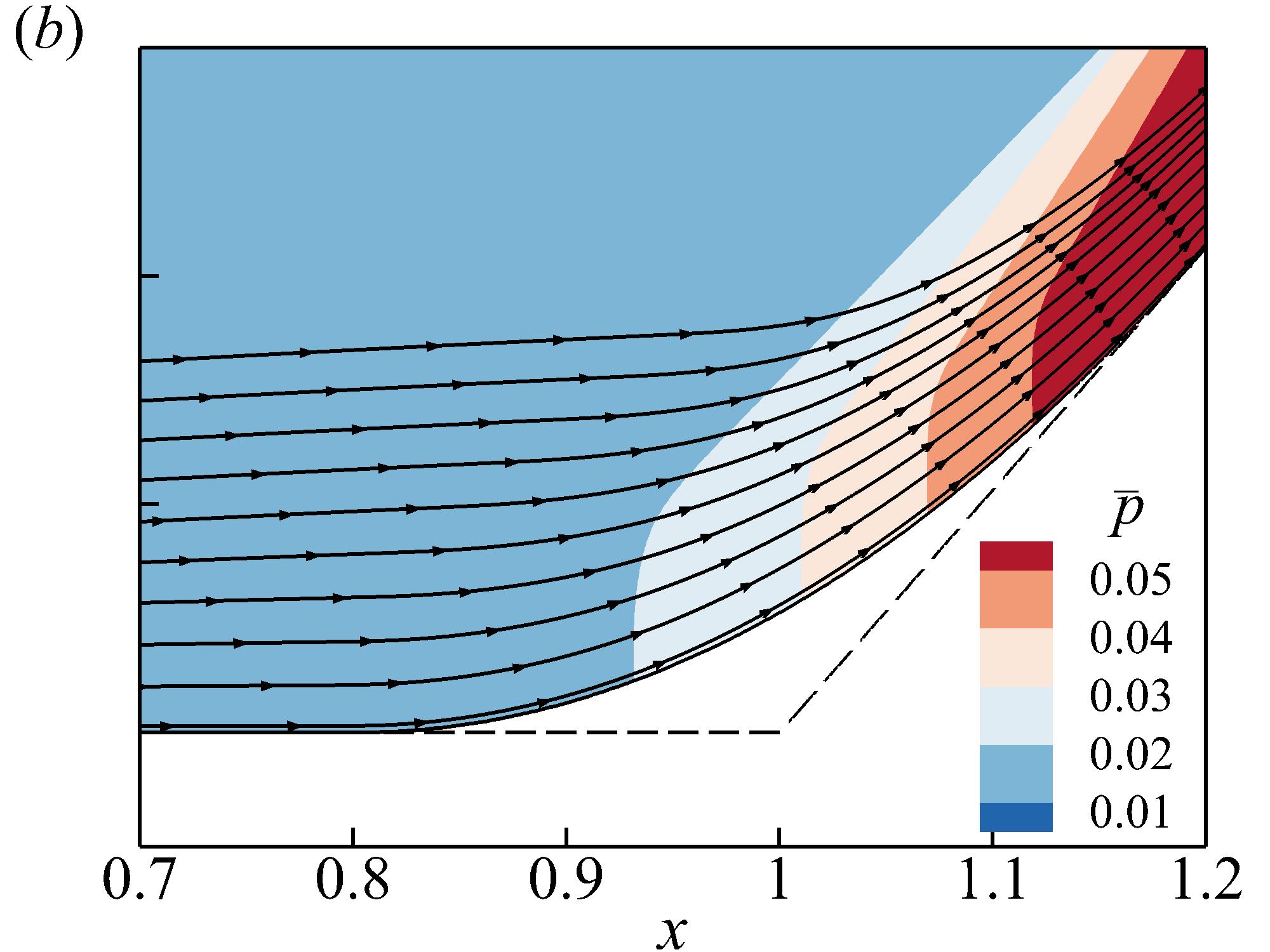}}
	\centering{ \includegraphics[height=4cm]{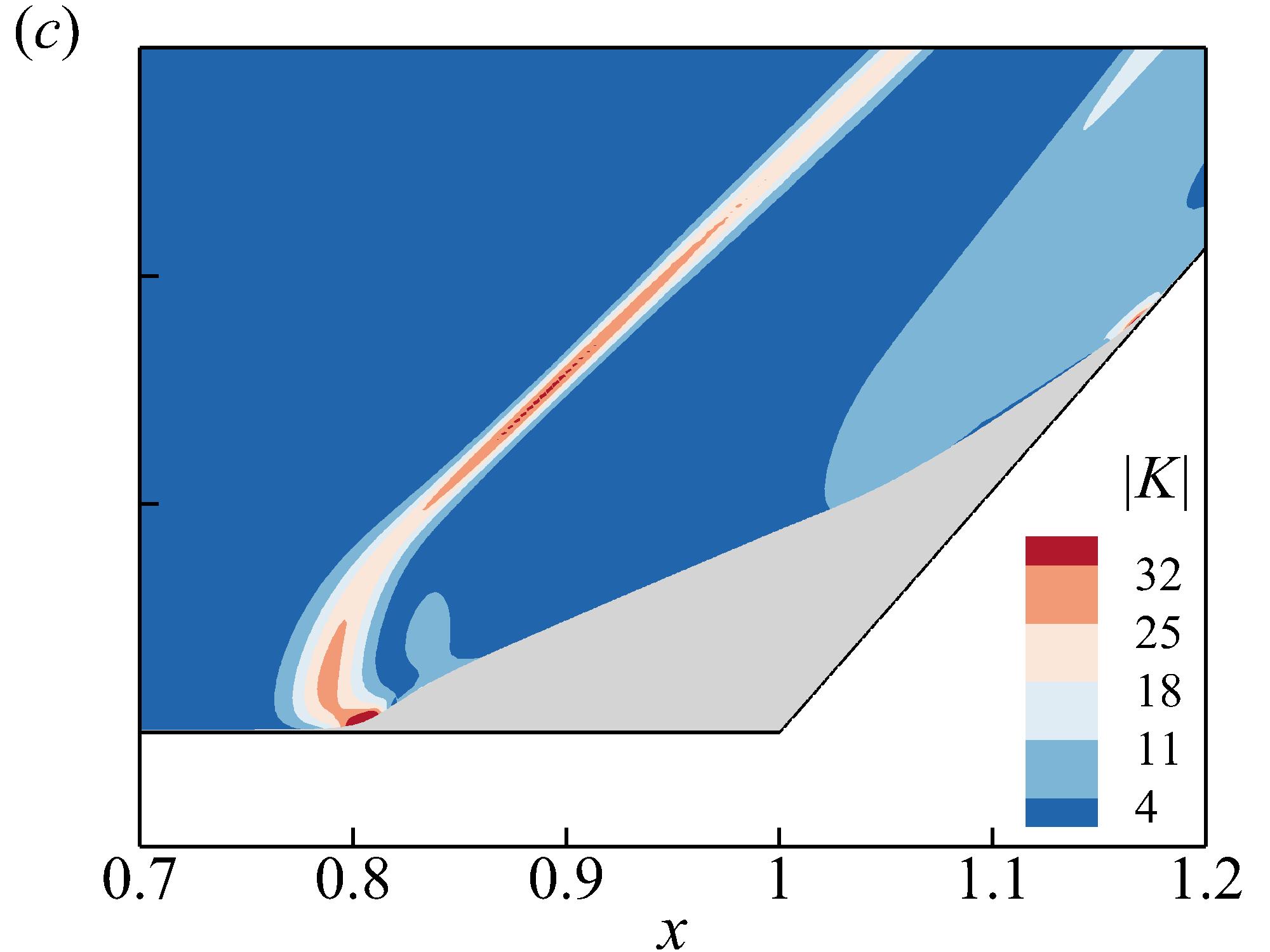}}
	\caption{Streamline and contour of the base pressure for cases (\textit{a}) without corner rounding and (\textit{b}) with rounding, and (\textit{c}) distribution of local curvature magnitude $|K|$ above the dividing line without rounding. Arrows in  (\textit{a}), separation and reattachment points.}
	\label{fig1}
\end{figure*}

A Cartesian coordinate system ($x$, $y$) is constructed with the origin at the leading edge, corresponding to the streamwise and wall-normal velocities ($u$, $v$). An orthogonal body-fitted coordinate system ($\xi$, $\eta$) is also defined, which is along the wall-tangent and normal directions, respectively. The surface temperature is assumed to be room temperature $T^{\ast}_w =293$ K. In this paper, the primitive variables are nondimensionalised by the corresponding freestream quantities except that the pressure $p$ is by the freestream $\rho^{\ast}_{\infty } {u}^{\ast 2}_{\infty }$, where $\rho$ represents density. The reference length scale for nondimensionalisation is $L^{\ast}$. The two-dimensional laminar base flow, obtained by our in-house Navier--Stokes (N--S) solver \cite{guo2023interaction,hao2020hypersonic,hao2023response}, is used for stability analyses.

The laminar flow field will be used for the present stability analysis. It should be noted that a transitional SWBLI subject to incoming disturbances may encounter the base flow distortion and the change of the bubble topology \cite{marxen2011effect,Lugrin2021}. As a result, a stability analysis of the laminar SWBLI flow is aimed to provide insights into how instability waves interact with the bubble at a low Reynolds number or in an early linear stage of the instability. However, the result will not be sufficient to account for the complete transition mechanism of the SWBLI problem. Fig. \ref{fig1}(\textit{a}) shows the streamline and the pressure field of the converged base flow. Based on the skin friction, the separation and reattachment points are located at $x_s = 0.805$  and $x_r = 1.17$, respectively. Between them, a recirculation separation bubble is formed. To isolate the effect of the separation bubble on instability waves, corner rounding with a radius of $r=1.9$ is also employed \cite{cao2023,hao2023response}. As shown in Fig. \ref{fig1}(\textit{b}), the separation is eliminated by corner rounding, while the effects of curvature and adverse pressure gradient are maintained as much as possible.

\section{Stability analysis tools}\label{sec:3}
\subsection{Resolvent analysis}\label{sec:3.1}

The present \replaced{Resolvent}{resolvent} analysis seeks the strongest response to an external forcing, which addresses the linear dynamics of the noise amplifier. The shapes of the forcing and the response are assumed to be on the $x$--$y$ plane. Starting from the LNSE for compressible flows, the \replaced{Resolvent}{resolvent} analysis gives evolutions of optimal disturbances as accurately as direct numerical simulation (DNS) in the linear stage \cite{guo2023interaction}. In detail, the disturbance vectors of both the conservative variable ${\bm{q}}=(\rho,\rho u,\rho v,\rho w,\rho e)^\text{T}$ and the forcing ${\bm{f}}=(f_1,f_2,f_3,f_4,f_5)^\text{T}$ are assumed to be in the form of  ${\bm{q}'}\left( {x,y,z,t} \right) = {\hat{\bm{q}}}\left( {x,y} \right)\exp \left( {{\rm{i}}\beta z - {\rm{i}}\omega t} \right) + {\rm{c}}{\rm{.c}}.$ Here, $w$, $e$ and the superscript `T' denote the spanwise velocity, the total energy per unit mass and the matrix transpose, respectively. The symbols $\omega$, $\beta$, $z$, $t$ and c.c. represent the circular frequency, the spanwise wavenumber, the spanwise coordinate, time and complex conjugate, respectively. The governing equation can be written as
\begin{equation}\label{eq1}
\partial {\bm{q}'}/\partial t = {\cal{A}}{\bm{q}'} + {\cal{B}}{\bm{f}'},
\end{equation}
where matrix ${\cal{A}}$ is the Jacobian matrix.\added{ To perform linearisation, a hand-derived continuous linear system rather than that by numerical discretisation is employed.} The matrix ${\cal{B}}$ is introduced to constrain ${\bm{f}'}$ at a streamwise station $x=x_i$, which performs as a Dirac delta function in the streamwise direction. One may also impose the forcing in the whole computational domain. In this work, we consider a constrained optimalisation problem without modifying the non-modality of the matrix ${\cal{A}}$. The employment of a localized forcing near the leading edge may be considered as an analogy to the situations frequently considered in high-fidelity numerical simulations and convective-type stability analyses. In these situations, the convective instability waves are excited upstream locally via external perturbations and then convected in the flow direction. This setup, previously applied in our other works \cite{cao2023,hao2023response,guo2023interaction} and also by Dwivedi et al. \cite{dwivedi2022oblique}, also makes it convenient for comparative studies between numerical simulations and other theoretical tools. For convective instabilities, \added{deep }connections\added{ exist} between `global' stability analysis of the entire flow field and `local' analysis of the local flow profile \deleted{can be thereby observed} \cite{huerre1990local}. The evolution of upstream instability waves in the streamwise direction will be the main concerned point. 

Substituting the form of ${\bm{q}'}$ and ${\bm{f}'}$ into (\ref{eq1}) yields
\begin{equation}\label{eq2}
{\bm{\hat q}} = {\cal R}{\cal B}{\bm{\hat f}},
\end{equation}
where
\begin{equation}
{\cal R} = {\left( { - {\rm{i}}\omega {\cal I} - {\cal A}} \right)^{ - 1}}.
\end{equation}
Here, $\cal I$ is the identity matrix. To seek the forcing and the response that maximise the energy amplification, an optimal gain $G$ is defined by
\begin{equation}
{G}^2\left( {\beta ,\omega } \right) = 
\mathop {\max }\limits_{{\bm{\hat f}}} \left\{ {{{{{\left\| {{\hat{\bm{q}}}} \right\|}_E}} \mathord{\left/{\vphantom {{{{\left\| {{\hat{\bm{q}}}}\right\|}_E}} {{{\left\| {{\cal B}{\hat{\bm{ f}}}} \right\|}_E}}}} \right.			\kern-\nulldelimiterspace} {{{\left\| {{\cal B}{\hat {\bm{f}}}} \right\|}_E}}}} \right\}.
\end{equation}
Here, ${\left\|  \cdot  \right\|_E}$ represents the Chu's energy norm defined in Refs. \cite{hao2023response} and \cite{guo2023interaction}. Subsequently, Eq. (\ref{eq2}) is transformed into an eigenvalue problem with respect to $G^2$. The eigenvalue problem is solved by ARPACK for a given $\beta$ and $\omega$, and the resulting eigenfunction gives the shape of the optimal forcing ${\hat {\bm{f}}}$. The shape of the optimal response is then calculated by Eq. (\ref{eq2}). In terms of the discretisation of $\cal A$, the modified Steger--Warming scheme \cite{maccormack2014numerical} is used for
inviscid fluxes near discontinuities detected by a modified Ducros shock sensor \cite{hendrickson2018improved}, and the second-order central scheme is used for
inviscid fluxes in smooth regions as well as viscous fluxes. Two grid resolutions $1000 \times 300$ and $1200 \times 350$ are tested for the compression ramp flow, which shows convergence of instability evolutions. The coarse mesh has been priorly verified by Ref. \cite{hao2023response} for the considered flow, and the finer mesh is finally adopted in this paper. 

\subsection{LST and PSE}\label{sec:3.2}

In addition to the `global' stability analysis, a parallel-flow LST starts from the `local' perspective to examine the local amplification of disturbances, which may constitute the global amplification in the global analysis \cite{huerre1990local}. Although the separation bubble seems to be nonparallel, the rationale for a locally parallel analysis has been clarified by Diwan \& Ramesh \cite{diwan2012relevance}. To further consider the nonparallelism, PSE is also included and provided here first \cite{herbert1997parabolized}. Assume the disturbance form to be
\begin{equation}
\bm{\phi}' = \bm{\psi}(\xi,\eta)\exp{\left[ {{\rm{i}}\left( {\int_{{\xi_0}}^{\xi} {\alpha (\check\xi)d\check{\xi}} +\beta z - \omega t} \right)} \right]}+ {\rm{c}}{\rm{.c}}{\rm{.}}
\end{equation}
Here, $\alpha$ is the complex wavenumber, $\xi_0$ corresponds to the initial point of marching, and $\bm{\psi}=(\hat{\rho}, \hat{u},\hat{v},\hat{w},\hat{T}$)$^\text{T}$ is the modal shape. The symbol $T$ represents temperature. The PSE can be expressed by
\begin{equation}
\left( {{{\cal L}_0} + {{\cal L}_1}} \right)\bm{\psi}  + {{\cal L}_2}\frac{{\partial \bm{\psi} }}{{\partial \xi}} + \frac{{d\alpha }}{{d\xi}}{{\cal L}_3}\bm{\psi}  = \bm{0},
\end{equation}
where ${\cal L}_0$ is the operator for parallel flows, and ${\cal L}_1$--${\cal L}_3$ represent the operators arising from nonparallel effects. The forms of ${\cal L}_0$--${\cal L}_3$ for a Cartesian coordinate system have been given by Paredes \cite{gonzalez2014advances}, and the transformation method into an orthogonal body-fitted coordinate system has been detailed by Chang et al. \cite{chang1993linear}. Solution is obtained via streamwise marching from a local profile that will be introduced later. With the assumption that the modal shape is streamwise slowly varying, an iterative scheme to update the wavenumber is given by
\begin{equation}\label{eq_normalization}
{\alpha_{{\textit{new}}}} = {\alpha_{{\textit{old}}}} - {\rm{i}}\frac{1}{E}\int_0^\infty  {\bar \rho \left( {{{\hat u}^\dag }\frac{{\partial \hat u}}{{\partial \xi }} + {{\hat v}^\dag }\frac{{\partial \hat v}}{{\partial \xi }} + {{\hat w}^\dag }\frac{{\partial \hat w}}{{\partial \xi }}} \right) d\eta },
\end{equation}
where the overbar indicates the base-flow quantity, the superscript `$\dag$' denotes complex conjugate, and $E$ is given by
\begin{equation}
E = \int_0^\infty  {\bar \rho \left( {{{\left| {\hat u} \right|}^2} + {{\left| {\hat v} \right|}^2} + {{\left| {\hat w} \right|}^2}} \right)d\eta }.
\end{equation}

By dropping ${\cal L}_1$--${\cal L}_3$, the nonparallel effect is neglected and $\bm{\psi}$ becomes a local solution with respect to $\eta$. The remaining equation ${\cal L}_0\bm{\psi}=\bm{0}$ can be also written in the form of
\begin{equation}\label{LSTequation}
{\cal H}_{yy}\frac{{\partial^2\bm{\psi} }}{{\partial\eta^2}}+{\cal H}_{y}\frac{{\partial\bm{\psi} }}{{\partial\eta}}+{\cal H}_{0}\bm{\psi}  = \bm{0},
\end{equation}
where the boundary condition is given by
\begin{equation}
\left\{ \begin{array}{l}
\hat u = \hat v = \hat w = \hat T = 0,\eta = 0\\[2pt]
\hat u = \hat v = \hat w = \hat T = 0,\eta \to \infty. 
\end{array} \right.
\end{equation}
The local equation is finally transformed to a complex eigenvalue problem of LST with respect to $\alpha$ \cite{malik1990numerical}. The operators of Eq. (\ref{LSTequation}), detailed in Ref. \cite{ren2014competition}, are related to $\alpha$, $\beta$, the local base flow and the metric factor $h_{1}(\eta)$, where $h_1=1+K\eta$ and $K$ is the curvature. The streamline curvature $K(\xi,\eta)$ in the whole domain is calculated to obtain the local $h_1$. Contour of $|K|$ for the case without rounding is shown in Fig. \ref{fig1}(\textit{c}). By setting $K$ equal to zero or the actual value, one can observe the effect of streamline curvature. For LST, the local growth rate $\sigma=-\alpha_i$ is positive if the mode is unstable. The LST and PSE analyses are performed by our in-house code, which has been well validated by benchmark cases compared to theoretical, simulation and experimental results \cite{guo2020linear,guo2021sensitivity,guo2022heat,guo2023interaction,cao2023,chen2023unified}. In the LST analysis, a \replaced{Chebyshev spectral collocation method}{global numerical method} is employed to \added{directly }obtain the complete spectrum\added{,} and  \replaced{an iterative fourth-order compact difference scheme}{a local method} is to improve the results of eigenmodes \cite{malik1990numerical}.\added{ The compact difference scheme, evidently less sensitive to the grid point than the collocation method, receives the initial guess from the discrete eigenmode of the spectrum for subsequent iterations. With the current collocation point number $N_y = 221$, the Euclidean norm of the $\alpha$ difference between the two numerical schemes leads to an error within 0.5 \%.} The mesh for \replaced{Resolvent}{resolvent} analysis is also used for LST and PSE. Convergence of each eigenvalue and eigenvector is confirmed by comparing calculation results of different mesh resolutions. 

\section{Results}\label{sec:4}
\subsection{Optimal travelling waves}\label{sec:4.1}

\begin{figure*}
	\centering{ \includegraphics[height=3cm]{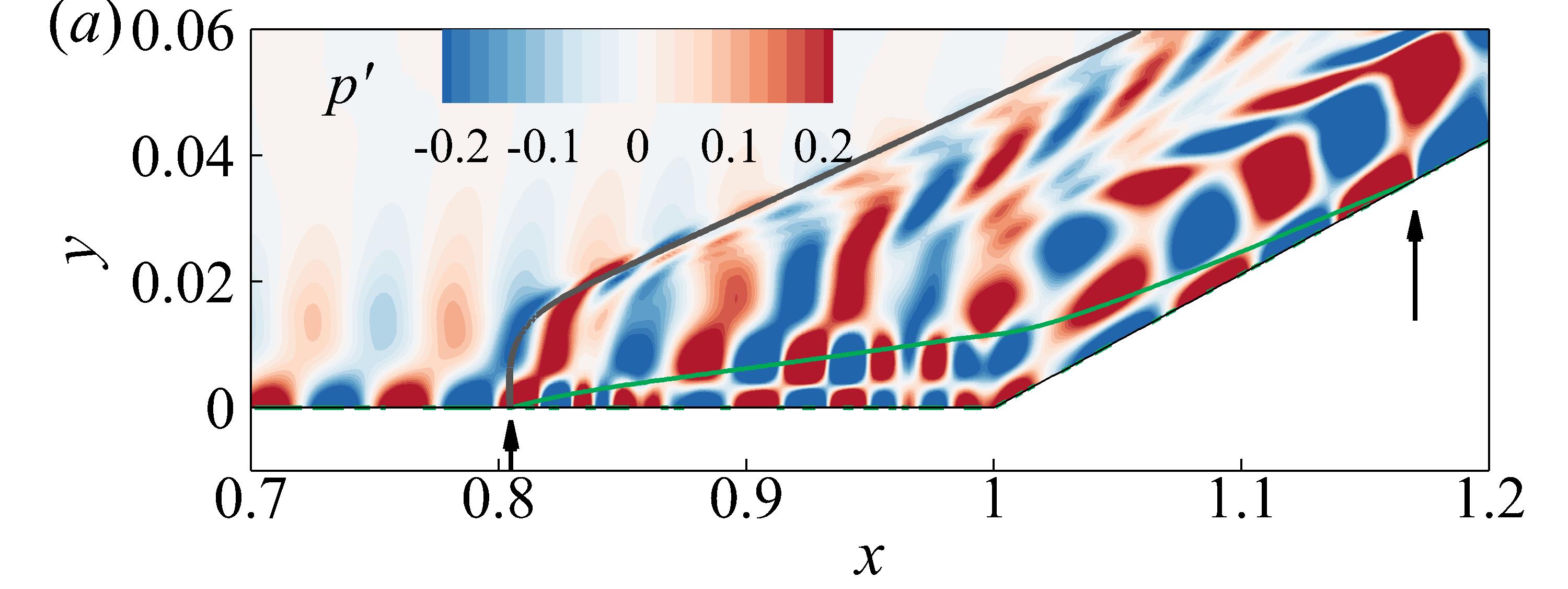}}
	\centering{ \includegraphics[height=3cm]{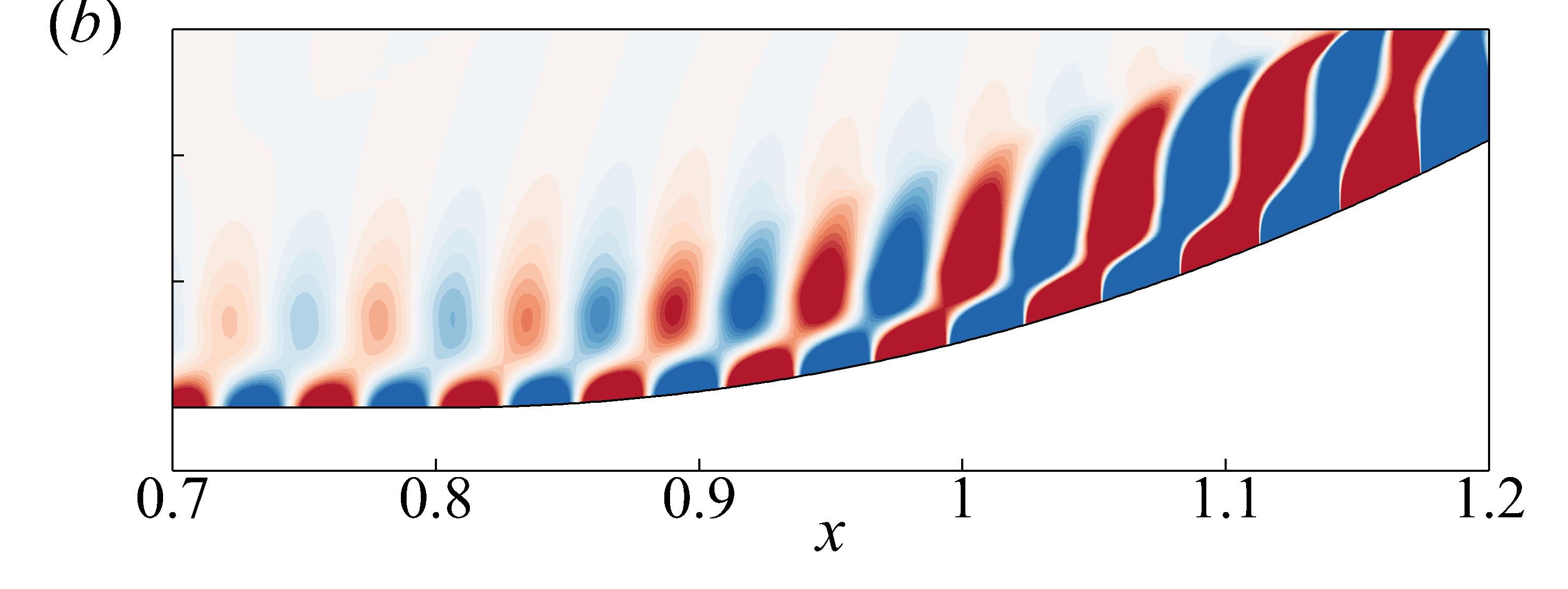}}
	\caption{Pressure fluctuation of the optimal disturbance $\omega=100$ for cases (\textit{a}) without and (\textit{b}) with corner rounding. Arrows in  (\textit{a}), separation and reattachment points. Gray line in (\textit{a}), the isoline $\bar{p} = 0.018$. Coloured line in (\textit{a}), dividing streamline.}
	\label{fig2}
\end{figure*}

\begin{figure*}
	\centering{ \includegraphics[height=3cm]{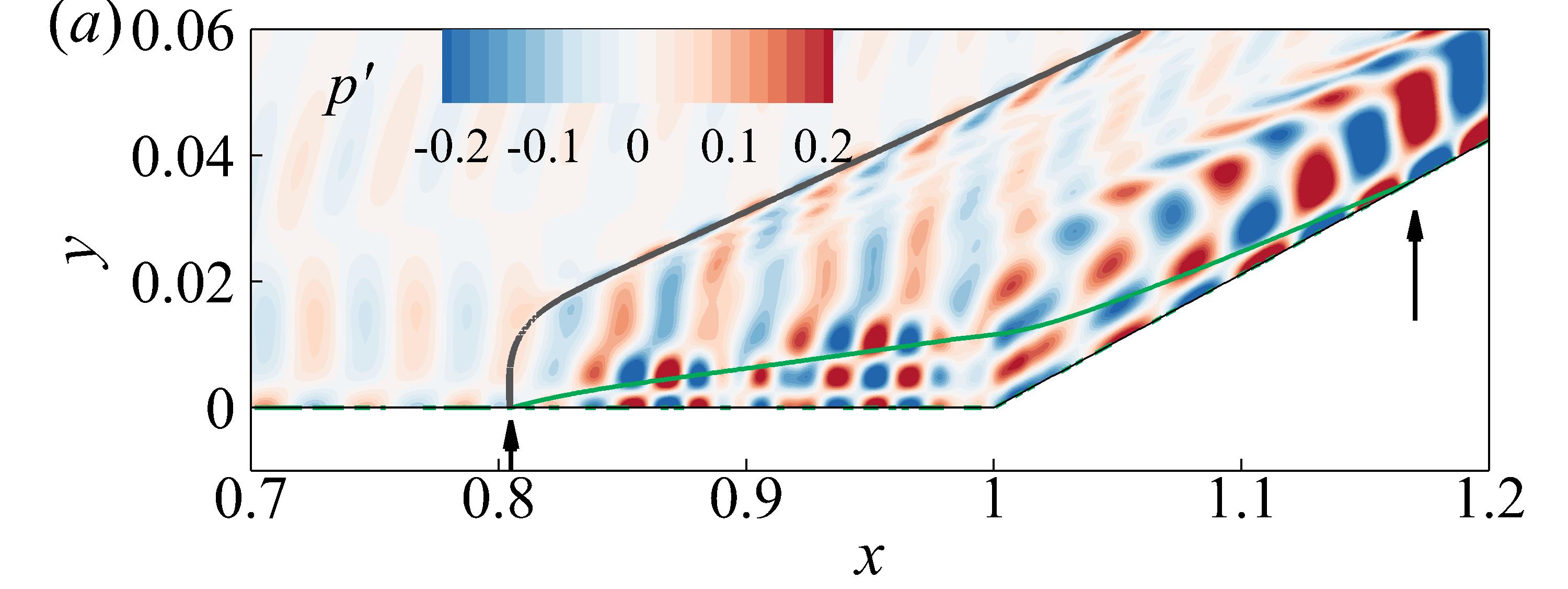}}
	\centering{ \includegraphics[height=3cm]{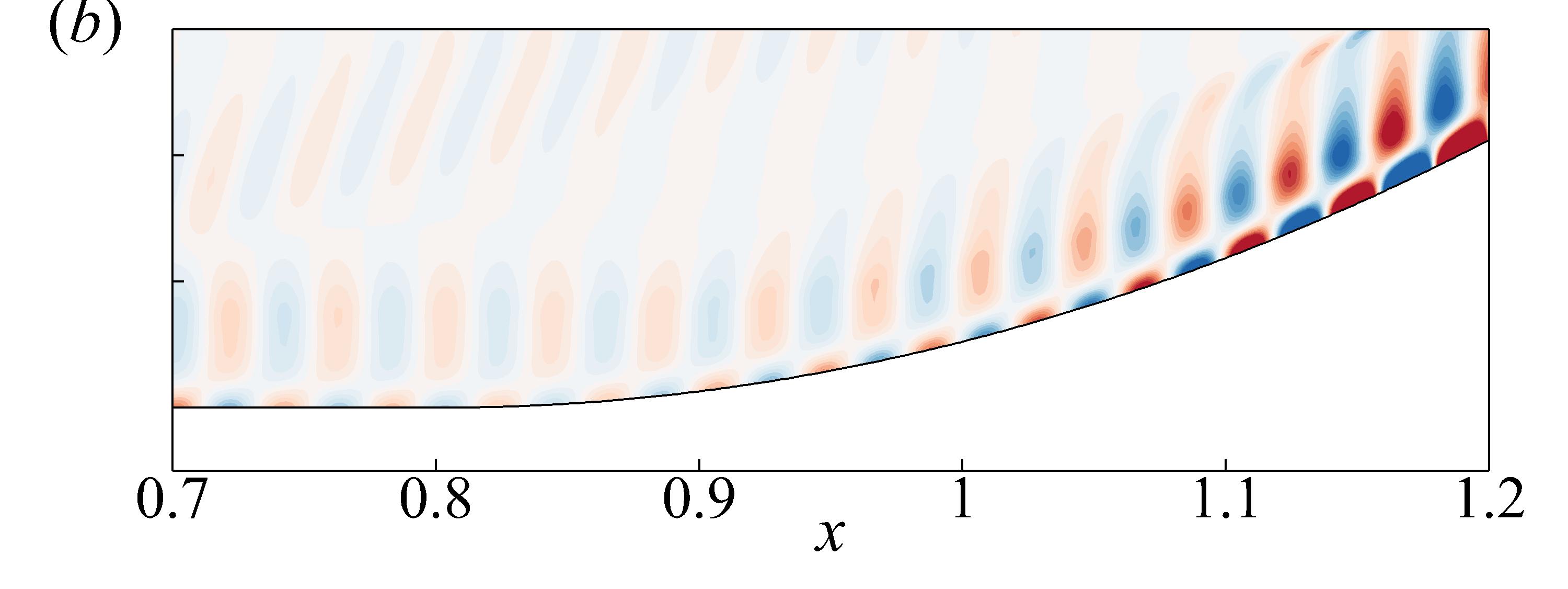}}
	\caption{Pressure fluctuation of the optimal disturbance $\omega=140$ for cases (\textit{a}) without and (\textit{b}) with corner rounding. Arrows in  (\textit{a}), separation and reattachment points. Gray line in (\textit{a}), the isoline $\bar{p} = 0.018$. Coloured line in (\textit{a}), dividing streamline.}
	\label{fig3}
\end{figure*}

Unless otherwise stated, the results of planar waves ($\beta=0$) are depicted to highlight the frequency selection of the separation bubble.\added{ This is because the 2-D disturbance is simpler to investigate and the most amplified for Mack modes. The role of the three-dimensionality will be later considered for non-Mack modes.} Firstly, the input forcing of the \replaced{Resolvent}{resolvent} analysis is placed at $x_i=0.2$, which is far upstream of the separation point $x_s$. As an example, the optimal disturbance $\omega=100$ is displayed, which corresponds to a dimensional frequency of $f^\ast\approx274.7$ kHz. The real parts of $\hat{p}$ for the two cases without and with corner rounding are shown in Fig. \ref{fig2}. The isoline $\bar{p} = 0.018$ which intersects the wall at the separation point is also displayed in Fig. \ref{fig2}(\textit{a}) to visualise the separation shock. With corner rounding in Fig. \ref{fig2}(\textit{b}), the typical double-cell structure of Mack second mode is maintained over the curved compression ramp. By contrast, in Fig. \ref{fig2}(\textit{a}), complicated structures of pressure fluctuations are observed in the separation bubble. Clearly, flow separation alters the disturbance shape significantly. For a further comparison, the optimal disturbance with $\omega=140$ is displayed in Fig. \ref{fig3}. As the frequency grows, the maximal cell number of the pressure fluctuation structure in the separation bubble is further increased. \deleted{Possibly, the existence of modes beyond the third mode is indicated.}

\begin{figure*}
	\centering{ \includegraphics[height=5.5cm]{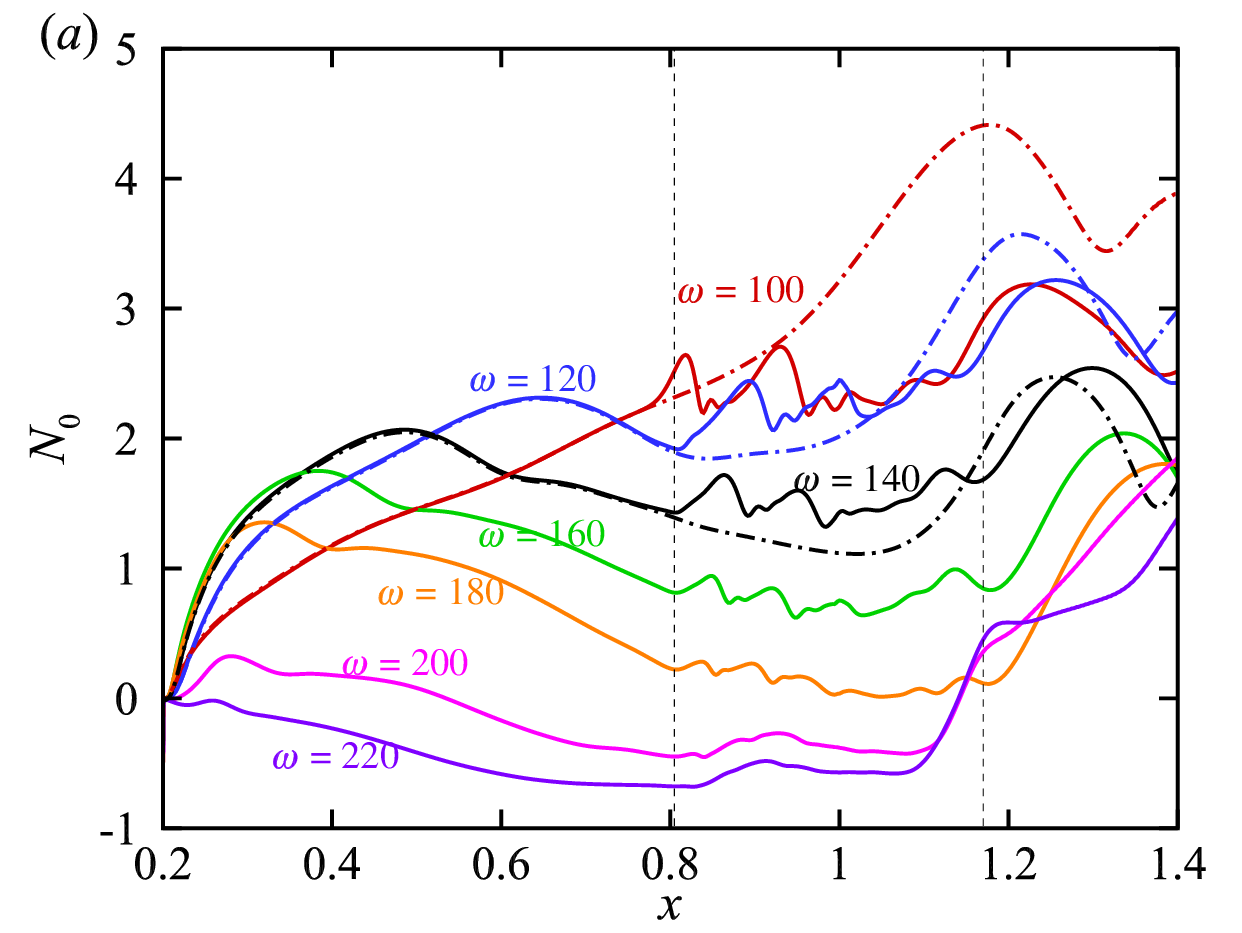}}
	\centering{ \includegraphics[height=5.5cm]{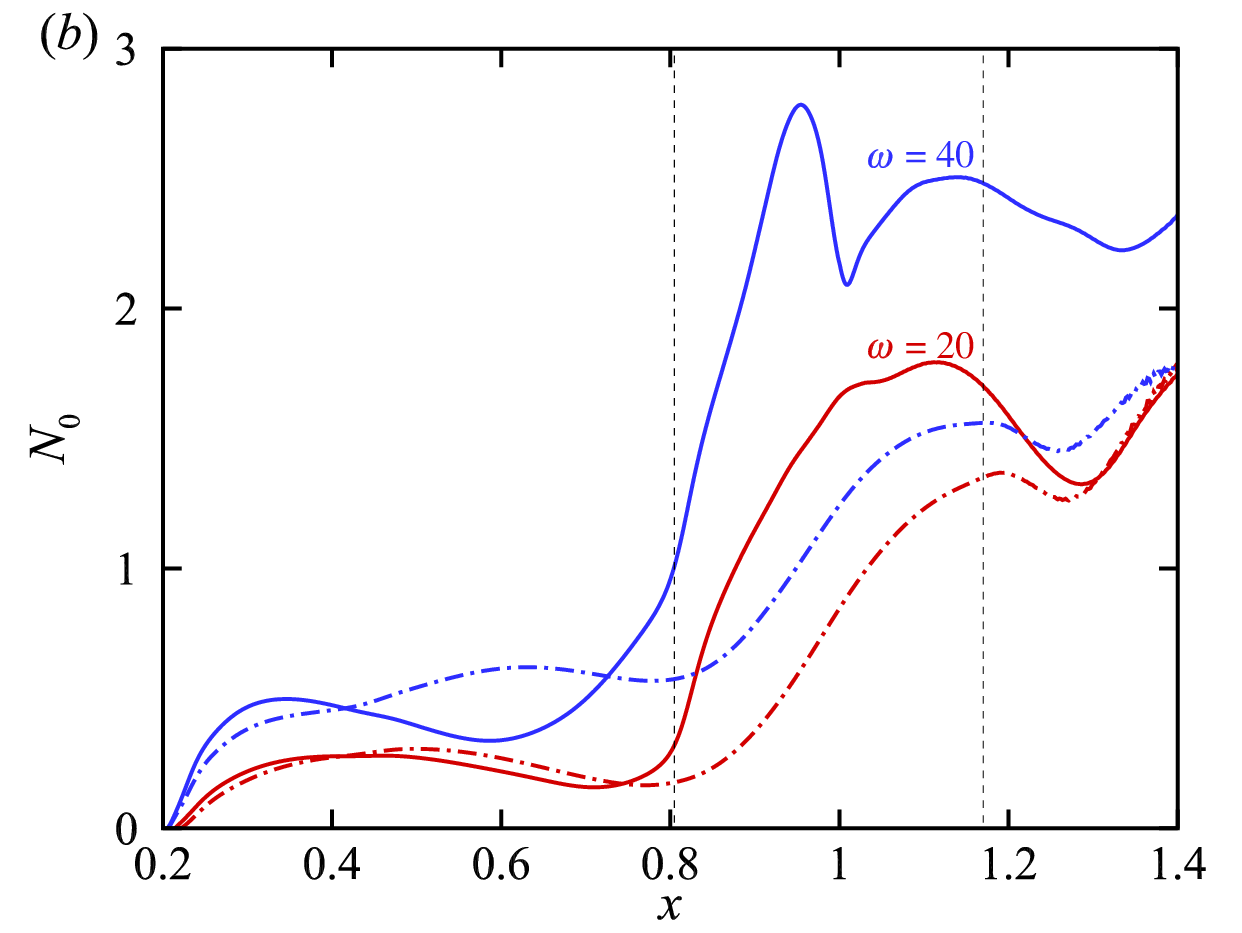}}
	\caption{$N$-factor based on $x_0=0.2$ of optimal disturbances for (\textit{a}) higher and (\textit{b}) lower frequencies. Solid lines, without corner rounding. Dashed dotted lines, with rounding\added{, where five representative cases $\omega=$ 20, 40, 100, 120 and 140 are shown}. Vertical dashed lines, locations of separation and reattachment.}
	\label{fig4}
\end{figure*}

To quantify the instability evolution, an $N$-factor is defined by
\begin{equation}
N_0 = 0.5\ln(E_\textit{Chu}/E_{\textit{Chu},0}),
\end{equation}
where $E_\textit{Chu}$ is Chu's energy density \cite{chu1965energy} locally integrated from the wall to infinity with respect to $\eta$, and $E_{\textit{Chu},0}$ is the value at a reference position $x=x_0$. The Chu's energy density consists of the kinetic energy and a positive definite thermodynamic energy, which has been extensively applied in stability analyses. Fig. \ref{fig4} shows the $N$-factor curves corresponding to $x_0=0.2$, whose performance can be categorised into two regimes. For high frequencies with hundreds of kilohertz in Fig. \ref{fig4}(\textit{a}), the travelling waves (solid lines) appear to be nearly neutral in the separation bubble between the two vertical dashed lines. This Mack-mode behaviour reported by \cite*{balakumar2005stability} seems to be independent of the upstream state whether the disturbance is amplified or decayed. Downstream of reattachment, the energy amplification rates $dN_0/dx$ with flow separation (solid lines) reduce to those without separation (dashed dotted lines).\added{ The displayed high-frequency range in Fig. \ref{fig4}(\textit{a}), probably associated with Mack modes, is also linked with the moderate peak along $\beta=0$ in the gain contour versus $\beta$ and $\omega$ in Fig. \ref{Appendix_gain_2D}.} For low frequencies with tens of kilohertz in Fig. \ref{fig4}(\textit{b}), however, the energy tends to be constantly amplified by the separation bubble. Therefore, the separation bubble presents a certain process of frequency selection. It should be noted that no unstable first mode is found in this state due to wall cooling and high Mach number \cite{hao2023response}. The \replaced{Resolvent}{resolvent} analysis shows that such low-frequency disturbances can be amplified in separated flows even though it is stable upstream.\added{ The sensitivity of the $N$-factor results to the input forcing location is also examined, where three choices $x_0=0.2$, $x_0=0.3$ and $x_0=0.4$ are considered. Downstream of the separation point, the resulting $N$-factor curves collapse into one or are close and parallel to each other. This observation indicates an unchanged conclusion subject to the variation in the forcing location.}

\subsection{Eigenmodal features of Mack modes}\label{sec:4.2}

\begin{figure*}
	\centering{ \includegraphics[height=5.4cm]{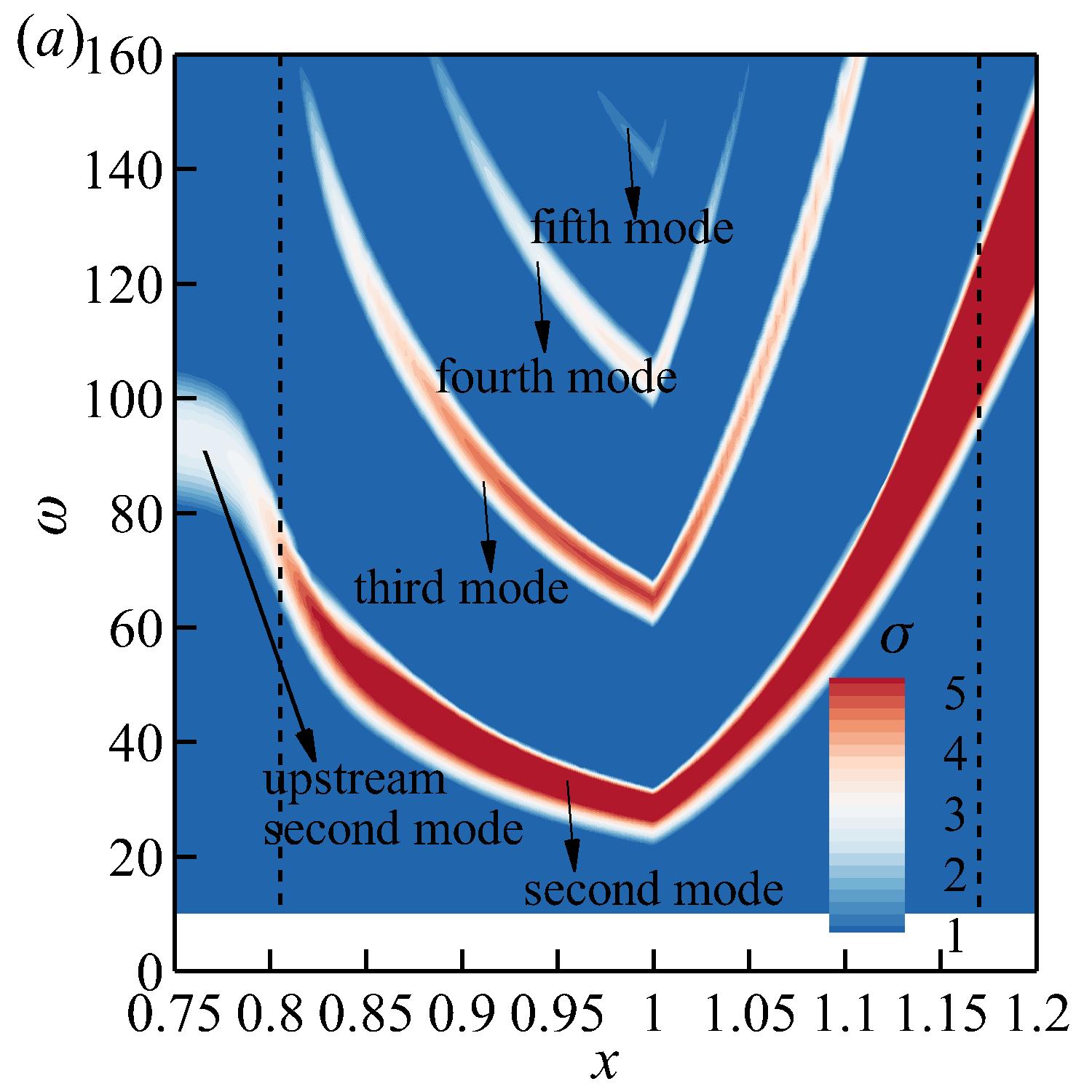}}
	\centering{ \includegraphics[height=5.4cm]{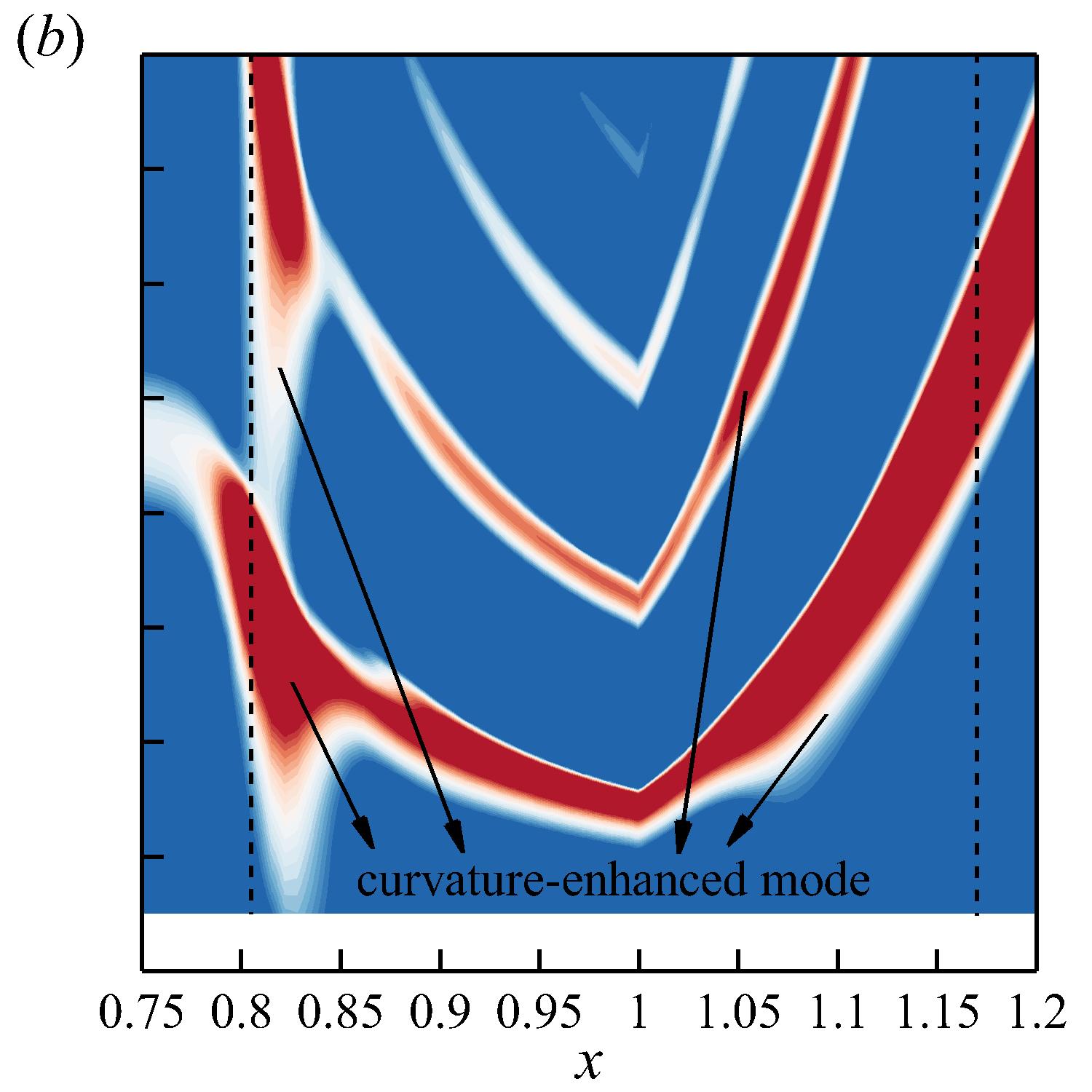}}
	\centering{ \includegraphics[height=5.4cm]{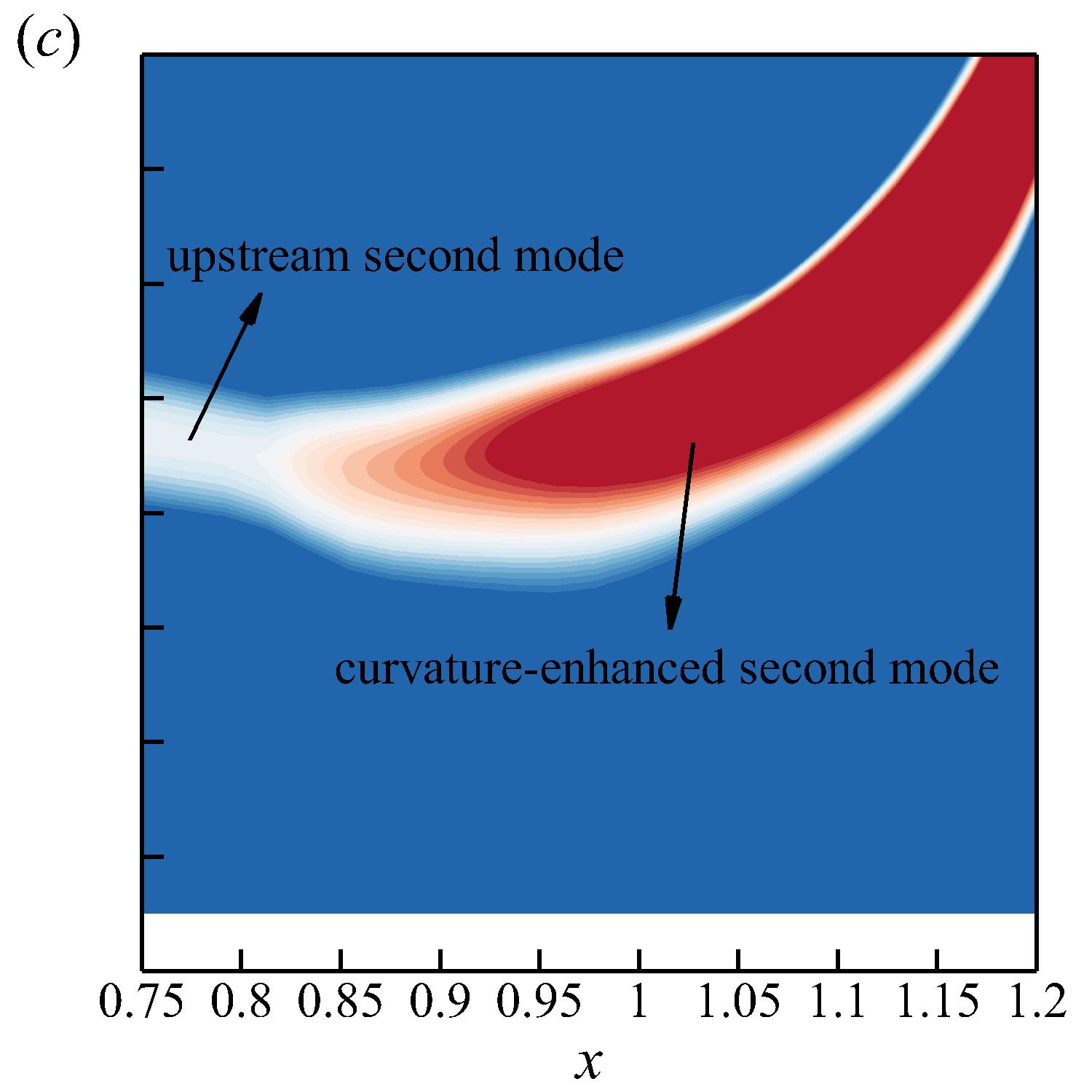}}
	\caption{Contour of $\sigma$ on the $x$--$\omega$ plane for the most unstable mode via LST: (\textit{a}) $K=0$ (without rounding), (\textit{b}) $K\ne0$ (without rounding) and (\textit{c}) $K\ne0$ (with rounding). Vertical dashed lines, locations of separation and reattachment.}
	\label{fig5}
\end{figure*}

To interpret the eigenmodal behaviour, the contour of the local growth rate $\sigma$ by LST is plotted versus $x$ and $\omega$.\added{ Similar plots were also given by Benitez et al. \cite{Benitez2020} in their Fig. 31 for a cone-cylinder-flare configuration.} In Fig. \ref{fig5}, a comparison between (\textit{a}) and (\textit{b}) reflects the effect of the streamline curvature, and a comparison between (\textit{b}) and (\textit{c}) mainly highlights the role of the separation bubble. With the occurrence of separation, unstable Mack modes up to the fifth mode are identified. The order of higher Mack modes is based on the relative magnitude of the frequency at a fixed $x$. A fixed-frequency disturbance may go through higher Mack-mode regions. For example, in Fig. \ref{fig5}(\textit{a}), the disturbance $\omega=100$ undergoes the second-mode instability once, the third-mode instability twice and then the second-mode instability again in the separation bubble. The alternating unstable and stable regions for a fixed $\omega$ agrees with the neutrally oscillating behaviour of the $N$-factor qualitatively. By contrast, only second-mode instability is identified in Fig. \ref{fig5}(\textit{c}) when separation is eliminated. Thus, flow separation is able to generate higher-order Mack modes. Moreover, increasing the frequency may give rise to higher-order Mack modes in the separation bubble in Fig. \ref{fig5}(\textit{a}). This tendency is consistent with the pressure fluctuation structure in the \replaced{Resolvent}{resolvent} analysis shown in Figs. \ref{fig2} and \ref{fig3}.\added{ The pressure eigenfunctions for higher-order Mack modes of LST are also displayed in Fig. \ref{Appendix_LST} in the Appendix for clarification.} In Fig. \ref{fig5}(\textit{b}), the centrifugal force effect induced by the curved streamline is to promote the instability in the vicinity of separation and reattachment points, where $|K|$ is large (see Fig. \ref{fig1}\textit{c}).

\begin{figure*}[h]
	\centering{ \includegraphics[height=4.6cm]{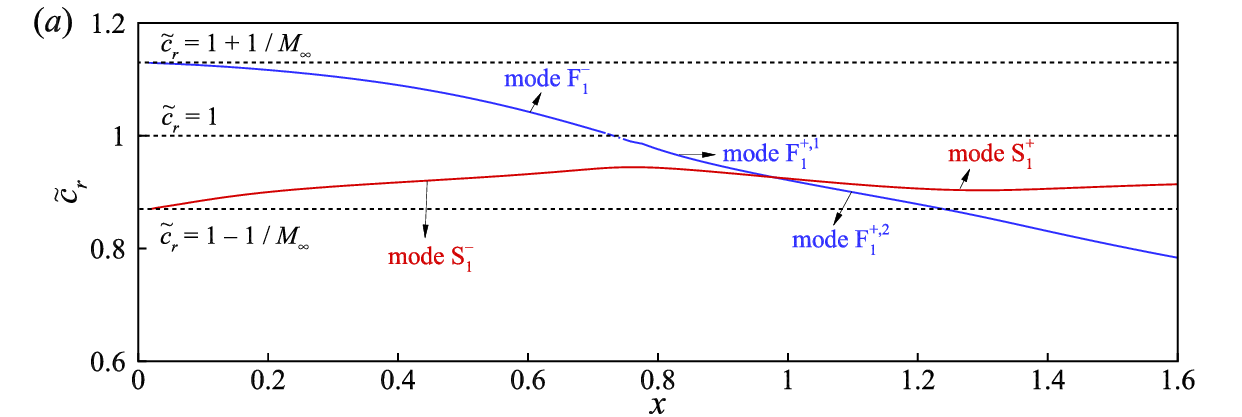}}
	\centering{ \includegraphics[height=4.6cm]{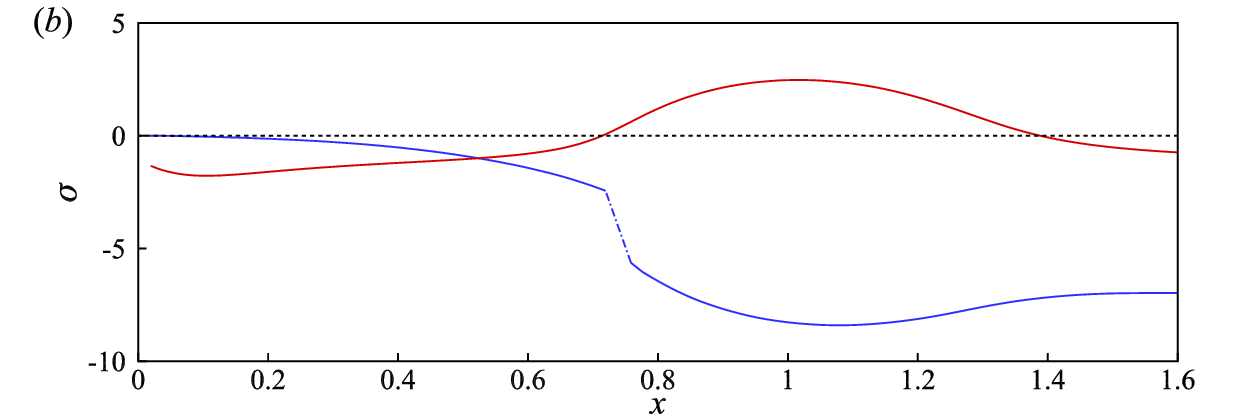}}
	\caption{Modal synchronisation with $\omega=80$ for a rampless boundary layer flow without pressure gradient: (\textit{a}) $\tilde{c}_r$--$x$ and (\textit{b}) $\sigma$--$x$.}
	\label{fig6}
\end{figure*}

\begin{figure*}
	\centering{ \includegraphics[height=4.6cm]{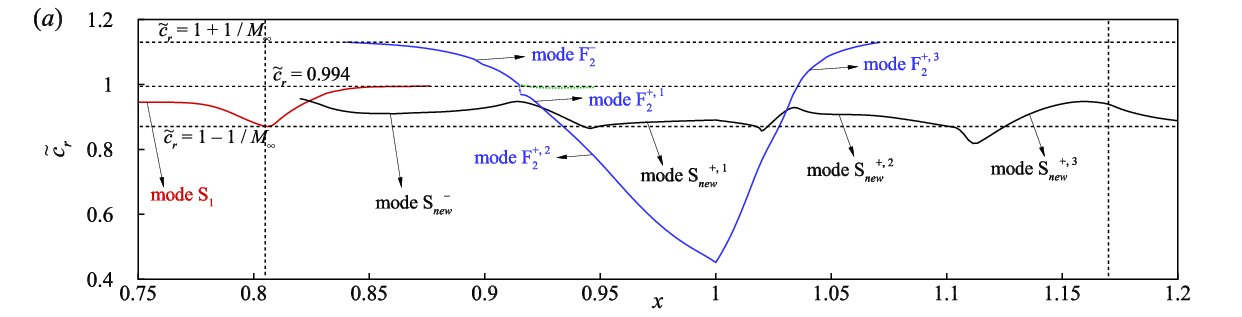}}
	\centering{ \includegraphics[height=4.6cm]{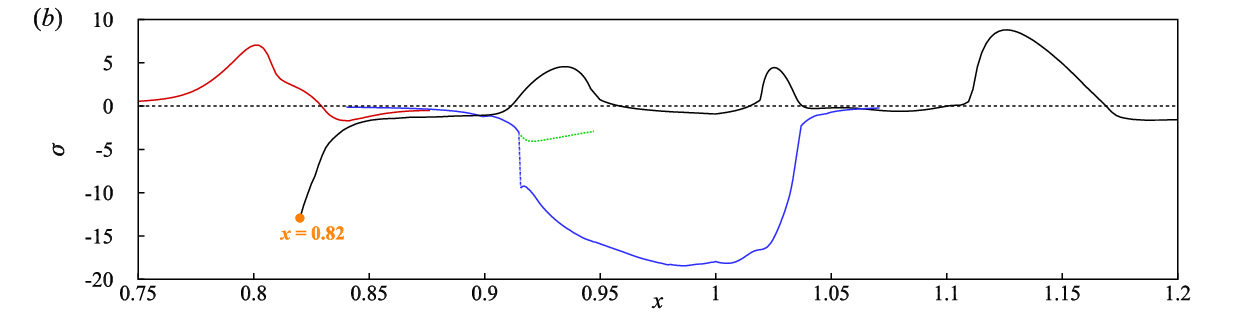}}
	\caption{LST results ($K\ne0$) of modal synchronisation at $\omega=80$ without rounding: (\textit{a}) $\tilde{c}_r$--$x$ and (\textit{b}) $\sigma$--$x$. Greed dashed line, one of the newly generated modes merging with the continuous spectrum. Vertical dashed lines, locations of separation and reattachment.}
	\label{fig7}
\end{figure*}

To further the understanding of Mack-mode instabilities in the separation bubble, multiple discrete modes including stable ones are identified by LST corresponding to $\omega=80$.\added{ This frequency is selected such that a moderate number of Mack modes (second and third modes) are involved and that the problem is not excessively troublesome. For higher frequencies, higher-order modes can be further included to complicate the discussion.} Before proceeding to the LST analysis of the ramp flow, a brief analysis of a rampless flat-plate boundary layer flow without pressure gradient is first performed under the same flow condition. The purpose is to establish a reference of the existing modes and the associated nomenclature for such a canonical problem. The base flow over the flat plate is also simulated by our N--S solver. The real part of the complex phase velocity $\tilde{c}$, i.e., the real part $\tilde{c}_r=\Re(\omega/\alpha)$ and the growth rate $\sigma$ are given versus $x$ in Fig. \ref{fig6}. Note that occasionally the difference between the real phase velocity $c_r=\omega/\alpha_r$ and the real part of the complex phase velocity $\tilde{c}_r$ is not ignorable if $|\alpha_r|\gg|\alpha_i|$ is not satisfied.\added{ This issue becomes apparent for mode F when it synchronises with the slow acoustic continuous spectrum in Figs. \ref{fig6} and \ref{fig7}.} For a hypersonic boundary layer without separation, the modal synchronisation between slow and fast discrete modes, originating from the slow and fast acoustic waves near the leading edge, gives rise to new branches related to unstable Mack modes \cite{ma2003receptivity,fedorov2011high}. As shown in Fig. \ref{fig6}(\textit{a}), near the leading edge, the phase-velocity limits of the fast mode F$^{-}_1$ and the slow mode S$^{-}_1$ are identical to those of the fast and slow acoustic waves, i.e., $\tilde{c}_r=1+1/M_{\infty}$ and $\tilde{c}_r=1-1/M_{\infty}$, respectively. Here, the superscripts `$-$' and `$+$' indicate that the mode is the branch on the left and right sides, respectively, of the first synchronisation point with other discrete or continuous spectra. The subscript `$1,2,...$' means that the mode is the first, second, etc. discrete mode originated from fast/slow acoustic waves. The superscript `$1,2,...$' indicates that the mode is the first, second, etc. new significant branch of the same family of the discrete mode, which arise from synchronisation with other spectra. As $x$ is increased, mode F$^{-}_1$ synchronises with the entropy/vorticity continuous spectrum with $\tilde{c}_r=1$ at $x=0.73$ and then with mode S$^{-}_1$ at $x=0.98$. The former synchronisation generates a new branch F$^{+,1}_1$ accompanying with a jump of the phase velocity and the growth rate, and the latter synchronisation generates branches F$^{+,2}_1$ and S$^{+}_1$. The synchronisation between the families of modes F and S is associated with the unstable peak of $\sigma$ in Fig. \ref{fig6}(\textit{b}) through an intermodal energy exchange mechanism \cite{fedorov2011high}. In the terminology framework of Mack, this unstable peak is related to the instability of Mack second mode. A low-frequency (or low-Reynolds-number) unstable limit belongs to the first-mode instability. In the considered flat-plate case, no unstable first mode is identified.

\begin{figure*}
	\centering{ \includegraphics[height=6.0cm]{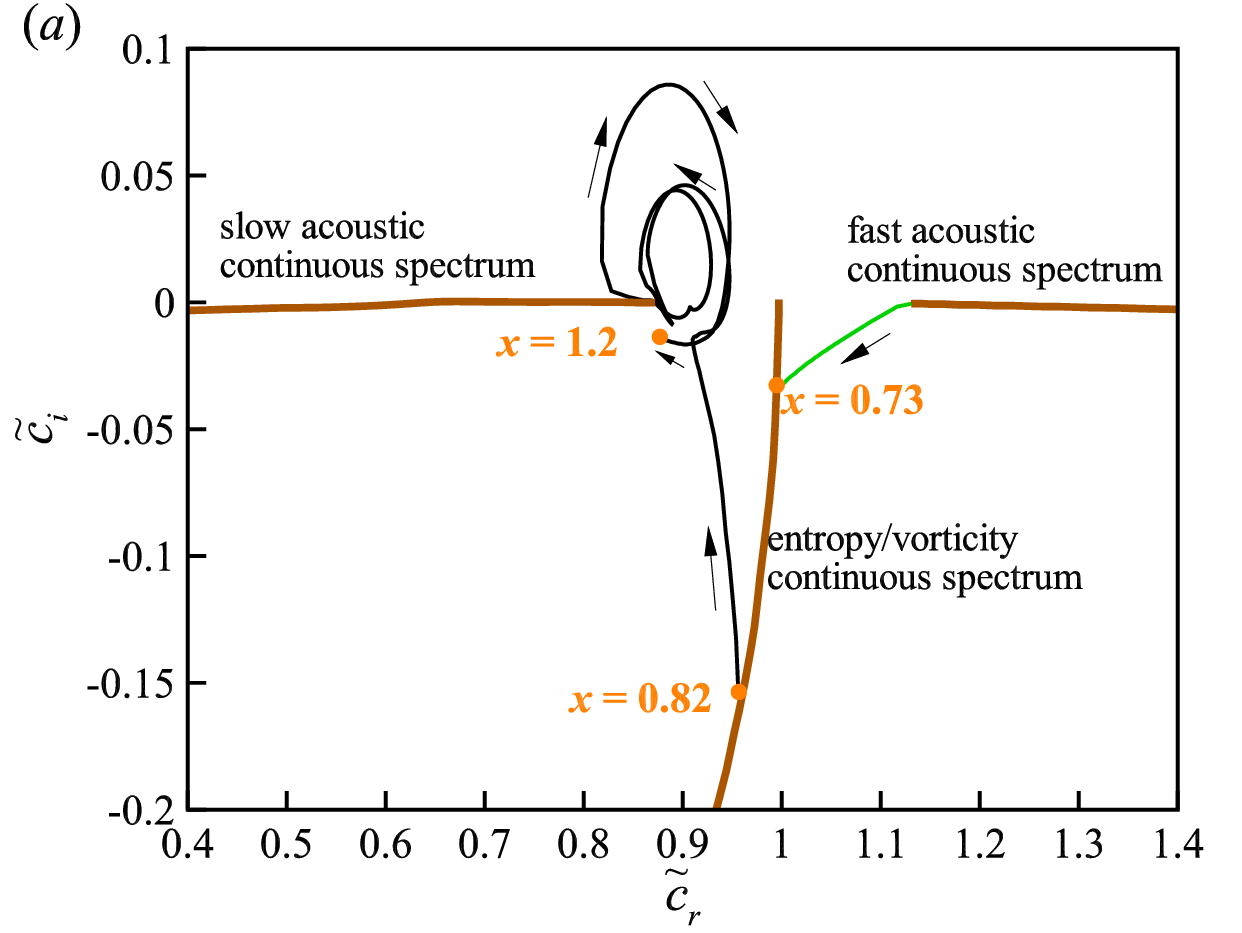}}
	\centering{ \includegraphics[height=6.0cm]{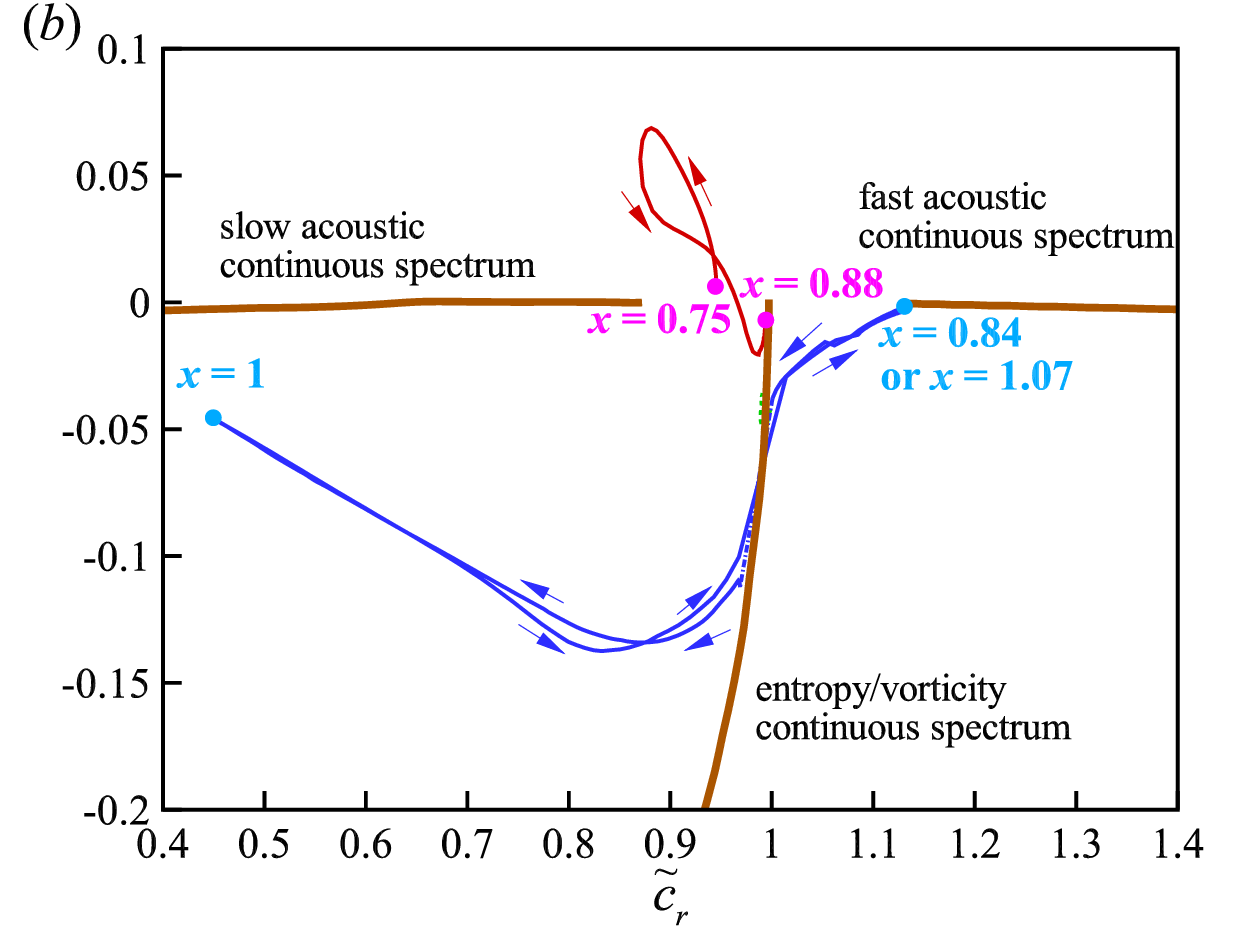}}
	\caption{LST results ($K\ne0$) of eigenvalue trajectories on the complex plane ($\tilde{c}_r,\tilde{c}_i$) with an increasing $x$ coordinate at $\omega=80$ without rounding: (\textit{a}) families of mode S$_\textit{new}$ (black line) and mode F$_1$ upstream of separation (green line), and (\textit{b}) families of mode S$_1$ (red line) and mode F$_2$ (blue line).}
	\label{fig8}
\end{figure*}

For the ramp flow, the main synchronisation process is depicted in Fig. \ref{fig7}. Not all of the new branches of discrete modes generated by synchronisations are marked (which are associated with the jump of the phase velocity), and only significant modes are displayed. All the discrete modes are tracked with a small grid spacing $\Delta x$\added{ $< 0.001$} as the $x$ coordinate is increased, such that the displacement of the complex phase velocity $\tilde{c}$ is minor during each step of tracking. \replaced{To seek the mode of the same family at the next grid point $x_{i+1}$, the eigenvalue variation, compared to the target mode at $x_i$, is minimised. Mathematically, the eigenvalue identifier $j$ at $x_{i+1}$ is sought with the objective $\text{min}_j \Vert \tilde{c}(x_i)-\tilde{c}_{j}(x_{i+1})\Vert $. Since $\Delta \tilde{c}$ of the same mode is minor}{As a result}, the mode can be continuously tracked and easily distinguished from other evolving discrete modes\replaced{ in most streamwise locations}{ via the global and local numerical methods by Malik \cite{malik1990numerical}}. To facilitate the understanding, the eigenvalue trajectories of the main modes on the complex plane ($\tilde{c}_r,\tilde{c}_i$) are displayed as $x$ is increased in Fig. \ref{fig8}. Note that a mode is unstable if $\tilde{c}_i>0$.

Fig. \ref{fig7} demonstrates that the synchronisation regime becomes evidently different when separation occurs. As shown in the growth rate contour of Fig. \ref{fig5}(\textit{b}), \added{for $\omega=80$, }the first peak of $\sigma$ near the separation point \deleted{for $\omega=80$} in Fig. \ref{fig7}(\textit{b}) is a continuation of the upstream second-mode instability. This growth rate peak is associated with mode S$_1$ in Fig. \ref{fig7}, which normally evolves from the leading-edge slow acoustic wave continuous spectrum. The continuation role of mode S$_1$ can be also concluded from the rampless results in Fig. \ref{fig6}(\textit{b}), where an unstable peak of the growth rate emerges at around $x=0.8$ for mode S$_1$. When penetrating the separation bubble, mode S$_1$ eventually merges with the entropy/vorticity continuous spectrum whose dimensionless $\tilde{c}_r$ is unity. This merge at around $x=0.88$ is also recorded by the eigenvalue trajectory (red line) in Fig. \ref{fig8}(\textit{b}). Near the separation point $x_s = 0.805$, another slow mode is newly identified at $x=0.82$, which is named mode S$^-_{\textit{new}}$. This new mode has a phase velocity lower than the freestream velocity. The subscript `${\textit{new}}$' is used to distinguish this new slow mode from the conventional slow mode\added{,} which evolve\added{s} from the leading edge receptivity. One may wonder if this new mode is a consequence of the upstream mode crossing the continuous spectrum. However, according to Fig. \ref{fig8}(\textit{a}), the upstream fast mode originated from the fast acoustic wave synchronises with the continuous spectrum at $x=0.73$, which is evidently upstream of the separation point $x_s = 0.805$. No other fast mode ($\tilde{c}_r>1$) is observed in the global spectrum between $x = 0.73$ and the location $x = 0.82$ where mode S$^-_\textit{new}$ first emerges. In fact, Fig. \ref{fig8}(\textit{a}) illustrates that mode S$^-_\textit{new}$ departs from the entropy/vorticity continuous spectrum at $x = 0.82$ possibly due to flow separation, which is initially very stable. \added{How mode S$^-_{\textit{new}}$ emerges is also depicted by overlaid eigenspectra in Fig. \ref{Appendix_fig-spectra} in the Appendix.}

\replaced{In}{Meanwhile, in} addition to mode F$_1$ evolving from the leading edge, a new branch of fast mode F$^-_2$ emerges in the separation bubble, which is originated from the fast acoustic wave ($\tilde{c}_r=1+1/M_{\infty}$). Subsequently, mode F$^-_2$ synchronises with the entropy/vorticity continuous spectrum, which gives rise to mode F$^{+,1}_2$ with a more negative $\sigma$, i.e., more stable. Following that, twice synchronisations between the families of modes F$_2$ and S$_\textit{new}$ occur, which generate the second and third peaks of the growth rate in Fig. \ref{fig7}(\textit{b}). The twice synchronisations are achieved by the non-monotonic evolution of $\tilde{c}_r$ of mode F$^{+,\text{2}}_\text{2}$ from decreasing on the plate ($x<1$) and increasing on the ramp ($x>1$). Fig. \ref{fig8}(\textit{b}) also shows that the family of mode F$_1$ (denoted by the blue line) first appears at around $x=0.84$ from the fast acoustic wave continuous spectrum, then undergoes a decrease and increase of the phase velocity, and finally merges with the fast acoustic wave spectrum at around $x=1.07$. Further downstream, after a synchronisation between mode S$^{+,\text{2}}_\textit{new}$ and the slow acoustic wave ($\tilde{c}_r=1-1/M_{\infty}$) near reattachment, the fourth peak of the growth rate is generated in Fig. \ref{fig7}(\textit{b}). Indicated by Fig. \ref{fig5}(\textit{b}), the first and fourth peaks approaching the separation and reattachment points in Fig. \ref{fig7}(\textit{b}) for $\omega=80$ belong to the second-mode instability, while the second and third peaks belong to the third-mode instability. However, the upstream second mode (first peak) is different from the downstream second mode (fourth peak). The upstream one is normally originated from mode S$_1$, whereas the downstream one originated from the family mode S$_\textit{new}$ is a complicated product due to flow separation.

\subsection{\added{Comparison between local and nonlocal methods for Mack modes}}\label{sec:4.3}

\begin{figure*}
	\centering{ \includegraphics[height=6cm]{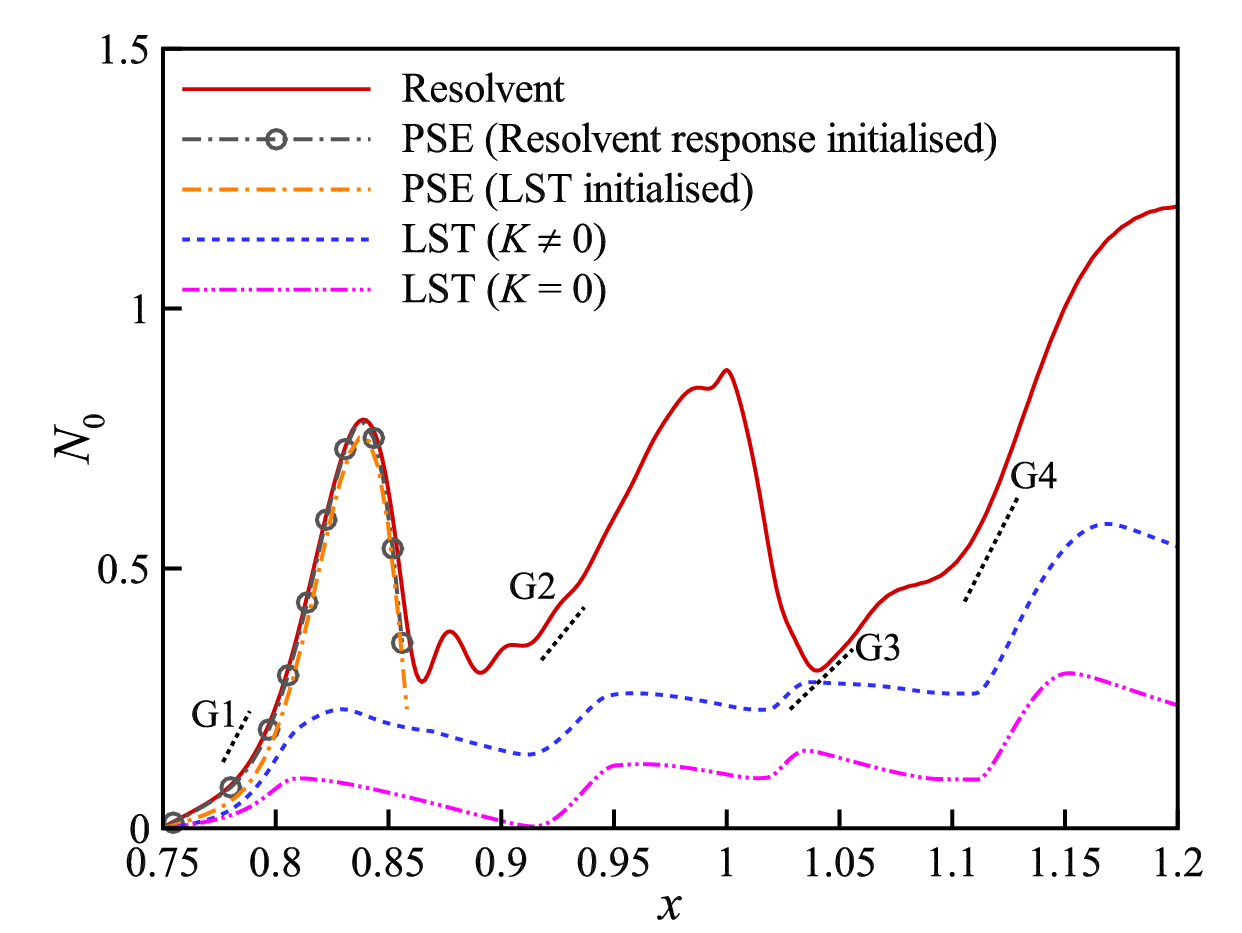}}
	\caption{$N$-factor based on $x_0=0.75$ of cases without corner rounding for planar waves with $\beta=0$ at $\omega=80$. Inclined dashed lines, approximate slopes of energy growth G1--G4.}
	\label{fig9}
\end{figure*}

The depicted complicated synchronisation regime gives rise to discontinuous regions of unstable modes in the separation bubble. In other places, the travelling wave is stable. The alternating unstable and stable regions are associated with the oscillating $N$-factor curves of high-frequency disturbances in the separation bubble. To evaluate how well LST can describe the eigenmodal mechanism, a comparison of the $N$-factor among \replaced{Resolvent}{resolvent} analysis, PSE and LST is shown in Fig. \ref{fig9}. Note that by imposing appropriate initial conditions, the LNSE system with non-normality can exhibit a large transient amplification of energy \cite{schmid2007nonmodal}. Meanwhile, PSE, marching from an initial condition, is found to be a good approximation of the (harmonic) LNSE in terms of the accuracy \cite{paredes2019nonmodal}. Accordingly, PSE results initialised by the \replaced{Resolvent}{resolvent} response are expected to achieve an approximate optimal amplification, which contains both nonmodal transient growth and possible modal growth \cite{guo2023interaction}. By contrast, the PSE result initialised by the LST solution is closely related to the normal-modal instability, which includes the nonparallel effect. The initialisation location for PSE is set to upstream at $x=0.25$. The initial streamwise wavenumber $\alpha$ is simply taken from the LST result. In fact, a different initial value of $\alpha$ has no effect on the downstream spatial marching of PSE, since the iterative procedure will quickly adjust the converged shape function $\bm{\psi}$ and wavenumber $\alpha$ closely downstream of the initialisation position.

\begin{figure*}
	\centering{ \includegraphics[height=5.2cm]{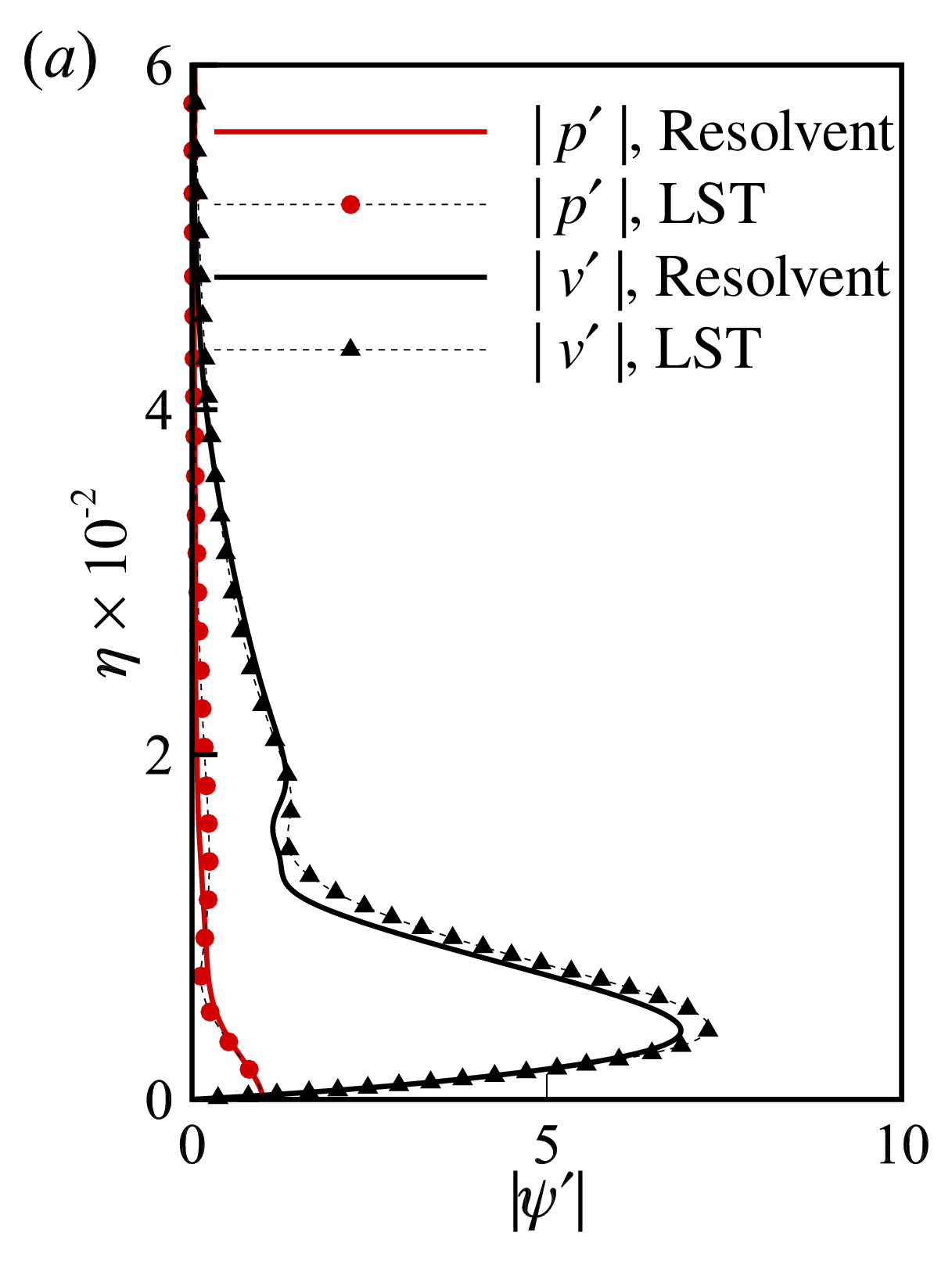}}
	\centering{ \includegraphics[height=5.2cm]{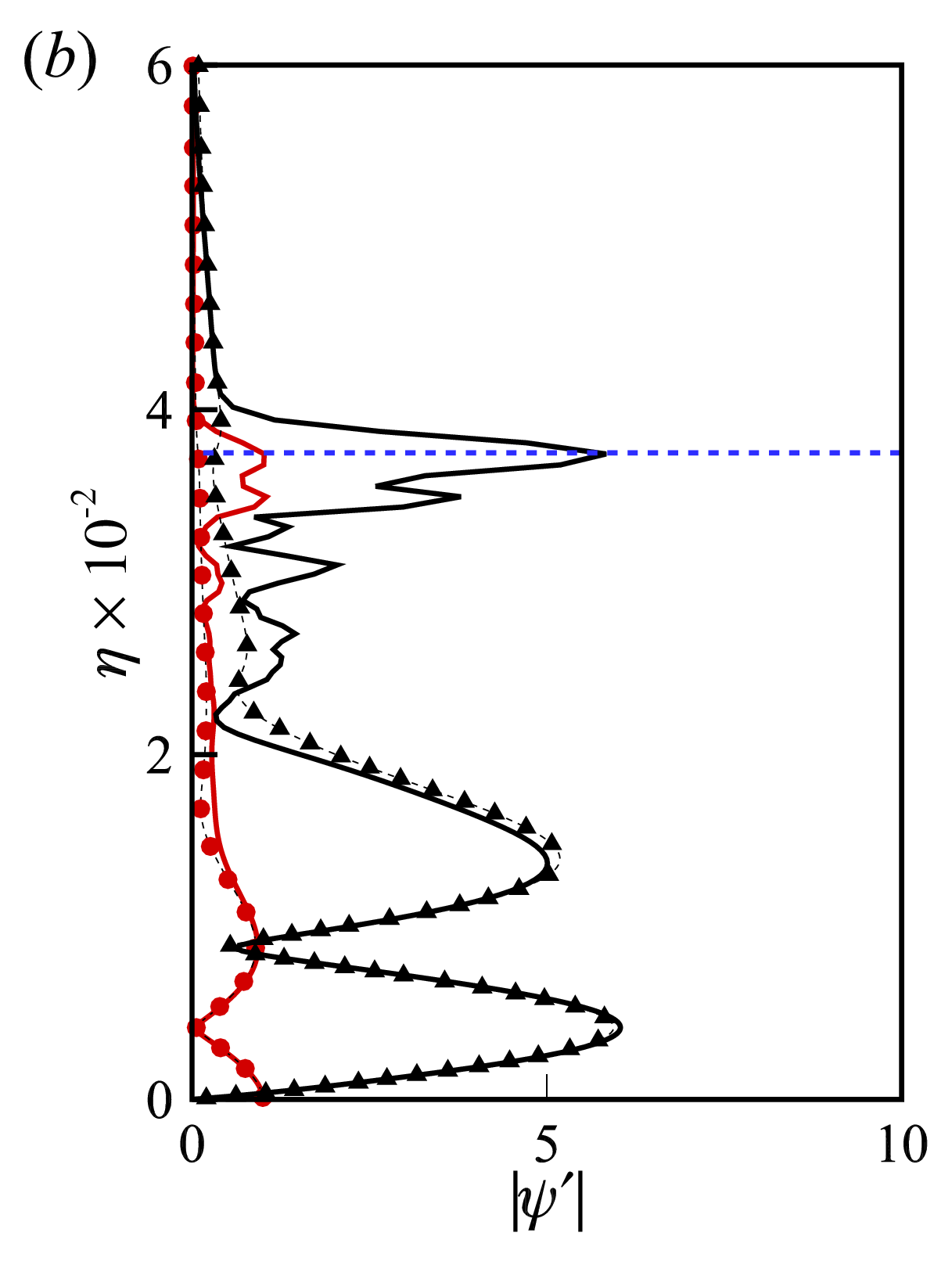}}
	\centering{ \includegraphics[height=5.2cm]{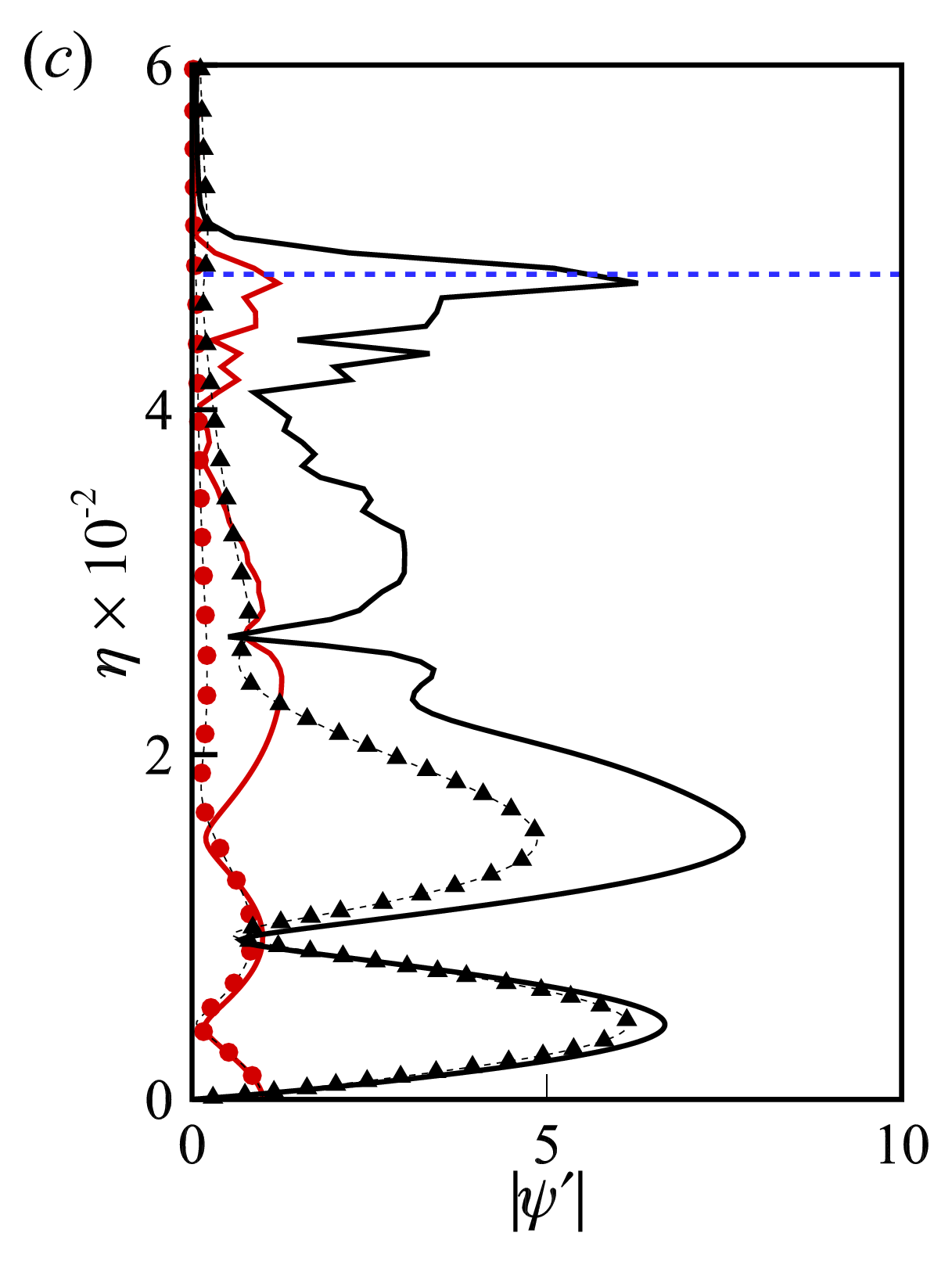}}
	\centering{ \includegraphics[height=5.2cm]{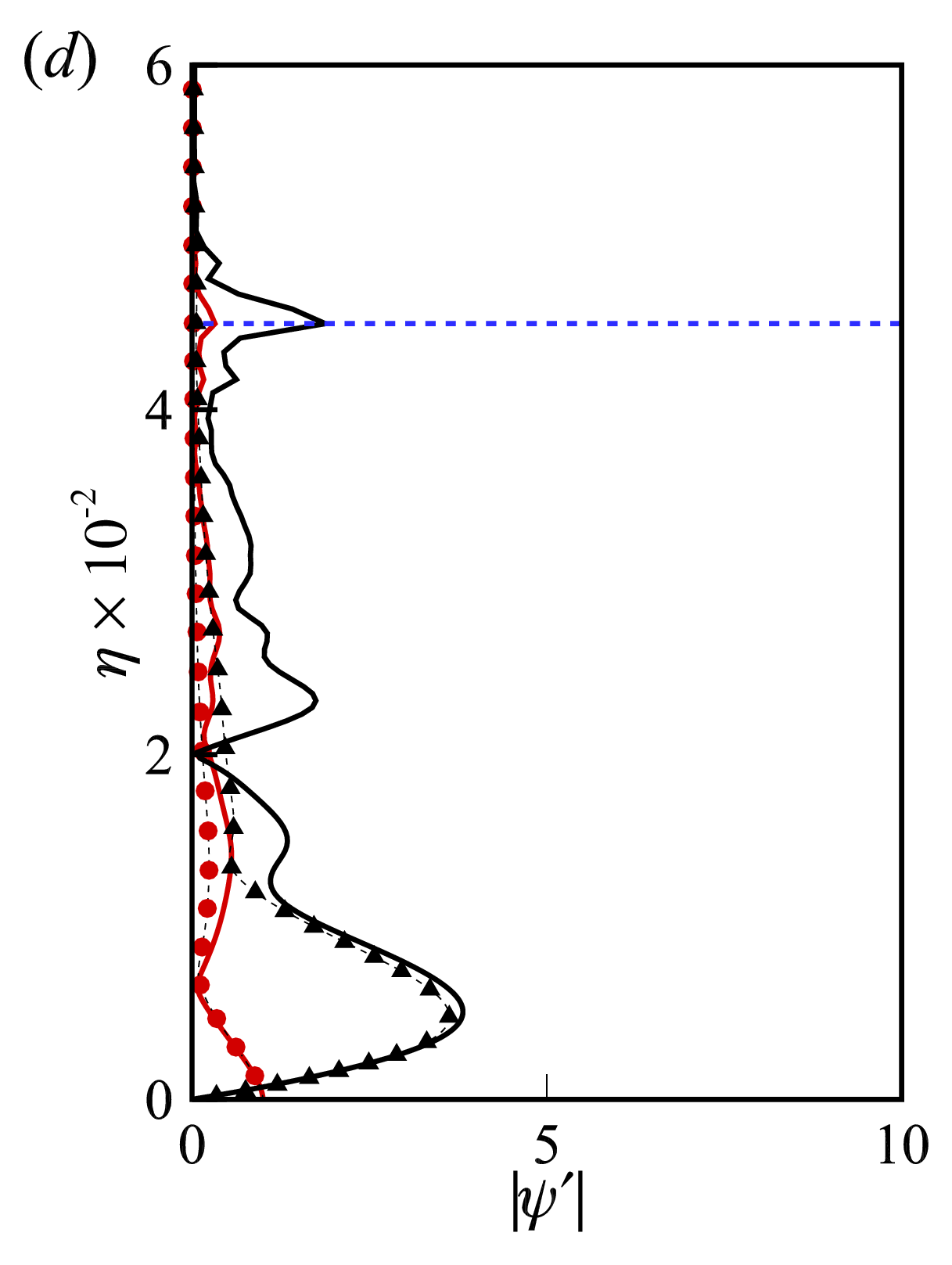}}
	\caption{Comparison of the disturbance shape for $\omega=80$ between the \replaced{Resolvent}{resolvent} response and the eigenfunction of the most unstable local mode. Four stations corresponding to the four peaks of $\sigma$ in Fig. \ref{fig7}(\textit{b}) are shown: (\textit{a}) $x=0.8$,  (\textit{b}) $x=0.934$,  (\textit{c}) $x=1.025$ and (\textit{d}) $x=1.125$. Horizontal dashed line, location of the separation shock. All the results are normalised by $|p'|$ at the wall.}
	\label{fig10}
\end{figure*}

It should be mentioned that the magnitude of the $N$-factor in Fig. \ref{fig9} is not large since the reference position is moved downstream to $x_0=0.75$. The far upstream region is not shown because we expect to highlight the instability evolution in the separation bubble. For $\omega=80$ in Fig. \ref{fig9}, the \replaced{Resolvent}{resolvent} analysis mainly shows four energy growths (G1--G4) in the separation bubble. The approximate amplification rate $dN_0/dx$ is marked by the inclined thin dashed line. Compared to the \replaced{Resolvent}{resolvent} analysis, the $N$-factor is underestimated by LST, which shows streamwise shorter growths of energy. However, the amplification rate (slope of curve) of G1--G4 as well as the oscillation behaviour of the $N$-factor curve is well captured by LST. Further considering the effect of streamline curvature ($K\ne0$), LST gives a closer amplification rate to the \replaced{Resolvent}{resolvent} analysis near the separation point $x_s = 0.805$. In general, the neutral oscillation of the high-frequency optimal disturbance is explainable by the local eigenmodal mechanism. Recall in Fig. \ref{fig7}(\textit{b}) that the first unstable peak (red line) is originated from another different family of the discrete mode compared to the remaining three peaks (black line). Therefore, PSE marching from upstream profiles can only capture the first growth G1 in Fig. \ref{fig9}(\textit{a}). However, an excellent agreement of the $N$-factor is reached between \replaced{Resolvent}{resolvent} analysis and PSE with two initialisation methods. Therefore, the underestimation in the $N$-factor by LST is caused by the nonparallel effect, and the modal amplification is dominant for the high frequency $\omega=80$. Fig. \ref{fig10} further shows the comparison of the wall-normal shape between the \replaced{Resolvent}{resolvent} response and the LST eigenfunction of the most unstable mode. Four stations corresponding to the four peaks of $\sigma$ in Fig. \ref{fig7}(\textit{b}) are chosen.\added{ Fig. \ref{fig10}(\textit{a}) corresponds to mode S$_1$, while Figs. \ref{fig10}(\textit{b}--\textit{d}) all correspond to mode S$_\textit{new}$.} The discrepancy in the disturbance shape is minor in the near-wall region, while it becomes larger in the outer region partly due to additional effects, such as the effects of the separation shock and nonparallelism. In general, the discontinuous energy growth of high-frequency Mack modes in the separation bubble is confirmed by both global and local analyses.

\subsection{Eigenmodal features of shear-layer modes}\label{sec:4.4}

\added{From the above analysis, flow separation can induce complicated multi-modal phenomena for high-frequency Mack modes. The relevant result naturally raises questions that merit considerations: do low-frequency components share the same multi-modal feature in a local analysis? If identified, is the local mode consistent with or linked with the global Resolvent mode? As aforementioned in the Introduction, high- and low-frequency components coexisted in the experiment of Butler and Laurence \cite{butler2021interaction}. Furthermore, the low-frequency signal with tens of kilohertz was visibly amplified in the separation bubble, which differed from the high-frequency counterpart. This amplification was mostly attributed to G{\"o}rtler instabilities in literature \cite{roghelia2017experimental,hao2023response}. However, the so-called G{\"o}rtler mode in such a situation was not carefully studied from a local analysis viewpoint in different local profiles. It remains unclear whether the mentioned energy amplification in the bubble is related to a single G{\"o}rtler mode or not. As a result, the growth mechanism, as stated in the second research question of the Introduction, is not fully clarified.}
	
\added{Here, in advance, we report here that multiple eigenmodes are also identified for low-frequency components. Moreover, at different stations, there are different dominant local unstable modes. They appear successively, contribute to the persistent energy amplification and agree with the global Resolvent analysis. The varying dominance of different eigenmodes can be clearly reflected in their growth rates along the streamwise direction. The newly identified multiple low-frequency modes are the main new findings in the subsection. In total, four low-frequency eigenmodes will be focused on in the following discussion. The four local modes are named by acronyms to manifest the respective emerging locations, i.e., modes SP-1 and SP-2 near the separation point, mode IB in the bubble, and mode RP near the reattachment point. Among them, mode SP-1 has a larger growth rate than mode SP-2, whereas both of them are included in the analysis.}

In \replaced{the following two subsections}{section}, we mainly focus on the low circular frequency $\omega=20$, whose 2-D local mode is almost stable in the separation bubble indicated by Fig. \ref{fig5}.\added{ Please note that the purpose here is not to select the most unstable low-frequency component for investigations. Instead, for tens of kilohertz, we intend to clarify the notable difference between the stable state by local analysis in Fig. \ref{fig5}(\textit{b}) and the unstable state by Resolvent analysis in Fig. \ref{fig4}(\textit{b}). As a result, $\omega=20$ appears to be a better choice than $\omega=40$, since the 2-D local mode of $\omega=20$ is generally stable across the separation bubble in Fig. \ref{fig5}(\textit{b}). One may be more interested in such a low-frequency component, which can be evidently amplified without local unstable mechanisms.}

\begin{figure*}
	\centering{ \includegraphics[height=5cm]{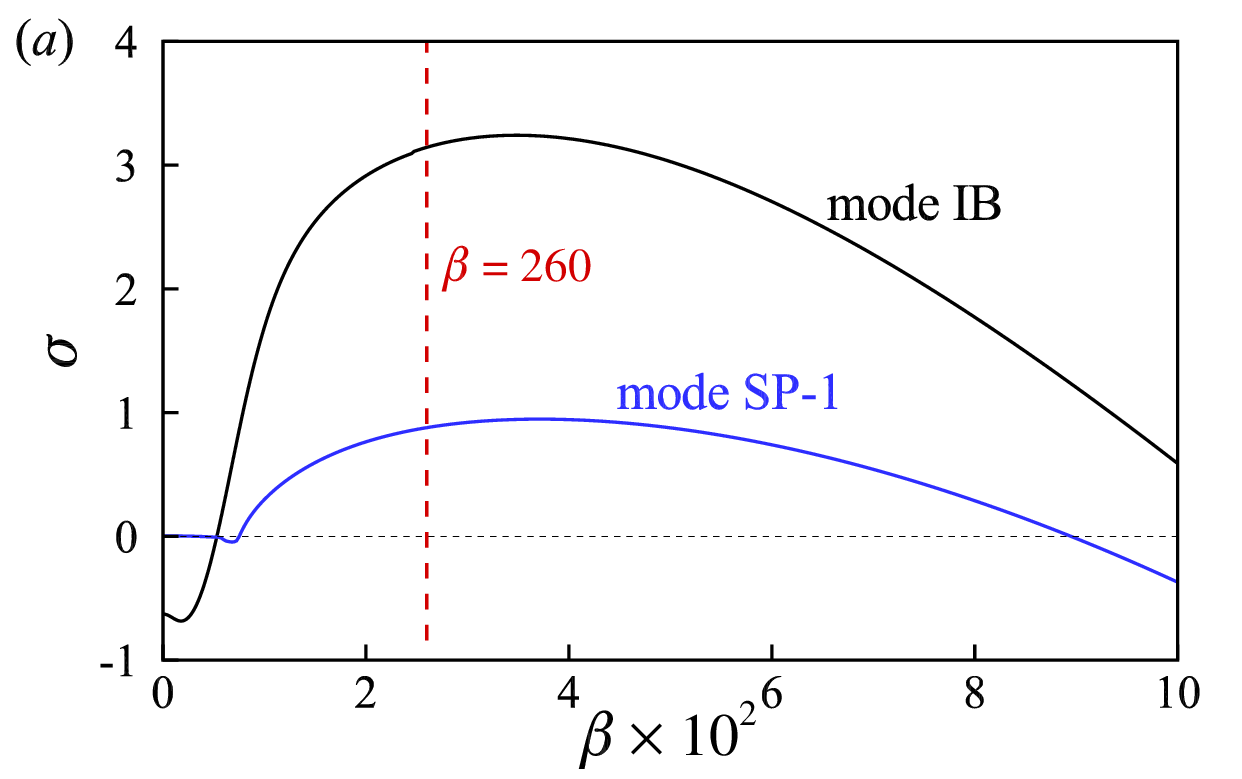}}
	\centering{ \includegraphics[height=5cm]{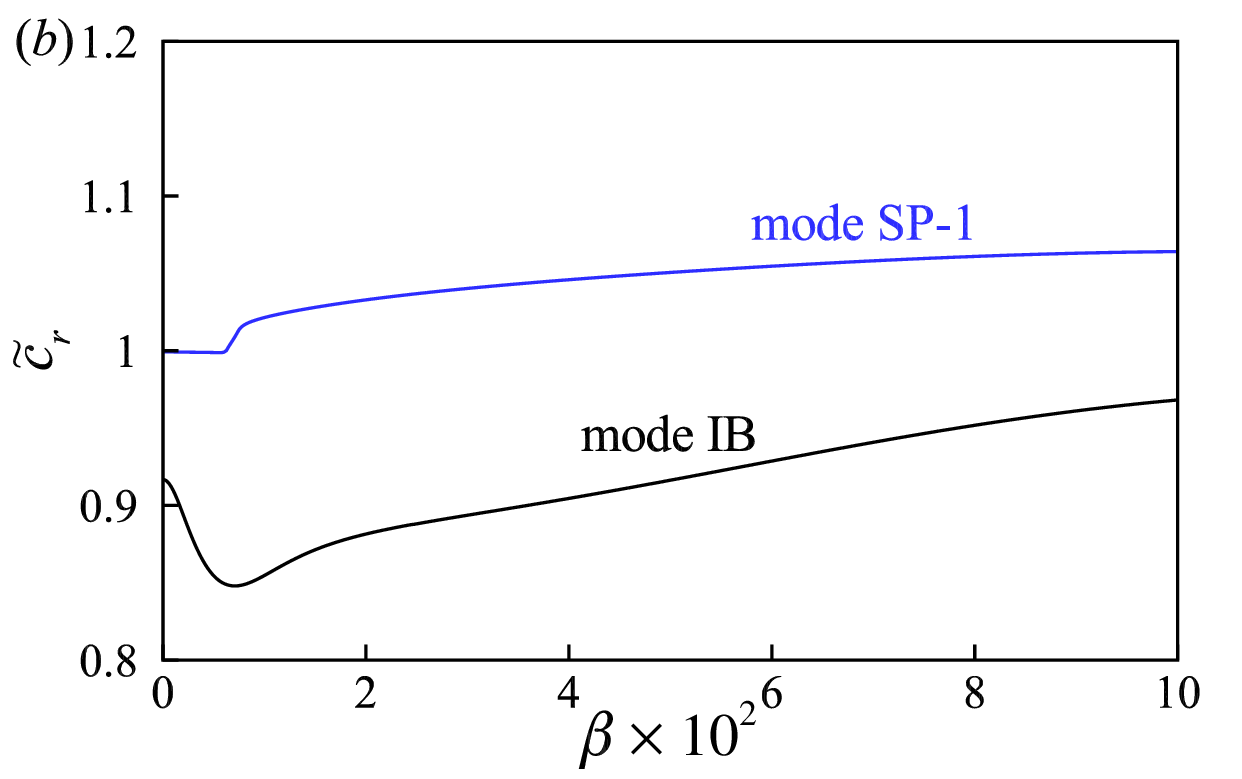}}
	\caption{(\textit{a}) Local growth rate  $\sigma$ and (\textit{b}) $\tilde{c}_r$ versus spanwise wavenumber $\beta$ for the unstable modes identified by LST ($K\ne0$) with $\omega=20$ at $x=0.9$ in the separation bubble. $\beta=260$ refers to the globally most amplified spanwise wavenumber with $\omega=20$ by \replaced{Resolvent}{resolvent} analysis.}
	\label{fig_growthrate_3D}
\end{figure*} 

As shown by \replaced{Resolvent}{resolvent} analysis of Hao et al. under the same flow condition \cite{hao2023response}, stationary streaks and low-frequency oblique waves tend to be the most amplified disturbances by the compression ramp flow. Low-frequency oblique waves can also trigger the three-dimensionality of flows and the transition to turbulence \cite{Lugrin2021,Caillaud2024Separation}. Therefore, in addition to planar waves, 3-D oblique waves are necessary to be considered for the $\omega=20$ case. With a fixed frequency $\omega=20$, \replaced{Resolvent}{resolvent} analysis for different spanwise wavenumbers reports that the optimal gain $G$ peaks at around $\beta=260$, i.e., the preferential wavenumber. The gain curve against the spanwise wavenumber is provided in Fig. \ref{figA_gain} in the Appendix.

\begin{figure*}
	\centering{ \includegraphics[width=8.0cm]{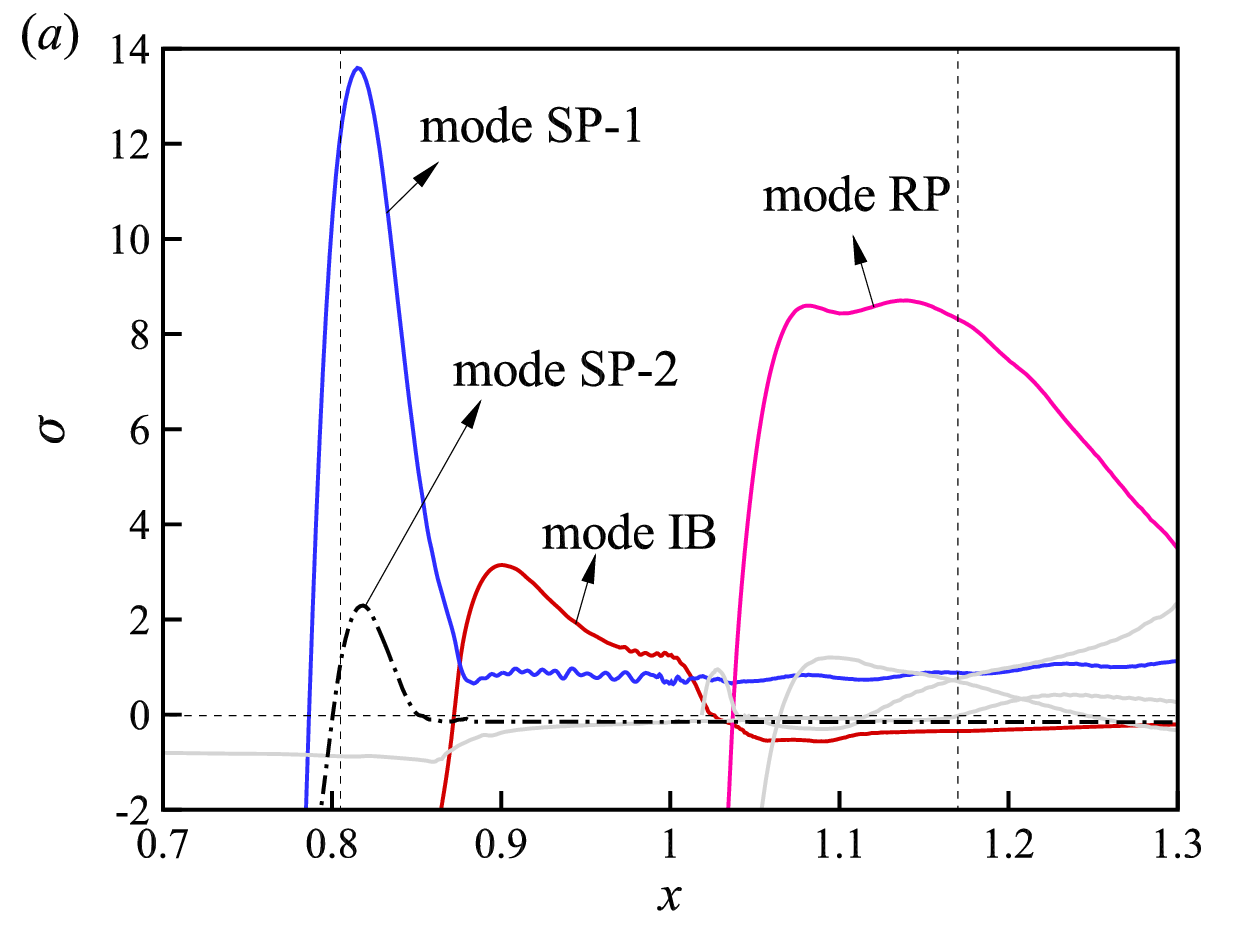}}
	\centering{ \includegraphics[width=8.0cm]{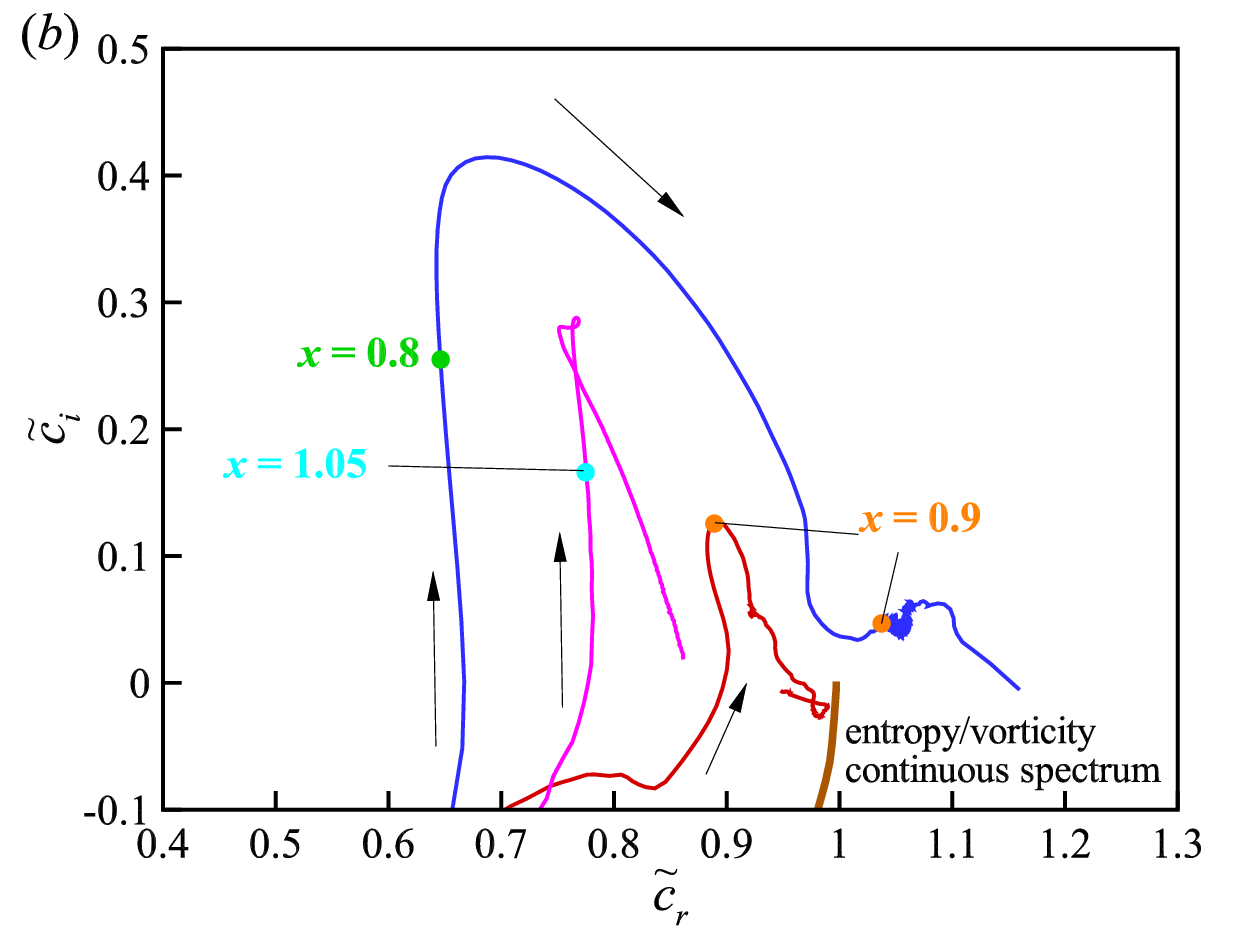}}
	\caption{(\textit{a}) Local growth rate $\sigma$ and (\textit{b}) eigenvalue trajectories with an increasing $x$ coordinate identified by LST ($K\ne0$) with $\omega=20$ and $\beta=260$. Solid trajectory lines in (\textit{b}) correspond to those in (\textit{a}) with the same line colour. Vertical dashed lines in (\textit{a}), locations of separation and reattachment. Gray lines in (\textit{a}), other lesser unstable modes near reattachment.}
	\label{fig_omega20_trajectory}
\end{figure*}

Before discussing the multi-modal feature of the 3-D low-frequency mode, we first show the parametric dependence of the mode on the spanwise wavenumber. Taking the streamwise location $x=0.9$ for example, we provide the growth rate and the phase velocity of the two unstable modes identified by LST in Fig. \ref{fig_growthrate_3D}. The eigenvalues of the two modes are tracked with a varying spanwise wavenumber and a constant $\omega=20$. The phase velocities of \replaced{modes SP-1 and IB}{modes I and II} are generally larger and smaller than unity, respectively. Both local modes possess a wide unstable range with respect to the spanwise wavenumber. Clearly, the globally most amplified wavenumber $\beta=260$ falls well in the unstable region of the oblique \replaced{modes SP-1 and IB}{modes I and II}. The value $\beta=260$ is also close to the most unstable wavenumber of the two local modes. It should be noticed that the oblique angle of such a wave with, say $c_r=1$, is $\theta=\tan^{-1}(\beta/\alpha_{r})=\tan^{-1}(260/20)\approx85.6^{\circ}$, and therefore the Mach number in the wave propagation direction is $\tilde{M}=M_{\infty}\cos\theta\approx0.6$. This low equivalent Mach number may account for why the high compressibility in this hypersonic state does not significantly suppress the shear-induced oblique mode, which is consistent with the conclusion of the compressibility study on the K--H instability \cite{Karimi2015thesis}. By contrast, the 2-D mode ($\beta=0$) with the same frequency is stable with a non-positive growth rate in Fig. \ref{fig_growthrate_3D}(\textit{a}).

With a fixed $\beta=260$, Fig. \ref{fig_omega20_trajectory} further depicts the growth rate evolution along the streamwise direction and the eigenvalue trajectory of the three dominant modes. \replaced{Modes SP-1, IB and RP}{Modes I, II and III} successively have the largest growth rate near the separation point, in the separation region and near the reattachment point. Near the separation point, another unstable \replaced{mode SP-2}{mode IV} is found, which is less important than \replaced{mode SP-1}{mode I}. In the vicinity of the reattachment point, three additional unstable modes are identified, whose growth rates are evidently smaller than that of \replaced{mode RP}{mode III}. Further based on the eigenvalue trajectories, \replaced{modes SP-1, IB and RP}{Modes I, II and III} independently evolve from very stable discrete modes, which appear successively on the stable half plane of the complex phase velocity. At the station $x=0.9$, the growth rate of \replaced{mode IB}{mode II} reaches its peak, while \replaced{mode SP-1}{mode I} is also unstable with a lower growth rate. In other words, unstable \replaced{modes SP-1 and IB}{modes I and II} coexist at $x=0.9$ in the separation bubble. It should also be noted that all these identified unstable modes almost become stable if the streamline curvature $K$ is set to zero everywhere. This observation suggests that these unstable low-frequency 3-D modes are sensitive to the curvature effect, which seemingly behave as unsteady G{\"o}rtler vortices. \deleted{The feature that several unstable modes emerge in the high-curvature region (near the separation and reattachment points) also appears to be similar to the multi-modal property of unstable stationary G{\"o}rtler modes reported by Ren \& Fu \cite{ren2014competition}.}

\begin{figure*}
	\centering{ \includegraphics[height=6cm]{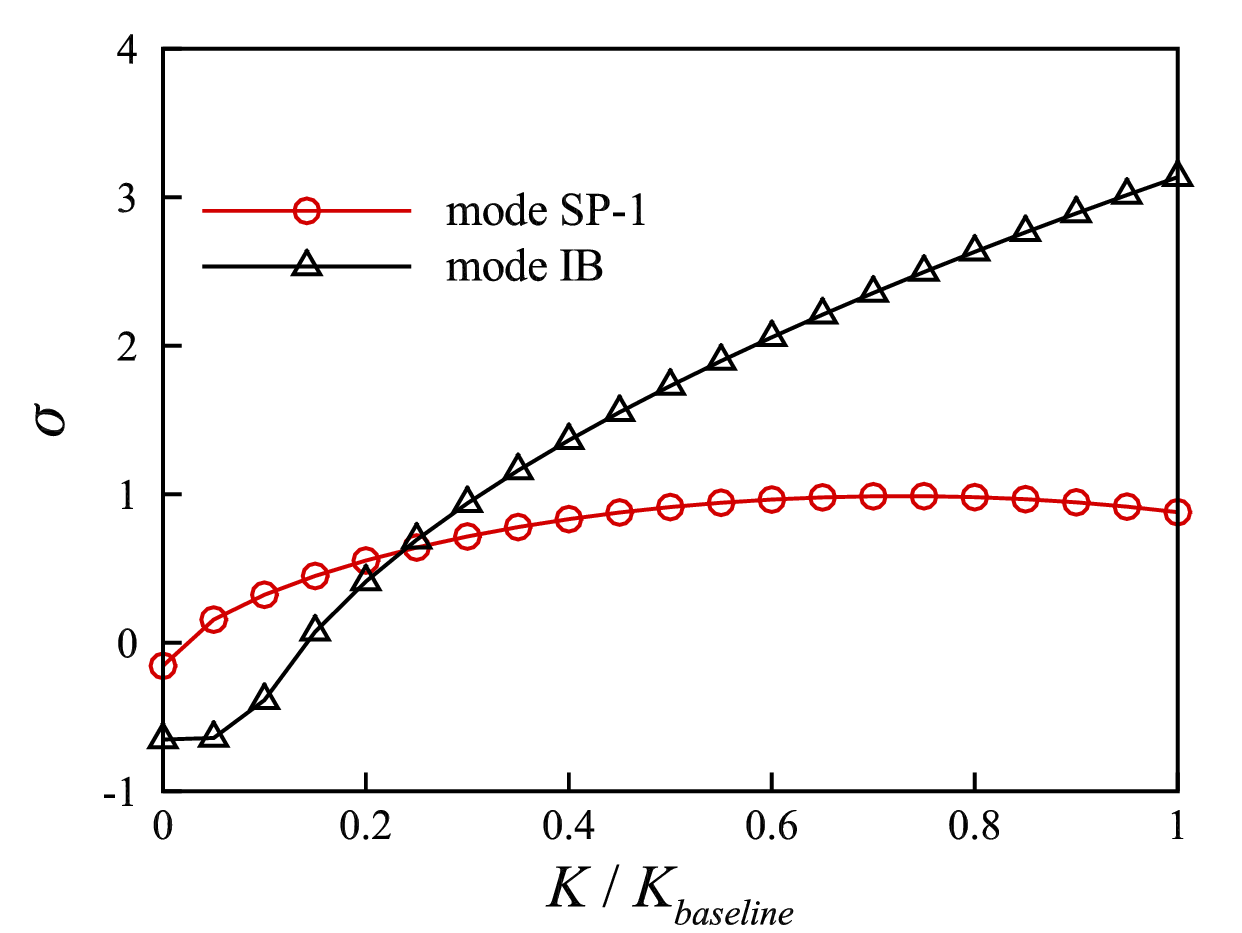}}
	\caption{Local growth rate versus the varying curvature ratio $K/K_\textit{baseline}$ for \replaced{modes SP-1 and IB}{modes I and II} with $\omega=20$ and $\beta=260$ at $x=0.9$, where $K_\textit{baseline}$ refers to the actual local curvature.}
	\label{fig_growthrate_K}
\end{figure*}
\begin{figure*}
	\centering{ \includegraphics[height=6.0cm]{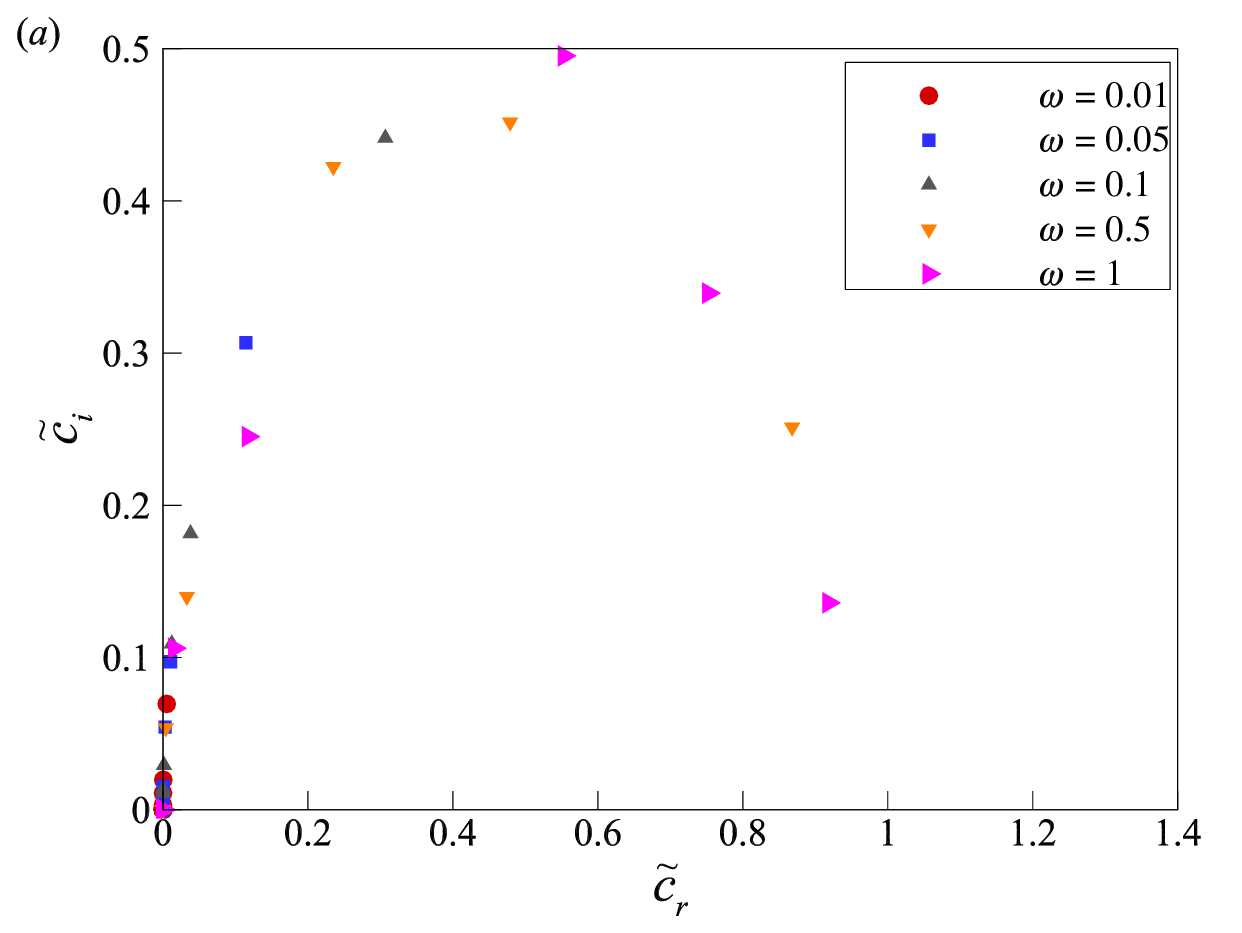}}
	\centering{ \includegraphics[height=6.0cm]{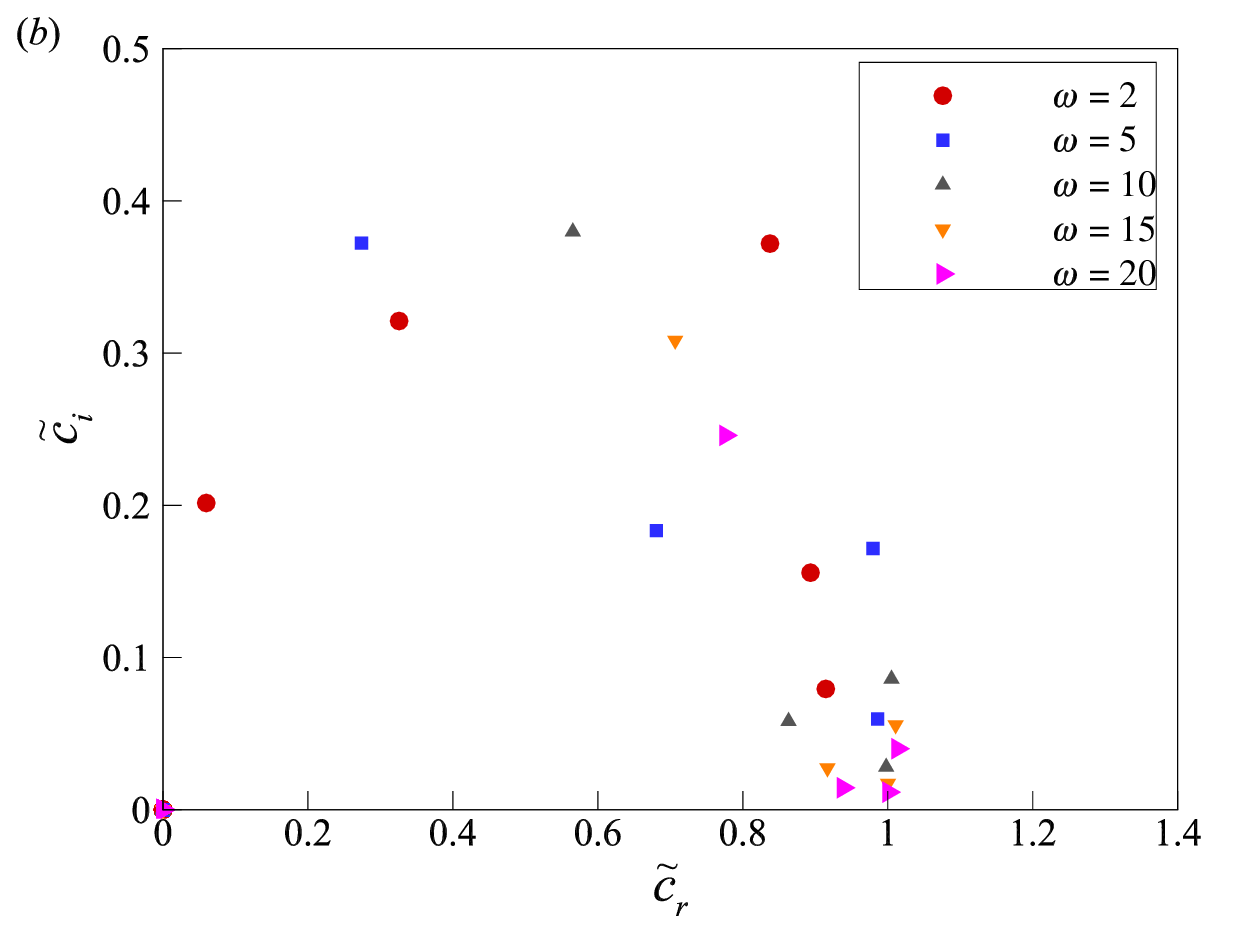}}
	\caption{Complex phase velocity of the unstable discrete modes for the case with corner rounding: (\textit{a}) $0.01\le\omega\le1$ and (\textit{b}) $2\le\omega\le20$. Other settings include $\beta=260$ and $x=1.1$.}
	\label{fig_eigenspectra_gortler}
\end{figure*}

To reveal the effect of the curvature on the multiple unstable modes, the value of $K$ is artificially reduced, starting from the actual value $K_\textit{baseline}$ for the baseline separated flow without corner rounding. Fig. \ref{fig_growthrate_K} illustrates that the originally unstable \replaced{modes SP-1 and IB}{Modes I and II} at $x=0.9$ gradually become stable when the curvature parameter $K$ is changed from $K_\textit{baseline}$ to zero. Therefore, the streamline curvature is necessary for the appearance of multiple unstable low-frequency 3-D modes in the separation region. The sensitivity to the curvature might be analogous to the stationary or unsteady G{\"o}rtler instabilities in a boundary layer over a concave wall. We further calculate the case with corner rounding under the settings $x=1.1$, $\omega=0$ and $\beta=260$, where the actual value of $K$ is used. Without flow separation, several unstable eigenmodes with $\alpha_{r}=0$ are identified, which is consistent with the observation of multiple stationary G{\"o}rtler modes by Ren \& Fu \cite{ren2014competition}. As $\omega$ is increased from 0.01 to 20, Fig. \ref{fig_eigenspectra_gortler} shows the evolution of $\tilde{c}$ for the identified unstable discrete modes. For the small value $\omega=0.01$, the real parts of the complex phase velocity $\tilde{c}_r$ of unstable modes are almost zero, which implies a stationary behaviour. As the angular frequency gradually grows, most of the unstable eigenmodes are moved away from the vertical axis $\tilde{c}_r=0$. The nonzero phase velocity indicates that the disturbance can be convected by the base flow. In this case, the unstable eigenmode with a small yet unignorable  angular frequency may be regarded as unsteady G{\"o}rtler modes.

\subsection{\added{Comparison between local and nonlocal methods for shear-layer modes}}\label{sec:4.5}

\begin{figure*}[t]
	\centering{ \includegraphics[height=13.0cm]{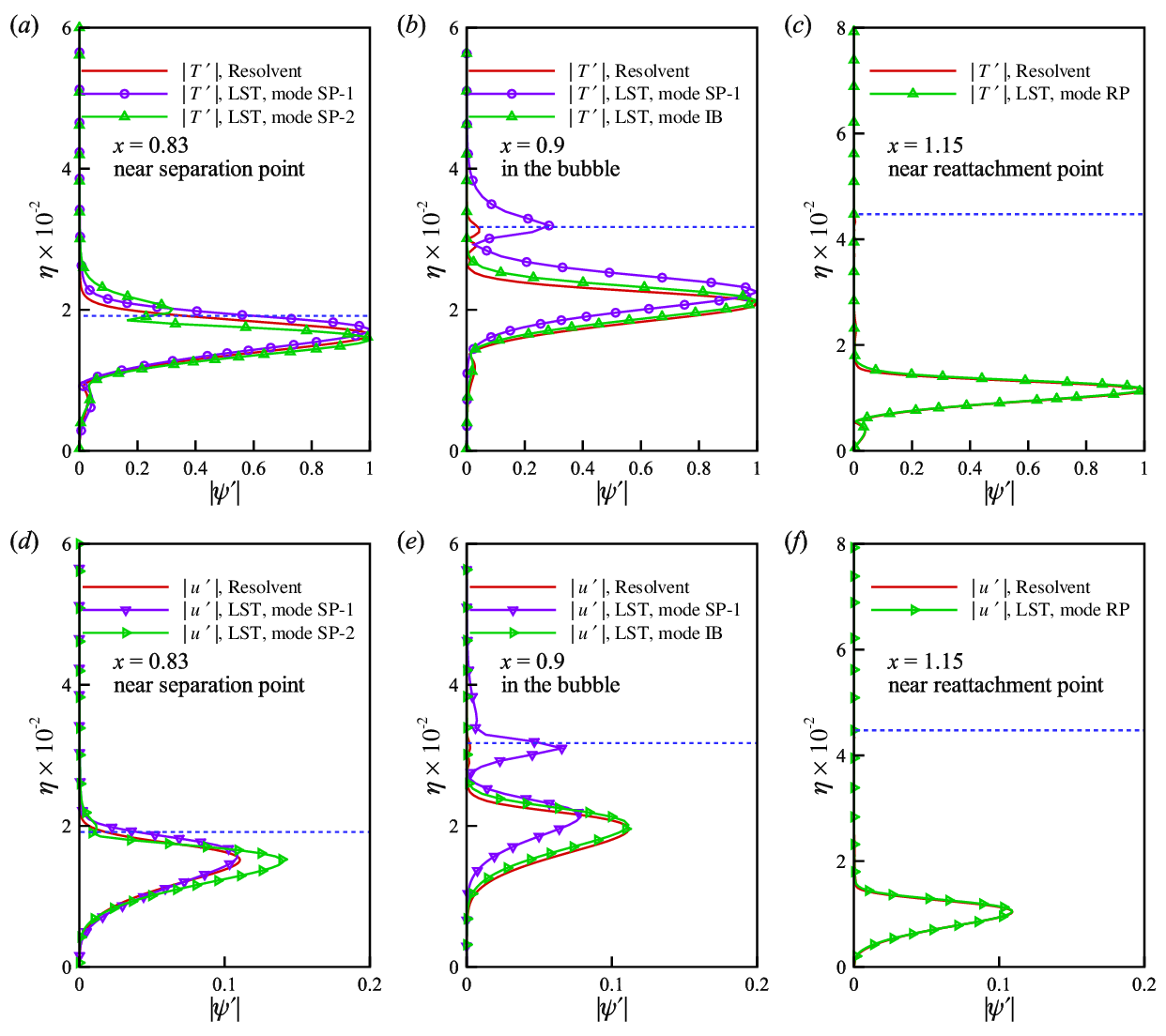}}
	\caption{Comparison of the disturbance shape for $\omega=20$ and $\beta=260$ between the \replaced{Resolvent}{resolvent} response and the LST eigenfunction\deleted{ at (\textit{a}) $x=0.83$ near the separation point, (\textit{b}) $x=0.9$ in the separation region, and (\textit{c}) $x=1.15$ near the reattachment point}. Horizontal dashed line, location of the separation shock. All the results are normalised by the maximum $|T'|$.}
	\label{fig_sl_mode_shape}
\end{figure*}

To further clarify the multi-modal finding, the disturbance shape is compared at three representation locations between \replaced{Resolvent}{resolvent} analysis and LST. In \replaced{Figs. \ref{fig_sl_mode_shape}(\textit{a},\textit{d})}{Fig. \ref{fig_sl_mode_shape}(\textit{a})}, \replaced{mode SP-1}{mode I} is predominant near the separation point at $x=0.83$, while \replaced{mode SP-2}{mode IV} is lesser. The shape of the \replaced{Resolvent}{resolvent} response seems to be jointly affected by the eigenfunctions of \replaced{modes SP-1 and SP-2}{mode I and IV}. The \replaced{Resolvent}{resolvent} response is generally more similar to \replaced{mode SP-1}{mode I}, because both of their profiles have only one peak of $|T'|$\added{ and they are closer in $|u'|$}. In \replaced{Figs. \ref{fig_sl_mode_shape}(\textit{b},\textit{e})}{Fig. \ref{fig_sl_mode_shape}(\textit{b})}, \replaced{mode IB}{mode II} evolves to be the dominant one at $x=0.9$, and the $|u'|$ and $|T'|$ of the \replaced{Resolvent}{resolvent} response become more similar to the eigenfunction of \replaced{mode IB}{mode II} rather than the less unstable \replaced{mode SP-1}{mode I}. Nevertheless, at $x=0.9$, the second peak of $|T'|$ of the unstable \replaced{mode SP-1}{mode I} near the separation shock (denoted by the horizontal dashed line) seems to affect the shape of the \replaced{Resolvent}{resolvent} response there, and thus $|T'|$ of the \replaced{Resolvent}{resolvent} response near the separation shock becomes visible in Fig. \ref{fig_sl_mode_shape}(\textit{b}). Meanwhile, the peaks of the eigenfunction modulus of both \replaced{modes SP-1 and IB}{modes I and II} are not very far away from the separation shock, where the streamline curvature is large in Fig. \ref{fig1}(\textit{c}). This observation may account for why the unstable \replaced{mode IB}{mode II}, not located at the separation and reattachment points, is still sensitive to the curvature effect.  In \replaced{Figs. \ref{fig_sl_mode_shape}(\textit{c},\textit{f})}{Fig. \ref{fig_sl_mode_shape}(\textit{c})}, \replaced{mode RP}{mode III} has an evidently larger growth rate than the remaining unstable modes near the reattachment point. An excellent agreement in the $|u'|$ and $|T'|$ is reached between the \replaced{Resolvent}{resolvent} response and the locally dominant \replaced{mode RP}{mode III}. 

\begin{figure*}[h]
	\centering{ \includegraphics[height=6.0cm]{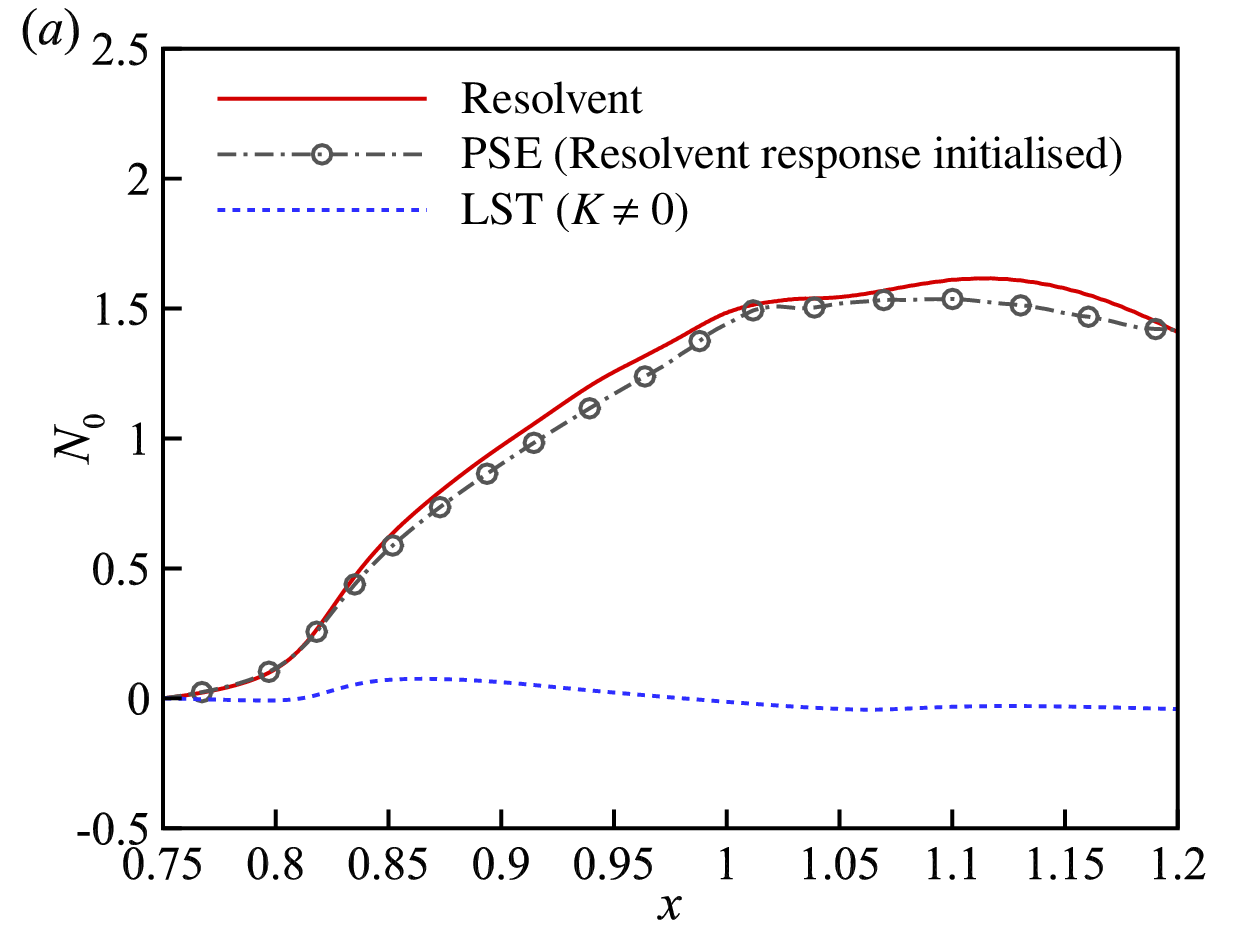}}
	\centering{ \includegraphics[height=6.0cm]{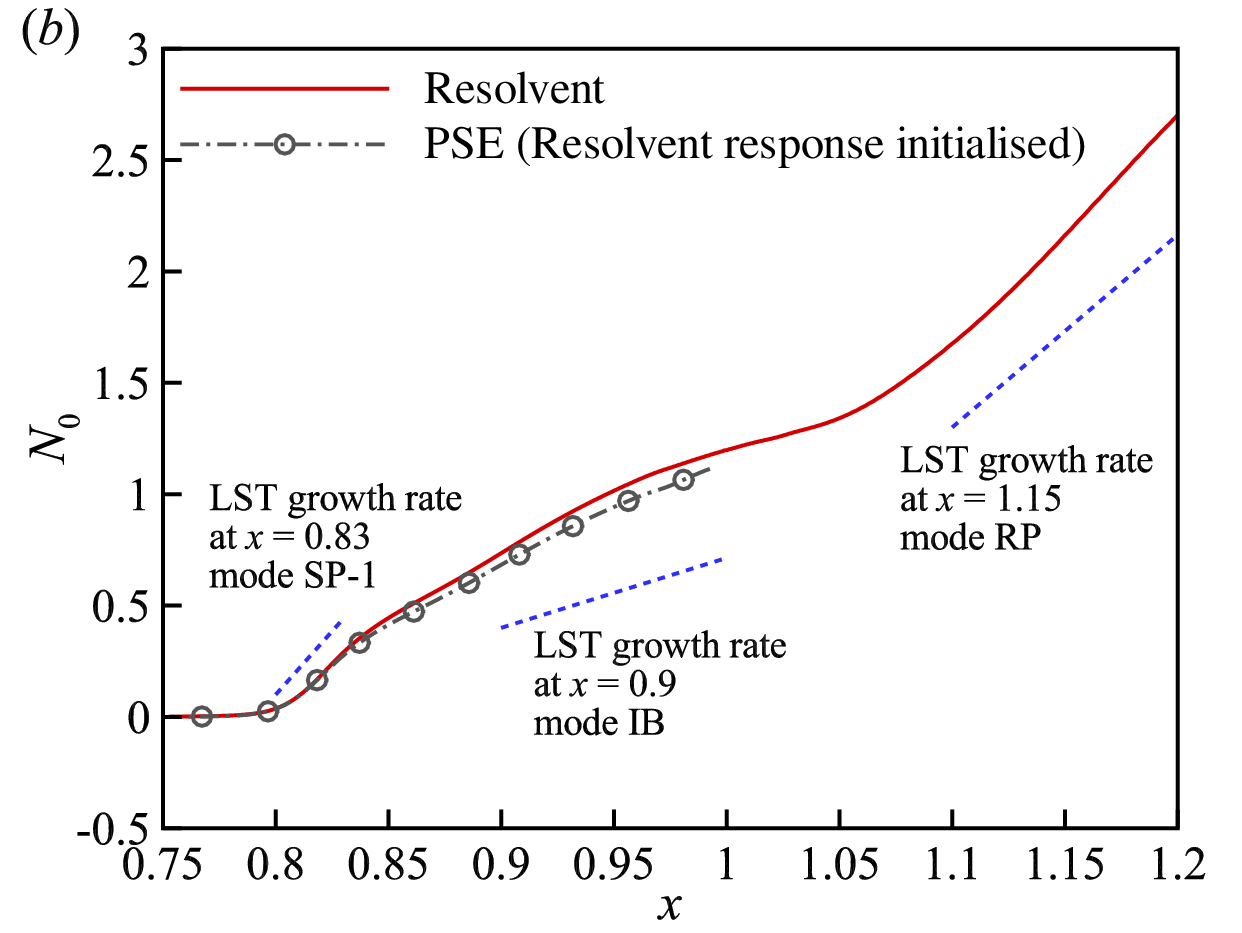}}
	\caption{$N$-factor based on $x_0=0.75$ of cases without corner rounding for $\omega=20$ with (\textit{a}) $\beta=0$  and (\textit{b}) $\beta=260$.}
	\label{fig_nfactor_omega20}
\end{figure*}

\begin{figure*}[h]
	\centering{ \includegraphics[height=2cm]{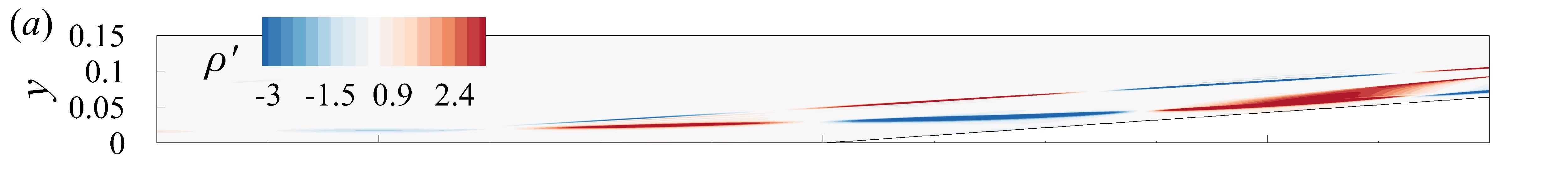}}
	\centering{ \includegraphics[height=2cm]{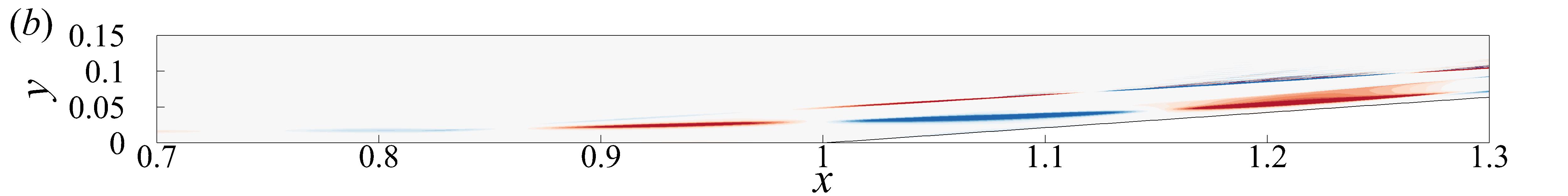}}
	\caption{Comparison of $\rho'$ for the planar shear-layer mode $\omega=20$ between  (\textit{a}) \replaced{Resolvent}{resolvent} analysis and (\textit{b}) PSE initialised by \replaced{Resolvent}{resolvent} response, all normalised by the maximum at $x=0.7$.}
	\label{fig_omega20_2Dshape}
\end{figure*}

Representative results of the $N$-factor and the density fluctuation for the low-frequency disturbance $\omega=20$ are shown in Figs. \ref{fig_nfactor_omega20} and \ref{fig_omega20_2Dshape}, respectively. The density fluctuation of the `shear-layer mode' is concentrated on the separated shear layer in Fig. \ref{fig_omega20_2Dshape}. \replaced{In Fig. \ref{fig_nfactor_omega20}(\textit{a}), the Chu's energy for $\omega=20$, calculated by LST, demonstrates that the 2-D low-frequency mode is not considerably amplified. This finding is consistent with Fig. \ref{fig5}(\textit{b}).}{Consistent with Fig. \ref{fig5}(\textit{b}), the LST result in Fig. \ref{fig_nfactor_omega20}(\textit{a}) also shows that Chu's energy of the 2-D local mode is not considerably amplified corresponding to $\omega=20$.} By contrast, PSE initialised by the \replaced{Resolvent}{resolvent} response nearly reproduces the optimal energy growth of the \replaced{Resolvent}{resolvent} analysis in Fig. \ref{fig_nfactor_omega20}(\textit{a}) and the disturbance shape in Fig. \ref{fig_omega20_2Dshape}(\textit{b}). Acceptable difference exists possibly because of dropping higher-order terms in the PSE. As a result, the \replaced{non-modal}{non-modality} nature of the operator may further participate in determining whether the response is amplified or attenuated in the separation bubble. The difference in the parallel LST operator and the full LNSE operator contributes to the fact that the 2-D mode is damped in a local analysis, whereas a maximal energy growth is achieved in the optimisation problem of the \replaced{Resolvent}{resolvent} analysis.

Corresponding to the preferential wavenumber $\beta=260$, the $N$-factor results  are shown in Fig. \ref{fig_nfactor_omega20}(\textit{b}). For this oblique wave, the iterative scheme Eq. (\ref{eq_normalization}) of PSE does not converge on the ramp ($x>1$) numerically. Nevertheless, the optimal growth on the plate region is still captured by PSE. The energy of the low-frequency 3-D wave is constantly amplified by the separation bubble. With regard to the local analysis, \replaced{the growth rates of the aforementioned three dominant local modes are also included for a comparison.}{we find multiple unstable 3-D modes in the separation bubble, among which three modes are dominant in different streamwise locations. These three modes are named in the order of appearance, including mode I, mode II and mode III. The three modes possess the largest local growth rate near the separation point, in the separation region and near the reattachment point, respectively.} Note that there is an equivalence between the local growth rate in LST and that of the defined $N$-factor, i.e., $\sigma=dN_0/dx$. As a consequence, the slope of the $N$-factor can be directly given by LST, which is marked by the dashed line in Fig. \ref{fig_nfactor_omega20}(\textit{b}). Good agreement in the growth rate between the \replaced{Resolvent}{resolvent} analysis and LST is reached\deleted{ near the separation and reattachment points}. \added{The agreement supports the dominance of the three local modes at different streamwise locations. In other words, the continuous amplification of low-frequency components is not contributed by a single G{\"o}rtler mode.} \deleted{In the separation region ($x=0.9$), the growth rate of the \replaced{Resolvent}{resolvent} analysis is slightly larger than that of LST. This is probably because mode II is not the only unstable mode with considerable dominance at $x=0.9$, while mode I and mode III are predominant unstable modes near the separation and reattachment points, respectively. The corresponding arguments will be shown later in Fig. \ref{fig_omega20_trajectory}(\textit{a}).}

To conclude, the growth of the optimal low-frequency 3-D wave can be well explained by the parallel-flow eigenmodal analysis. Under the considered conditions, the separation bubble tends to select 3-D unstable low-frequency waves while freezing the 2-D high-frequency Mack modes.

\section{Conclusion}\label{sec:Conclusions}
In the present paper, the linear dynamics of the separation bubble as a noise amplifier in a compression ramp flow is investigated. The main contribution is that an explanation for the neutral Mack mode and the possibly amplified low-frequency `shear-layer' mode in the separation region is provided. The high-frequency components related to Mack modes are nearly neutral due to several modal synchronisations, which generate discontinuous unstable regions in the separation bubble. Based on a local analysis, the stability diagram of multiple Mack modes is clearly presented, and multiple newly generated discrete modes are identified and discussed. The eigenfunctions in the local analysis agree well with the disturbance shapes in the \replaced{Resolvent}{resolvent} analysis. Furthermore, the upstream and downstream Mack second modes originate from different branches of discrete modes due to flow separation. Although local analysis reports several discontinuous instabilities with high frequencies, it fails for the 2-D low-frequency `shear-layer mode'. However, several unstable 3-D eigenmodes are identified by LST, which are sensitive to the curvature effect. Among them, three dominant unstable modes (I, II and III) successively appear near the separation point, in the separation bubble and near the reattachment point. Good agreement is reached between the local mode and the \replaced{Resolvent}{resolvent} response regarding the disturbance shape and the growth rate of energy. The globally preferential spanwise wavenumber falls in the unstable range of the local unstable modes. At a fixed streamwise location, multiple unstable 3-D low-frequency modes may coexist, which is not observed in the calculation of high-frequency Mack modes. These multiple unstable modes constitute a continuous unstable region for the 3-D low-frequency disturbance in the bubble, which supports the constant amplification of energy as shown by \replaced{Resolvent}{resolvent} analysis. By contrast, Mack modes do not provide the eigenmodal foundation for a constantly amplification of high-frequency disturbances in the bubble. The present study indicates that the parallel-flow eigenmodal analysis is able to provide new insights into the convective-instability mechanism of separated flows.\\

\section*{Appendix: Other results of \replaced{Resolvent}{resolvent} analysis\added{ and LST}}
\begin{figure*}[htbp]
	\centering{ \includegraphics[height=5cm]{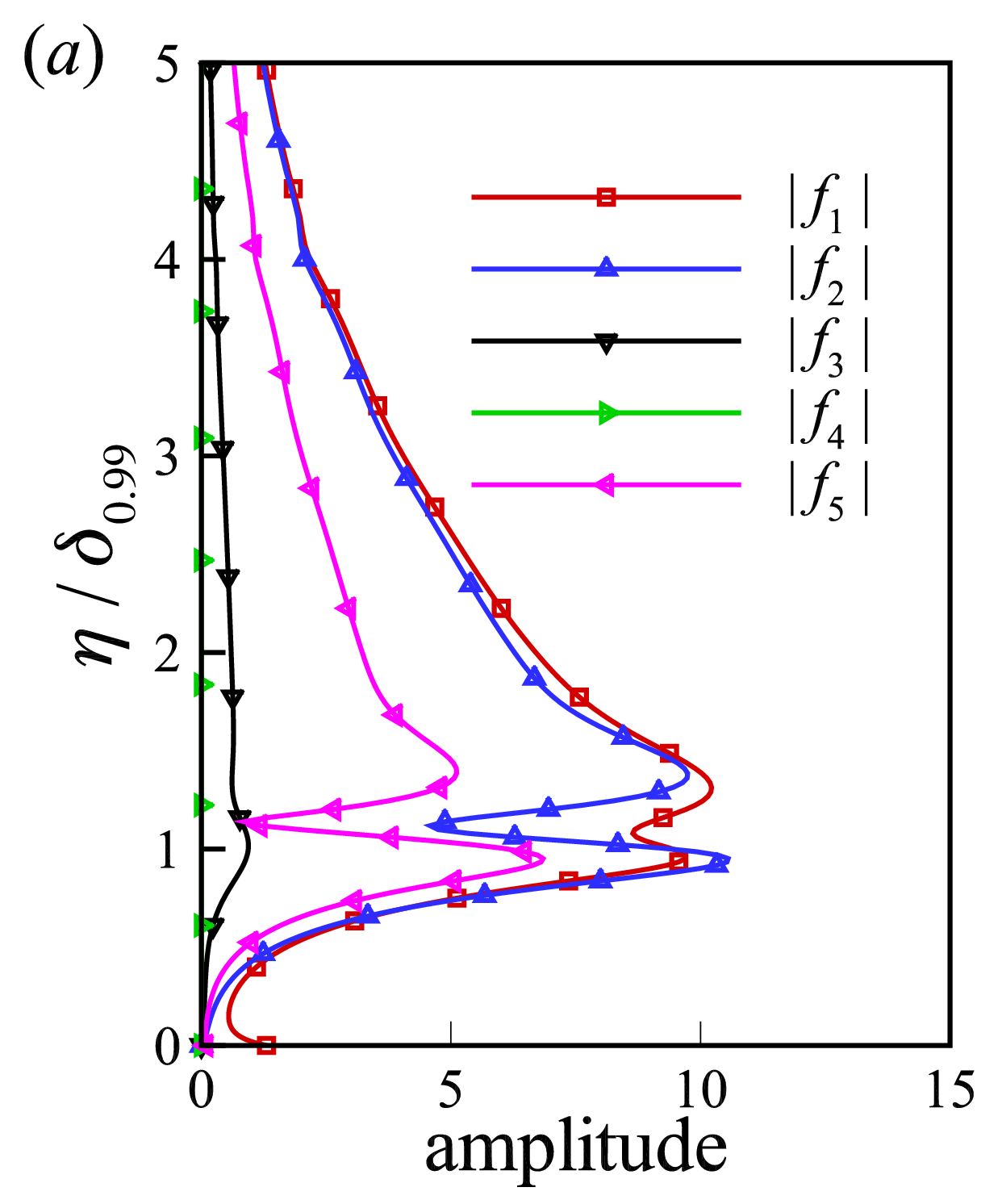}}
	\centering{ \includegraphics[height=5cm]{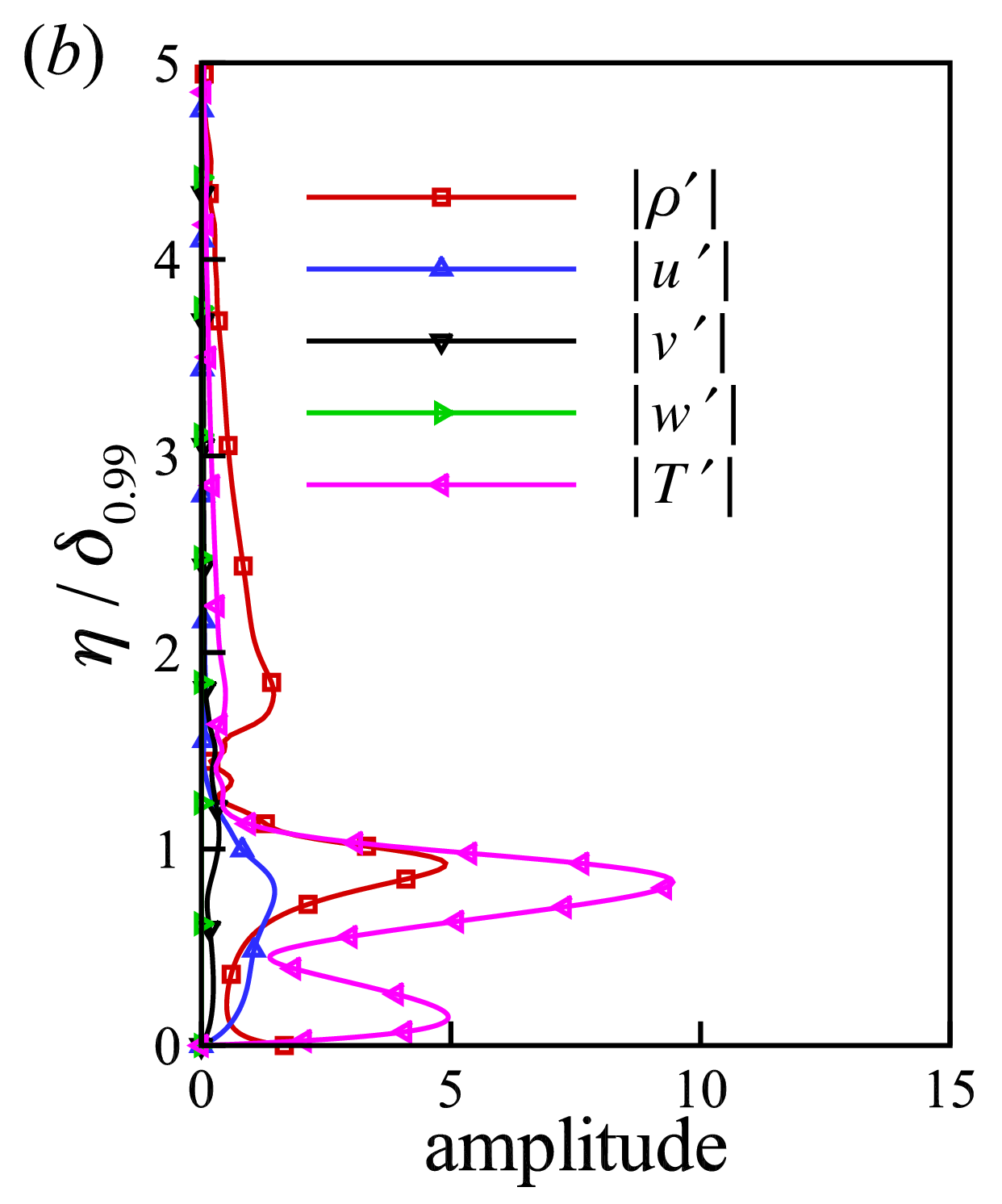}}
	\caption{Amplitudes of the input forcing at $x=0.2$ and optimal response at $x=0.25$ with $\omega=80$ and $\beta=0$. Here, $\delta_{0.99}$ refers to the local boundary layer thickness.}
	\label{figA_omega80}
\end{figure*}

\begin{figure*}[htbp]
	\centering{ \includegraphics[height=5cm]{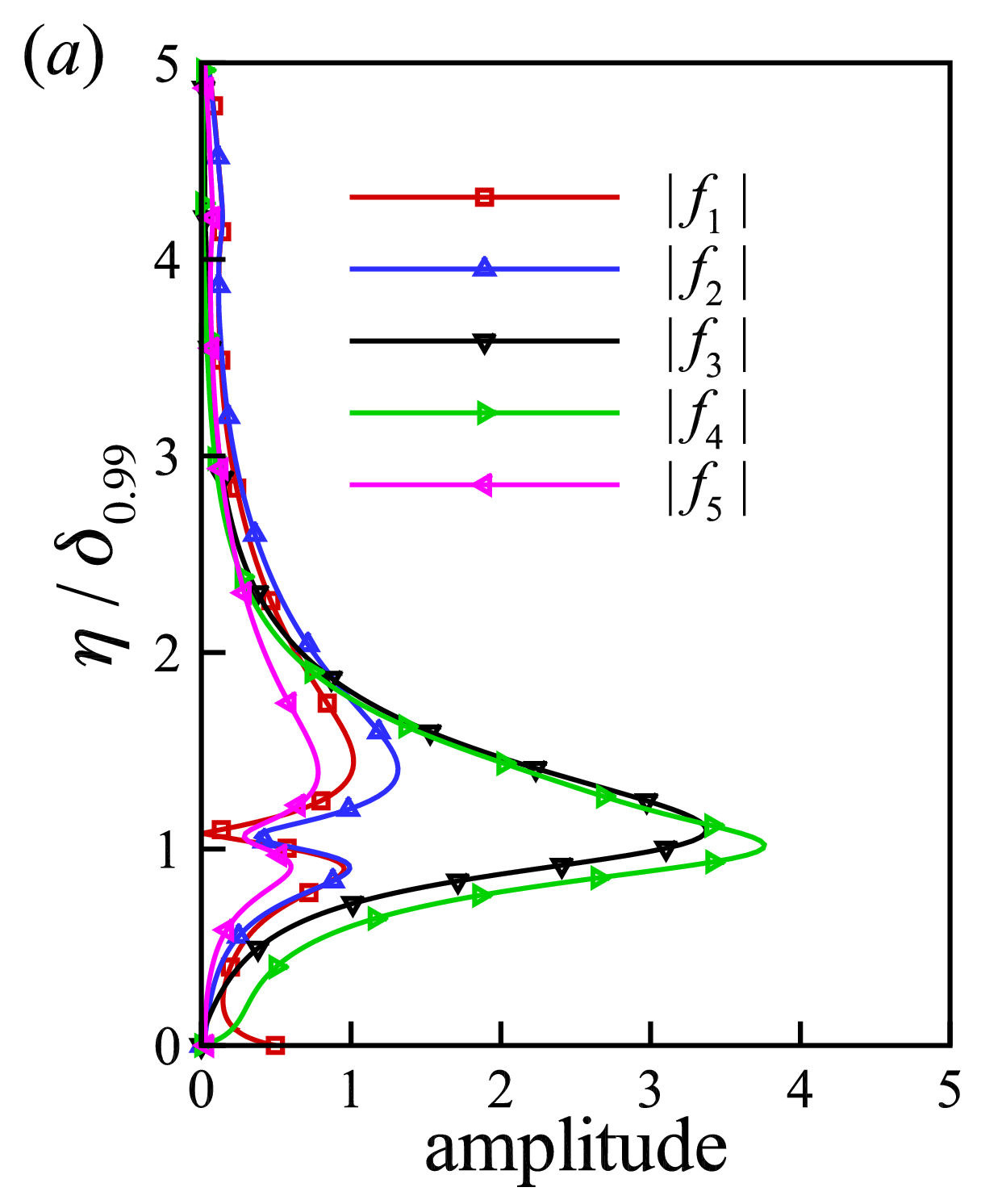}}
	\centering{ \includegraphics[height=5cm]{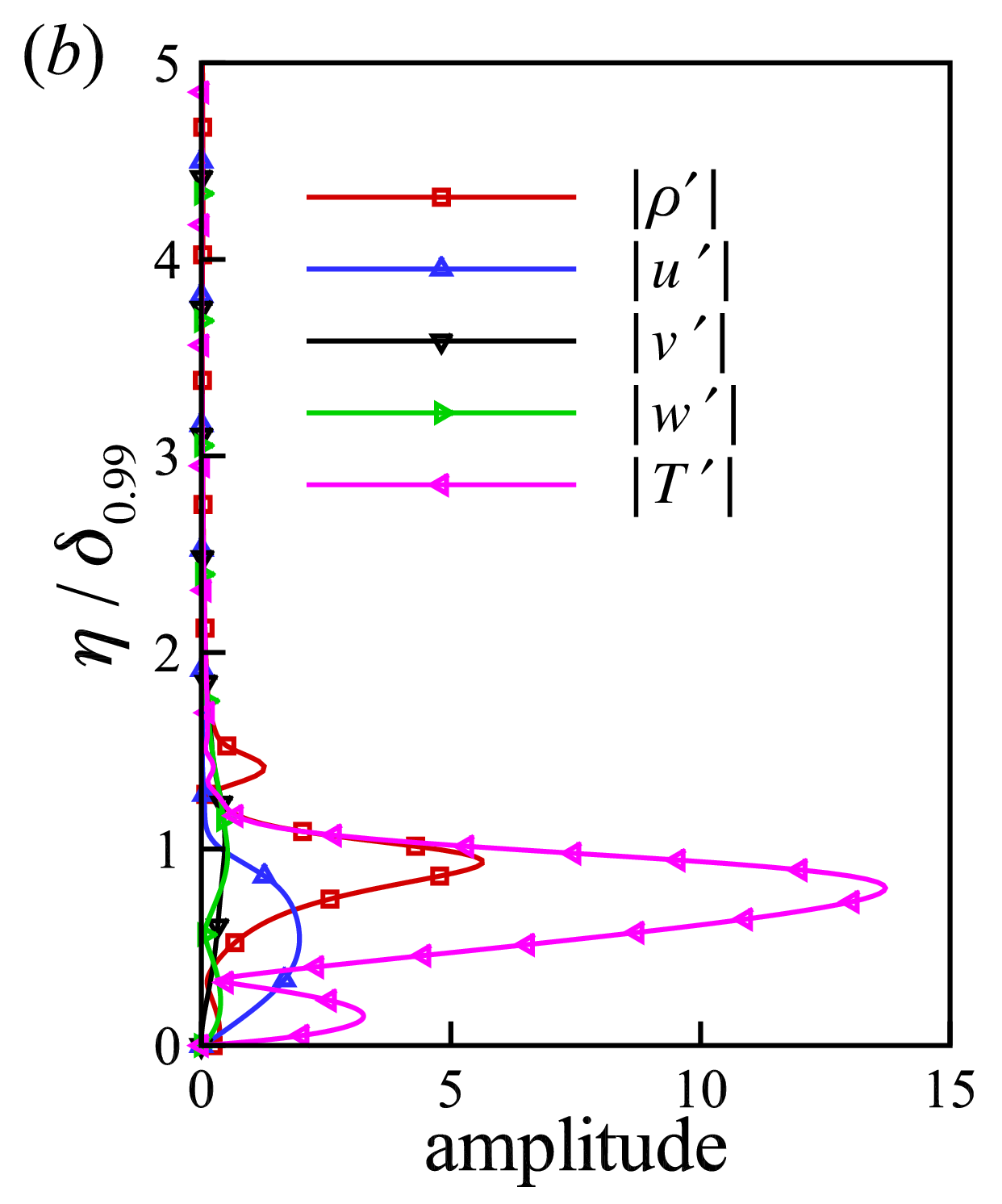}}
	\caption{Amplitudes of the input forcing at $x=0.2$ and optimal response at $x=0.25$ with $\omega=20$ and $\beta=260$. Here, $\delta_{0.99}$ refers to the local boundary layer thickness.}
	\label{figA_omega20}
\end{figure*}

\begin{figure}[htbp]
	\centering{ \includegraphics[height=5.5cm]{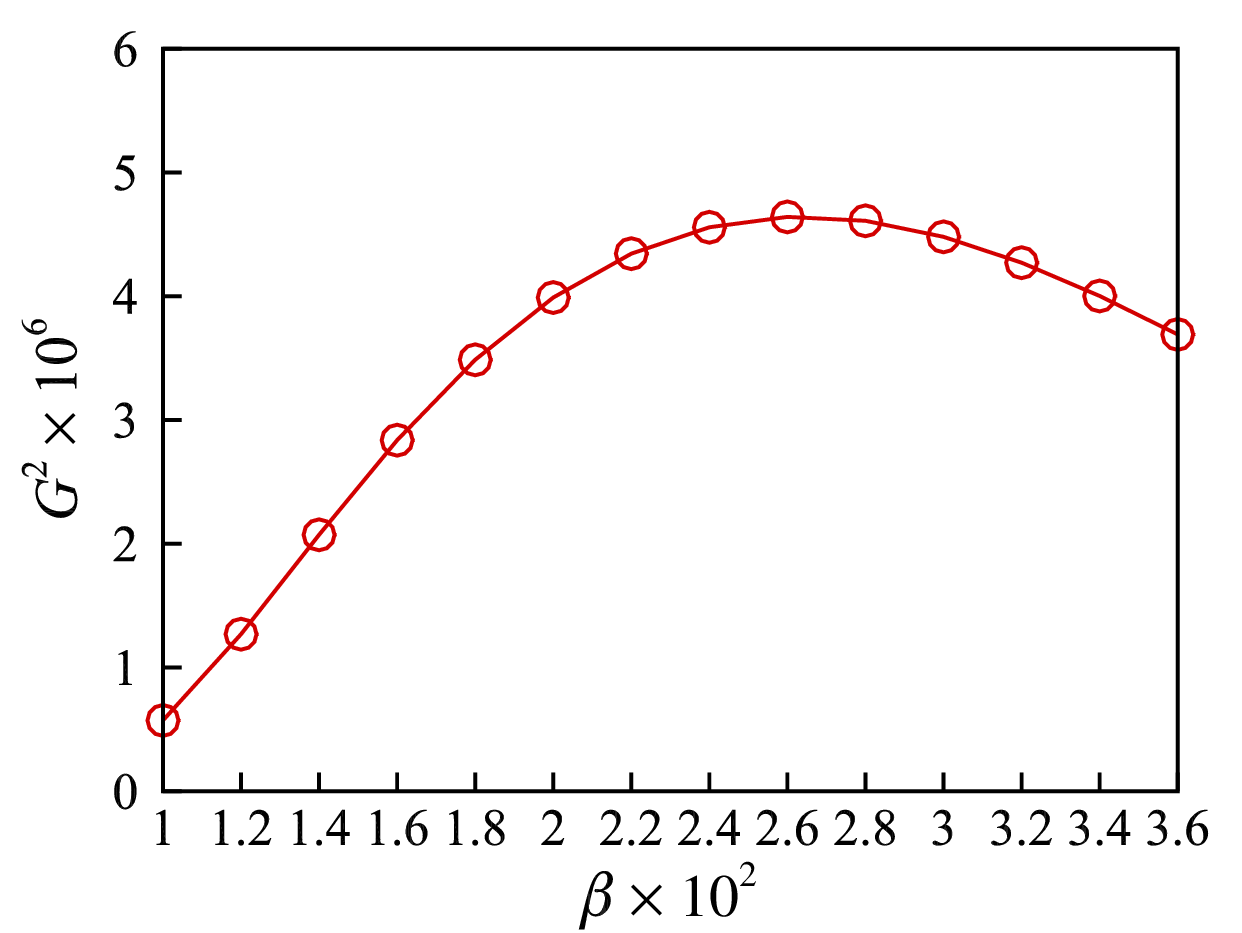}}
	\caption{Optimal gain  versus the spanwise wavenumber $\beta$ of the compression ramp flow for $\omega=20$.}
	\label{figA_gain}
\end{figure}

\begin{figure}[htbp]
	\centering{ \includegraphics[height=6cm]{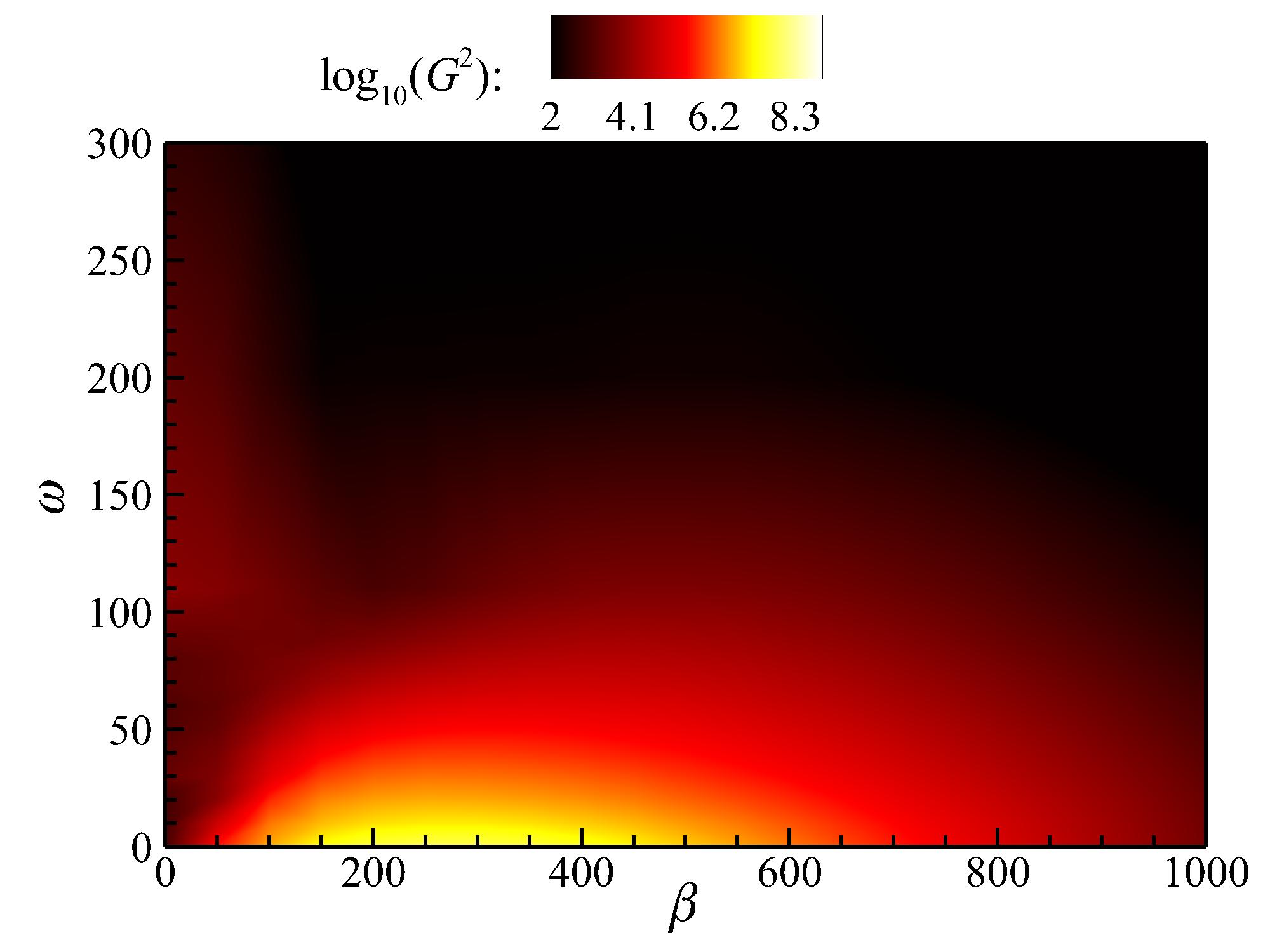}}
	\caption{\added{Common logarithm of the optimal gain  versus the spanwise wavenumber $\beta$ and the angular frequency $\omega$ of the compression ramp flow.}}
	\label{Appendix_gain_2D}
\end{figure}

\begin{figure}[htbp]
	\centering{ \includegraphics[width=8cm]{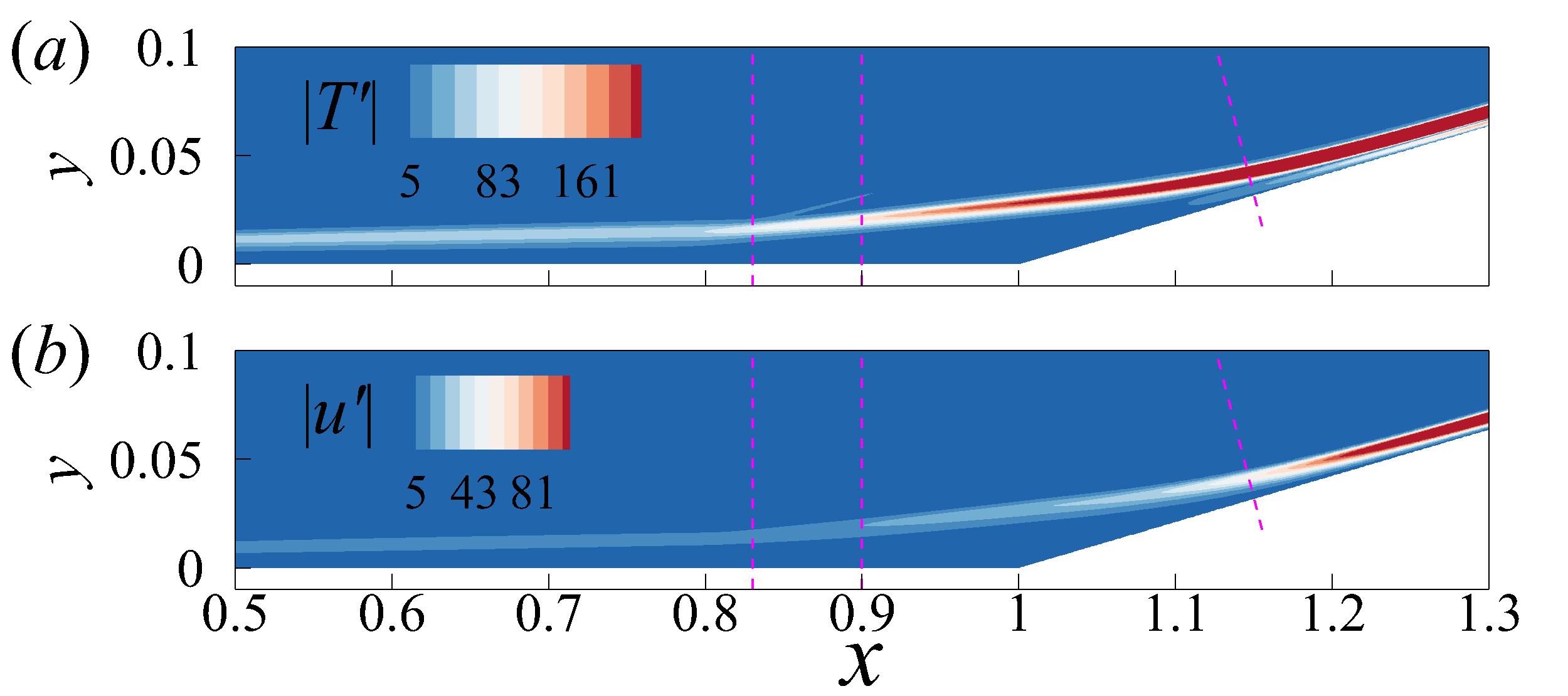}}
	\caption{\added{Modulus of the optimal response with respect to temperature and velocity disturbances for $\omega=20$ and $\beta=260$. Dashed lines, locations corresponding to $x=0.83$, $x=0.9$ and $x=1.15$ in Fig. \ref{fig_sl_mode_shape}.}}
	\label{omega20_beta260_output}
\end{figure}

Typical results of the optimal planar wave corresponding to 2-D Mack modes ($\omega=80$ and $\beta=0$) and of the optimal oblique wave corresponding to 3-D `shear-layer modes' ($\omega=20$ and $\beta=260$) are shown in this section. No corner rounding is applied here. Figs. \ref{figA_omega80} and \ref{figA_omega20} give the amplitudes of the input forcing and the closely downstream output response for the optimal planar wave and oblique wave, respectively. While the forcing in the $w$- momentum equation and the response $w'$ are zero for the planar wave, they are nonzero for the oblique wave. Meanwhile, although the input forcing has a pronounced amplitude outside the boundary layer, the energy of the downstream response is mainly concentrated inside the boundary layer. To reveal the preferential spanwise wavenumber of the 3-D `shear-layer mode', the optimal gain is plotted against $\beta$ in Fig. \ref{figA_gain} for the studied low frequency $\omega=20$. The most amplified wavenumber is found to be around $\beta=260$.\added{ The gain contour versus $\beta$ and $\omega$ is also given in Fig. \ref{Appendix_gain_2D}. For the considered 12$^\circ$ ramp angle, the optimal stationary streak present on the horizontal axis $\omega=0$ has an evidently larger gain than the optimal planar wave on the vertical axis $\beta=0$. Nonetheless, considering other factors such as the receptivity, the transition may be induced either by streaks or Mack modes (see our recent work \cite{Cao_Hao_Guo_2024}). In addition, the optimal planar wave has a moderate level of gain ($\log_{10}(G^2)\ge3.2$) with $\omega$ ranging from 70 to 200, which is probably associated with Mack modes. As a result, this frequency range is mostly discussed in Fig. \ref{fig4}(\textit{a}). The modulus of the optimal response is also shown in Fig. \ref{omega20_beta260_output} to record the streamwise evolution of the mode shape in Fig. \ref{fig_sl_mode_shape}.}

\begin{figure}[htbp]
	\centering{ \includegraphics[height=5.2cm]{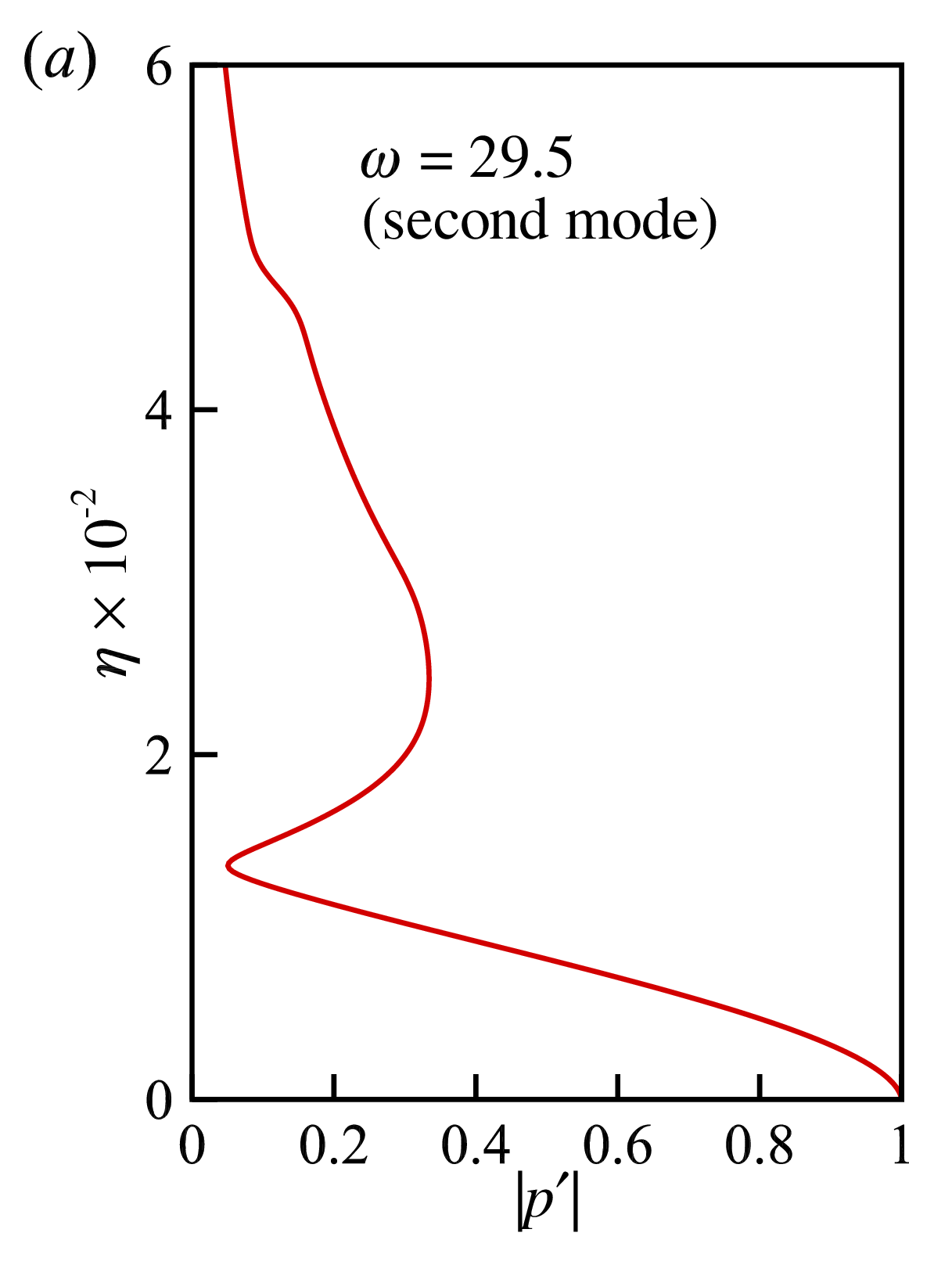}}
	\centering{ \includegraphics[height=5.2cm]{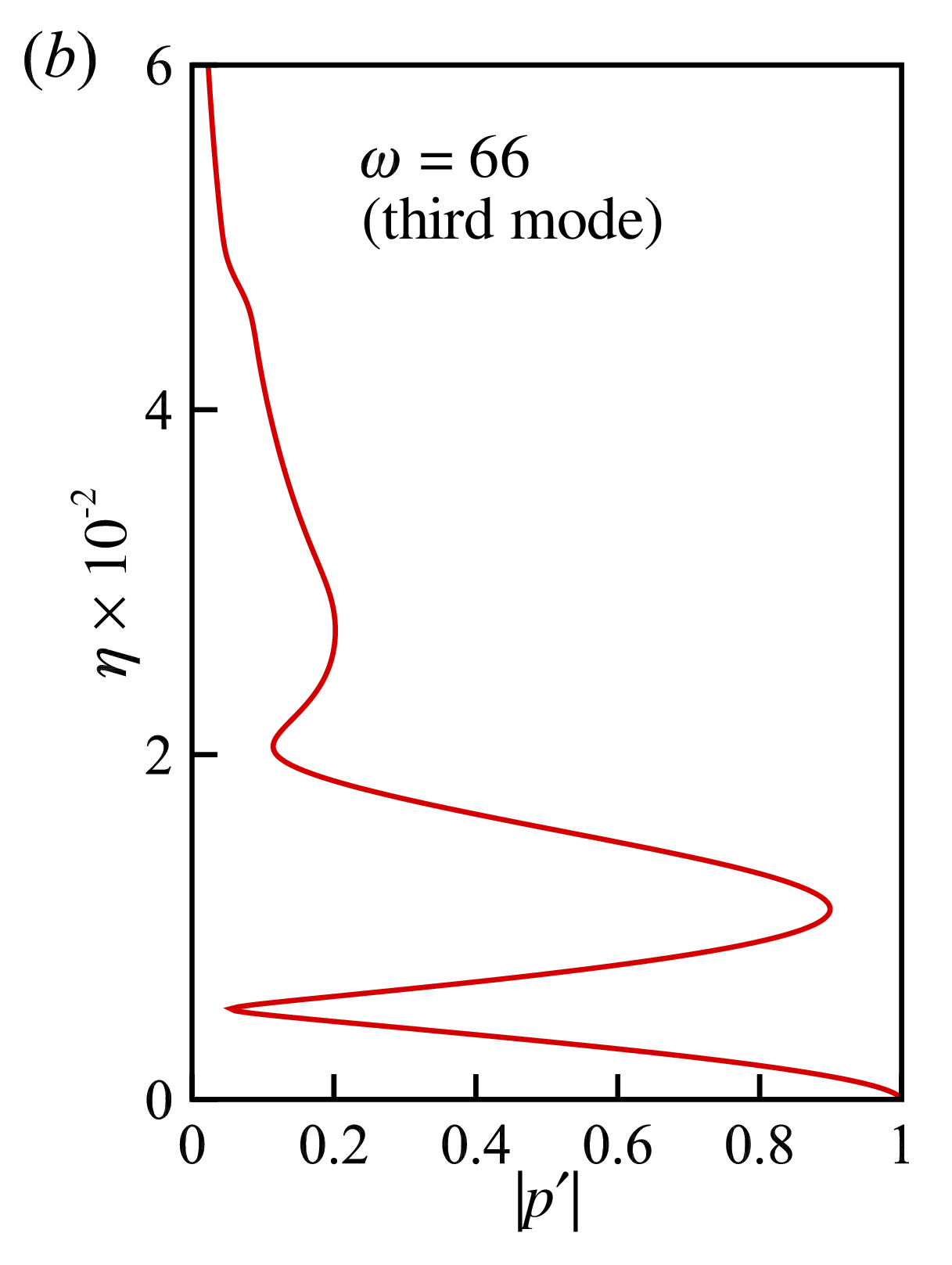}}
	\centering{ \includegraphics[height=5.2cm]{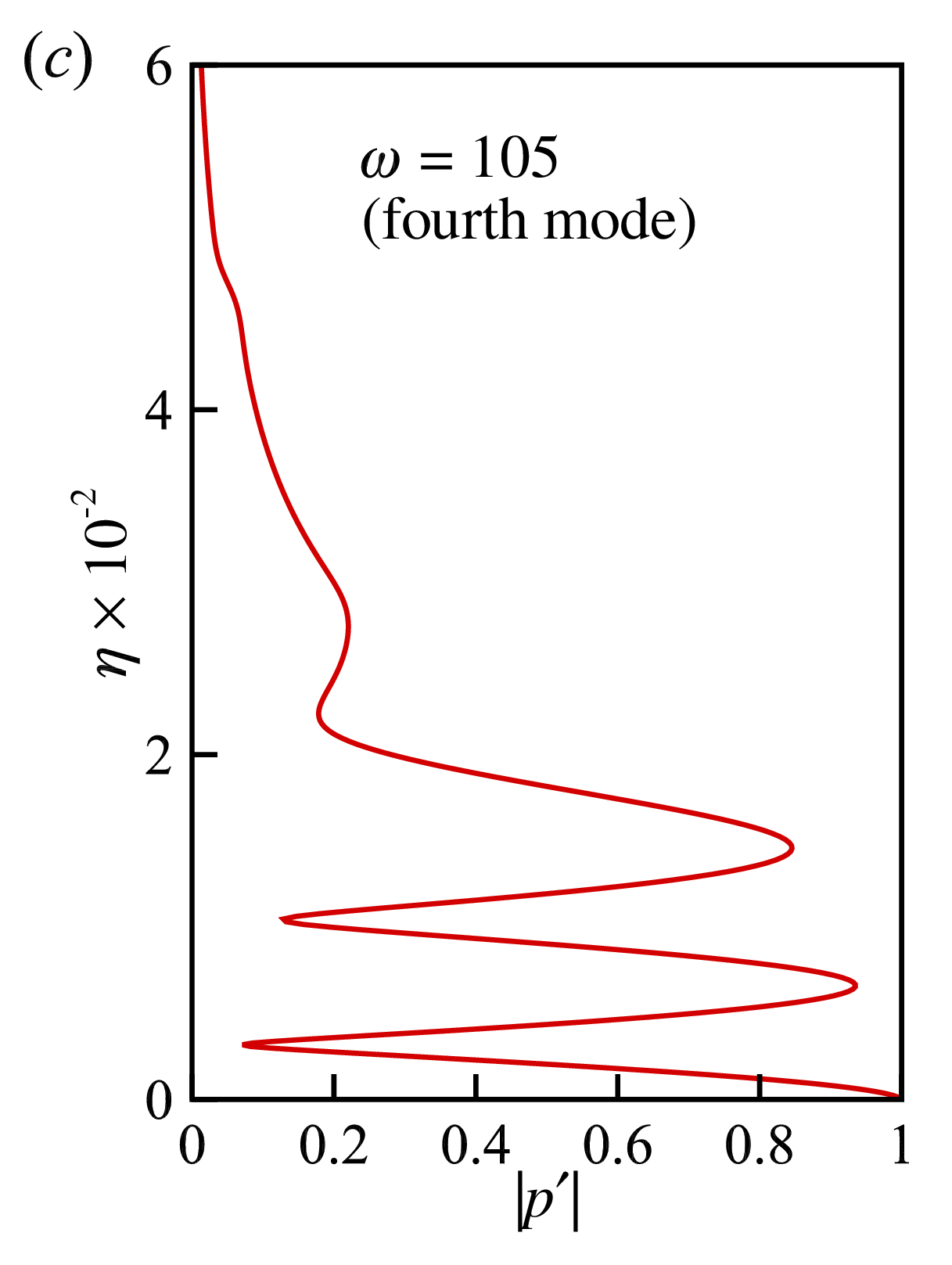}}
	\centering{ \includegraphics[height=5.2cm]{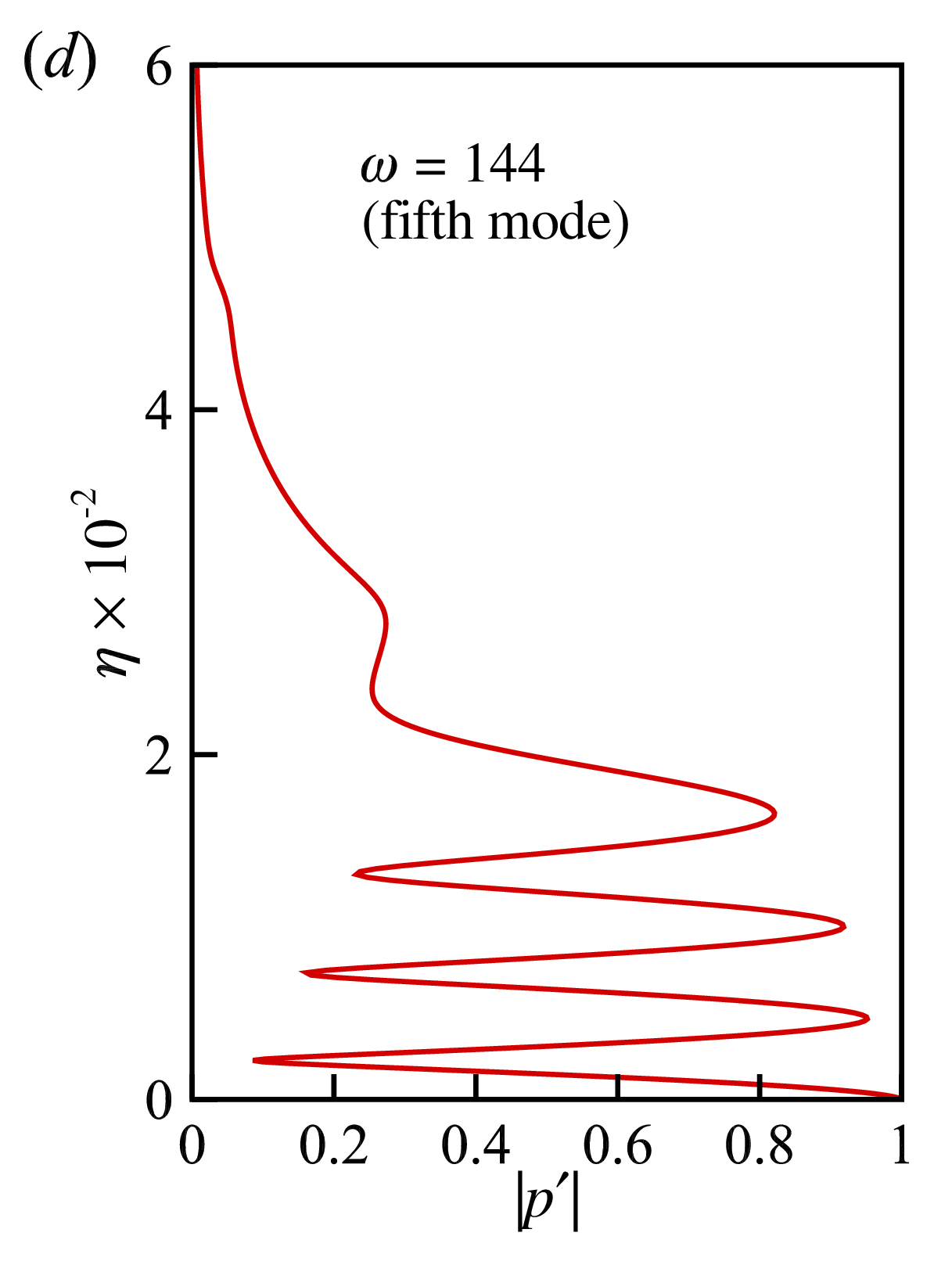}}
	\caption{\added{Moduli of pressure eigenfunctions of the most unstable modes via LST with different frequencies for $x=0.99$ and $\beta=0$ in Fig. \ref{fig5}(\textit{b}). No without corner rounding is applied.}}
	\label{Appendix_LST}
\end{figure}

\added{For clarification purposes, the moduli of pressure eigenfunctions of Mack second, third, fourth and fifth modes are provided in Fig. \ref{Appendix_LST}. The streamwise location is fixed at $x=0.99$ in Fig. \ref{fig5}(\textit{b}). Clearly, the main peak numbers of pressure eigenfunctions agree with the order of Mack modes described in Fig. \ref{fig5}.}

\begin{figure}[htbp]
	\centering{ \includegraphics[height=6cm]{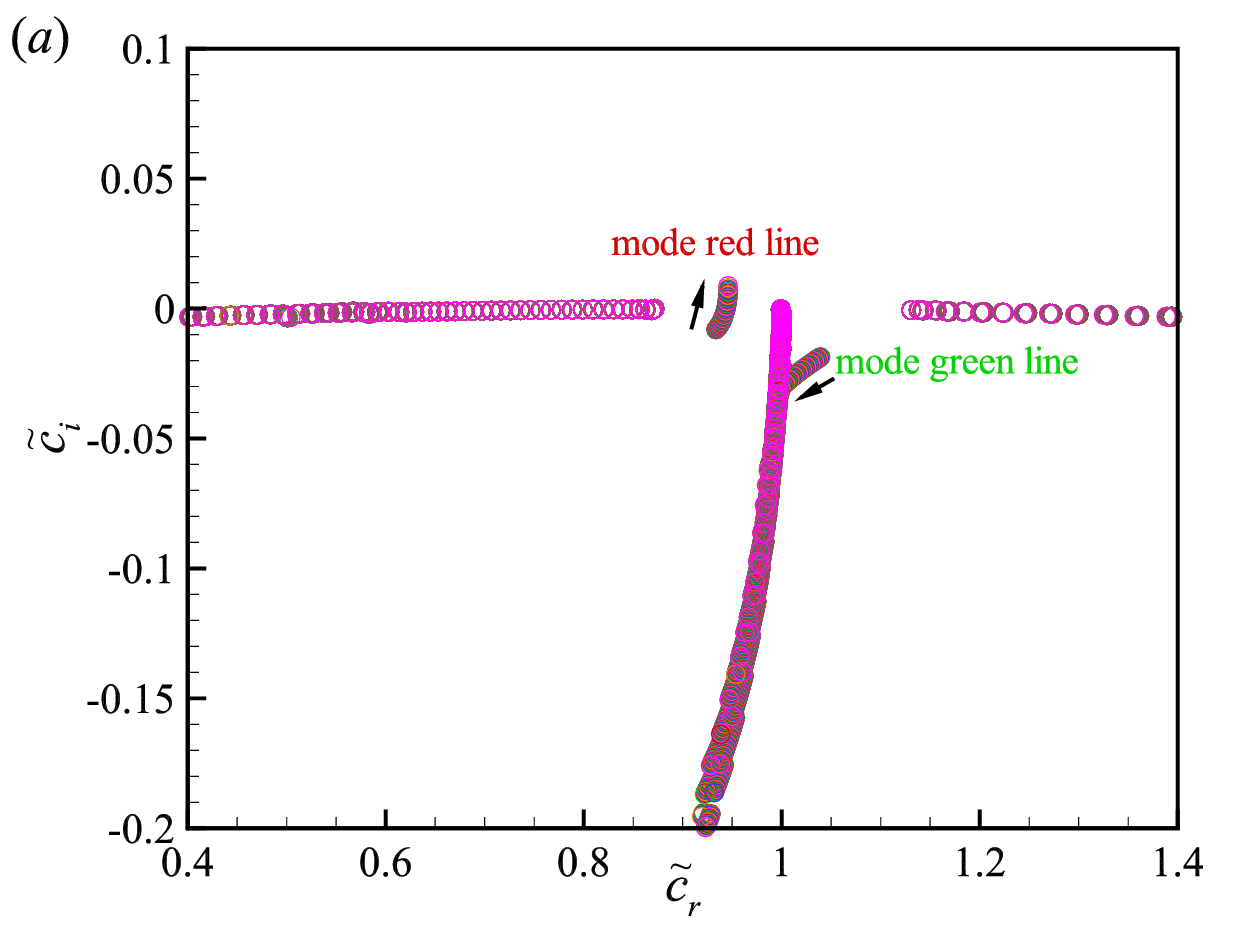}}
	\centering{ \includegraphics[height=6cm]{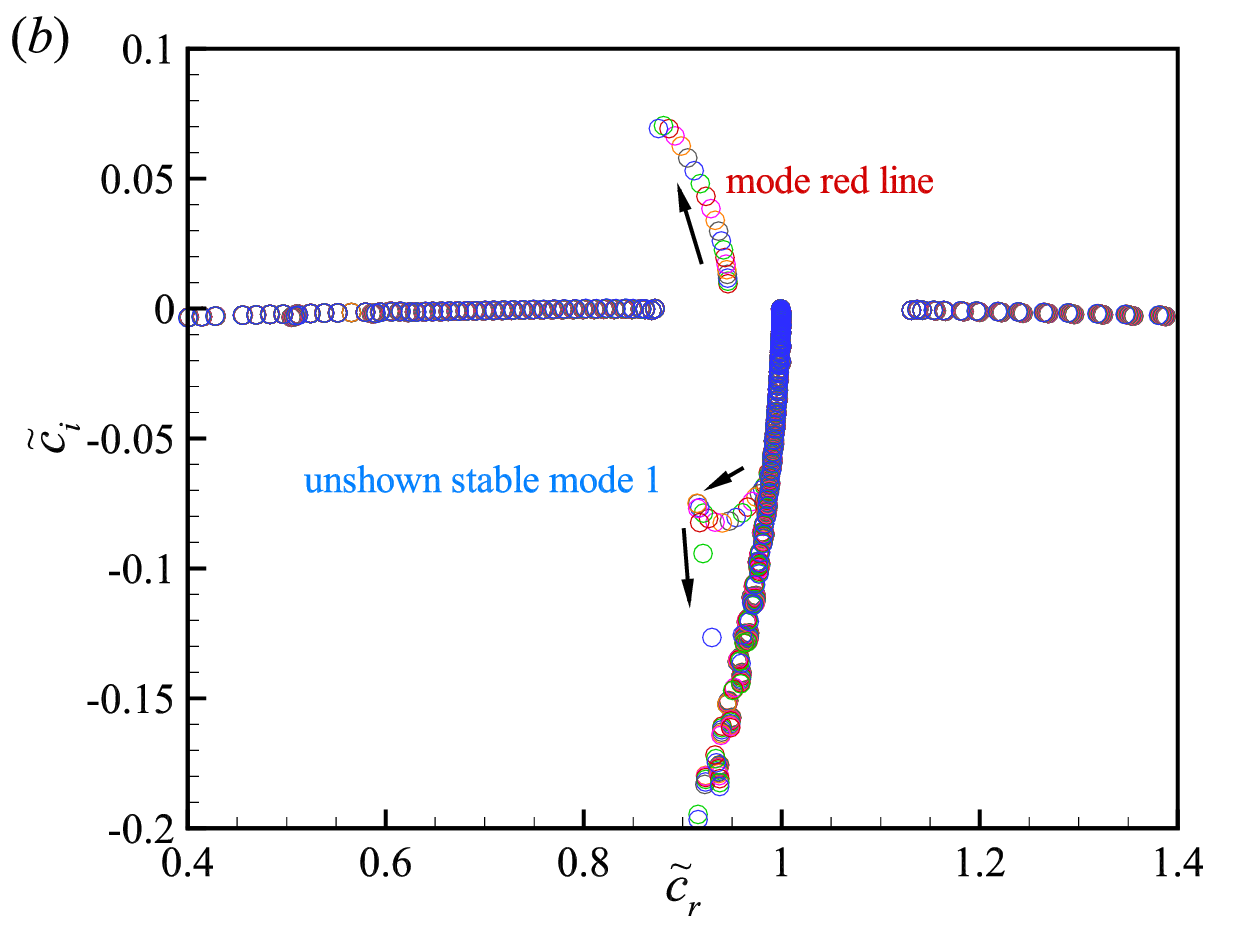}}
	\centering{ \includegraphics[height=6cm]{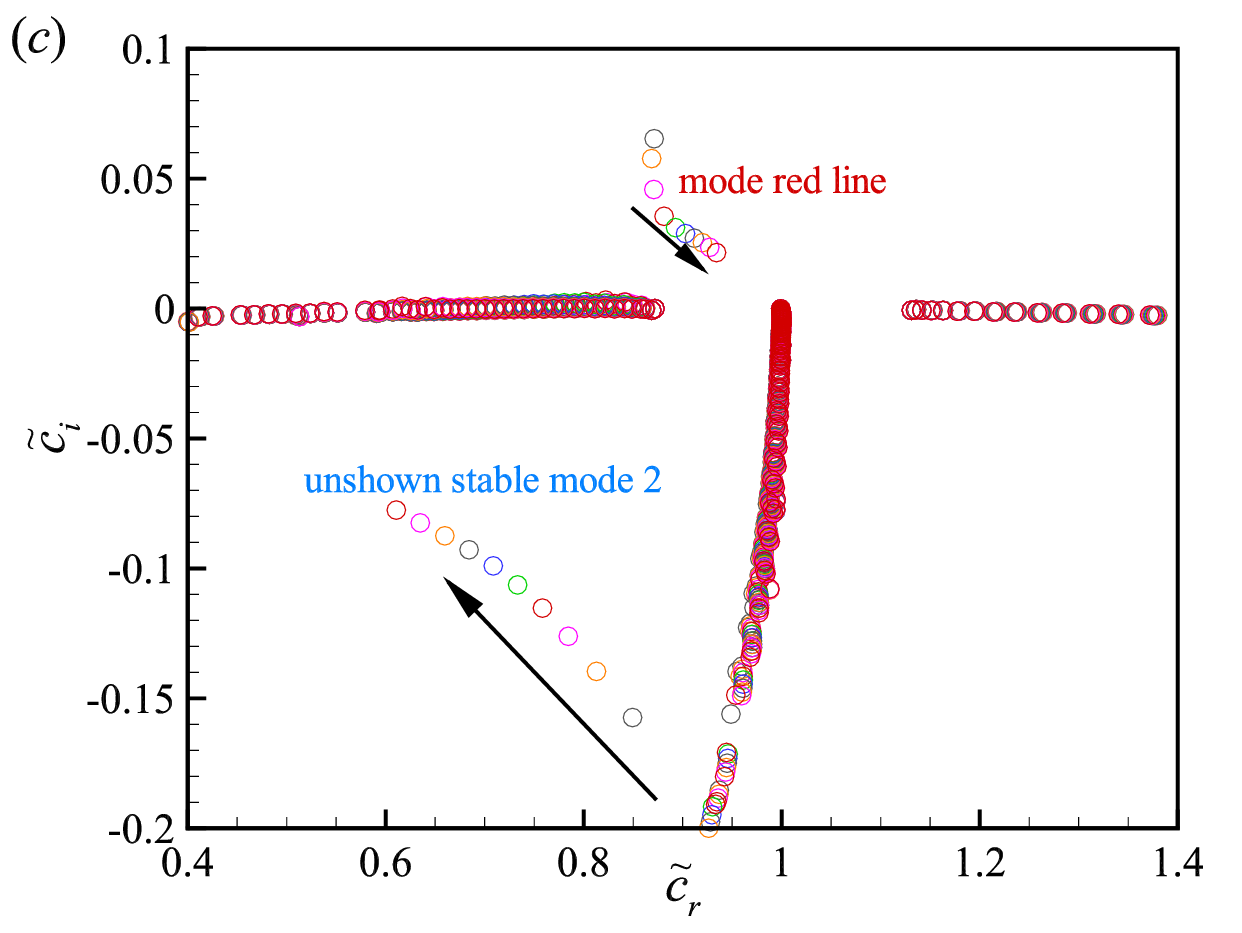}}
	\centering{ \includegraphics[height=6cm]{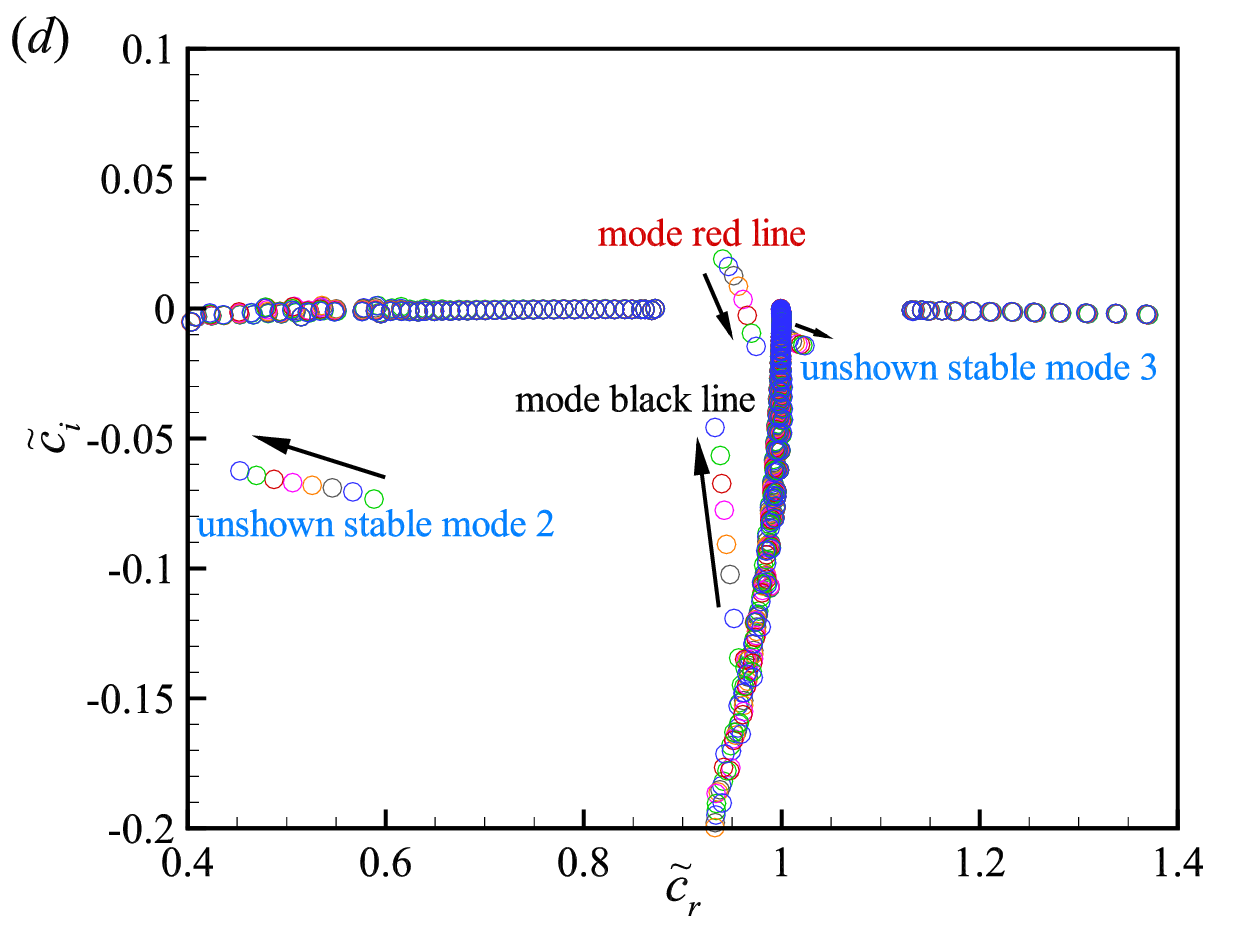}}
	\caption{\added{Entire overlaid eigenspectra corresponding to Figs. \ref{fig7} and \ref{fig8}: (\textit{a}) $x \in [0.61, 0.76]$, (\textit{b}) $x \in (0.76, 0.802]$, (\textit{c}) $x \in (0.802, 0.82]$, and (\textit{d}) $x \in (0.82, 0.834]$. Arrows mark the directions of eigenvalue trajectories with an increasing $x$.}}
	\label{Appendix_fig-spectra}
\end{figure}

\added{In addition, to clarify how mode S$^-_{\textit{new}}$ emerges, the entire eigenspectra from $x = 0.61$ to $x = 0.834$ are overlaid. As shown in Fig. \ref{Appendix_fig-spectra}, we divide the results into four stages: (\textit{a}) $x \in [0.61, 0.76]$, (\textit{b}) $x \in (0.76, 0.802]$, (\textit{c}) $x \in (0.802, 0.82]$, and (\textit{d}) $x \in (0.82, 0.834]$. Obviously, the eigenspectra become more complicated in the separation bubble. In Fig. \ref{Appendix_fig-spectra}(\textit{a}) before separation, two discrete modes are identified, i.e. mode in green line with $c_r > 1$ and mode in red line with $c_r < 1$ corresponding to Fig. \ref{fig8}. The mode in green line with $c_r > 1$ corresponding to Fig. 8 merges into the continuous spectrum at $x = 0.73$. When $x \in [0.73, 0.76]$, there is no discrete mode with $c_r > 1$. The mode in red line gradually becomes unstable in this stage. In Fig. \ref{Appendix_fig-spectra}(\textit{b}) immediately after separation, the unstable mode in red line still exists, while another stable mode 1 detaches from and then merges into the continuous spectrum. This stable mode 1 is not shown in the main text since it is unimportant. In Fig. \ref{Appendix_fig-spectra}(\textit{c}), in addition to the unstable mode in red line, another unshown stable mode 2 appears. In Fig. \ref{Appendix_fig-spectra}(\textit{d}), in addition to the unshown stable mode 2 and the unstable mode in red line, mode in black line emerges and becomes less and less stable. Another unshown stable mode 3 is also depicted. Finally, the mode in black line moves into the unstable upper half plane, which is the so-called mode S$^-_{\textit{new}}$.}
	
\added{In fact, there are multiple stable modes departing from the continuous spectrum, which emerge just downstream of the separation point. However, most of them do not become unstable in the whole stage, and they are not important. The discussed mode S$^-_{\textit{new}}$ becomes unstable finally, and thus it is concerned in the discussion. Moreover, mode S$^-_{\textit{new}}$ directly departs from the continuous spectrum without obvious connections with discrete modes upstream of the separation. The location where mode S$^-_{\textit{new}}$ appears is $x = 0.82$, and the location where the upstream mode in green line disappears is $x = 0.73$. The distance between the two locations is not small. In our view, the departure of mode S$^-_{\textit{new}}$ from the continuous spectrum may be conceivable, because the linear stability of the separated flow is not analogous to that in the attached boundary layer.}

\newpage
\section*{Funding Sources}

This research is supported by the Hong Kong Research Grants Council (no. 15216621, no. 15217622, no. 15204322 and no. 25203721) and the National Natural Science Foundation of China (no. 12102377)\added{ and the Start-up Fund for RAPs by the Hong Kong Polytechnic University}. 


\bibliography{sample}

\end{document}